\documentclass{aa}
\usepackage{txfonts}
\usepackage{amsmath}
\usepackage{graphicx}
\usepackage{stfloats}
\usepackage{natbib}
\bibpunct{(}{)}{;}{a}{}{,} 
\usepackage{rotating}
\usepackage{pdflscape}
\usepackage{booktabs}
\usepackage{longtable}
\usepackage{float}
\usepackage{makecell}
\usepackage{siunitx}
\usepackage{afterpage}
\usepackage{color}
\usepackage[usenames, dvipsnames]{xcolor}

\usepackage{hyperref}
\hypersetup{
        colorlinks=true,
        breaklinks=true,%
        linkcolor=blue,
        citecolor=blue,
        filecolor=blue,
        urlcolor=blue,%
        unicode=false,
        pdftoolbar=true,
        pdfmenubar=true,
        pdffitwindow=false,
        pdfstartview={Fit},
        pdftitle={Revisited catalog of YSOs in Orion\,A},
        pdfauthor={Josefa},
        pdfsubject={OrionA},%
        pdfcreator={Josefa},
        pdfkeywords={},%
        pdfnewwindow=true,
        pdfdisplaydoctitle=true
}

\begin{document}

\title{VISION - Vienna survey in Orion\thanks{Full Table~C.1 is only available in electronic form at the CDS via anonymous ftp to cdsarc.u-strasbg.fr (number) or via http://cdsarc.u-strasbg.fr/viz-bin/qcat?J/A+A/vol/page}}
\subtitle{III. Young stellar objects in Orion\,A}
\titlerunning{Young stellar objects in Orion\,A}

\author{Josefa~Elisabeth~Gro\ss schedl\inst{1}
          \and
          Jo\~ao~Alves\inst{1,2,3}
          \and
          Paula~S.~Teixeira\inst{1,4,5}
          \and
          Herv\'e~Bouy\inst{6} 
          \and
          Jan~Forbrich\inst{1,7,8} 
          \and
          Charles~J.~Lada\inst{8}
          \and
          Stefan~Meingast\inst{1}
          \and 
          \'Alvaro~Hacar\inst{1,9} 
          \and
          Joana~Ascenso\inst{10} 
          \and 
          Christine~Ackerl\inst{1}
          \and 
          Birgit~Hasenberger\inst{1}
          \and
          Rainer~K\"ohler\inst{1}
          \and
          Karolina~Kubiak\inst{1}
          \and 
          Irati~Larreina\inst{1}
          \and
          Lorenz~Linhardt\inst{11} 
          \and
          Marco~Lombardi\inst{12} 
          \and 
          Torsten~M\"oller\inst{11,13}
        }%
  
   \authorrunning{J.~Gro\ss schedl}

   \institute{
            Universit\"at Wien, Institut f\"ur Astrophysik, T\"urkenschanzstrasse 17, 1180 Wien, Austria, \\
            \email{josefa.elisabeth.grossschedl@univie.ac.at}
        \and
		University of Vienna, Faculty of Earth Sciences, Geography and Astronomy, Data Science @ Uni Vienna
	\and
		Radcliffe Institute for Advanced Study, Harvard University, 10 Garden Street, Cambridge, MA 02138, USA
	\and
            Institut de Ci\`{e}ncies del Cosmos (ICCUB), Universitat de Barcelona (IEEC-UB), Mart\'{i} i Franqu\`{e}s 1, 08028 Barcelona, Spain
        \and
            Scottish Universities Physics Alliance (SUPA), School of Physics and Astronomy, University of St.\,Andrews, North Haugh, St.\,Andrews, Fife KY16 9SS, UK
        \and
            Laboratoire d`Astrophysique de Bordeaux, Universit\'e de Bordeaux, All\'ee Geoffroy Saint-Hilaire, CS 50023, 33615 PESSAC CEDEX, France
        \and
            Centre for Astrophysics Research, University of Hertfordshire, College Lane, Hatfield AL10 9AB, Hertfordshire, UK
        \and
            Harvard-Smithsonian Center for Astrophysics, 60 Garden St., Cambridge, MA 02138, USA 
        \and
            Leiden Observatory, Leiden University, P.\,O.\,Box 9513, 2300-RA Leiden, The Netherlands 
        \and
            Universidade do Porto, Dep.~de Engenharia F\'isica da Faculdade de Engenharia, Rua Dr.\,Roberto Frias, P-4200-465, Porto, Portugal
        \and    
            Universit\"at Wien, Fakult\"at f\"ur Informatik, W\"ahringer Stra\ss e 29/S6, 1090 Wien, Austria 
        \and
            University of Milan, Department of Physics, via Celoria 16, 20133 Milan, Italy 
        \and
        	    	University of Vienna, Faculty of Computer Science, Data Science @ Uni Vienna
             }%

\date{Received January 3, 2018; accepted September 27, 2018; updated \today}

\abstract{We have extended and refined the existing young stellar object (YSO) catalogs for the Orion\,A molecular cloud, the closest massive star-forming region to Earth. This updated catalog is driven by the large spatial coverage ($\SI{18.3}{deg^2}$, $\SI{\sim950}{pc^2}$), seeing limited resolution ($\SI{\sim0.7}{\arcsec}$), and sensitivity ($K_s<\SI{19}{mag}$) of the ESO-VISTA near-infrared survey of the Orion\,A cloud (VISION).
Combined with archival mid- to far-infrared data, the VISTA data allow for a refined and more robust source selection. We estimate that among previously known protostars and pre-main-sequence stars with disks, source contamination levels (false positives) are at least $\sim$6.4\% and $\sim$2.3\%, respectively, mostly due to background galaxies and nebulosities. 
We identify 274 new YSO candidates using VISTA/\emph{Spitzer} based selections within previously analyzed regions, and VISTA/\emph{WISE} based selections to add sources in the surroundings, beyond previously analyzed regions. The \emph{WISE} selection method recovers about 59\% of the known YSOs in Orion\,A's low-mass star-forming part L1641, which shows what can be achieved by the all-sky \emph{WISE} survey in combination with deep near-infrared data in regions without the influence of massive stars.
The new catalog contains 2980 YSOs, which were classified based on the de-reddened mid-infrared spectral index into 188 protostars, 185 flat-spectrum sources, and 2607 pre-main-sequence stars with circumstellar disks. 
We find a statistically significant difference in the spatial distribution of the three evolutionary classes with respect to regions of high dust column-density, confirming that flat-spectrum sources are at a younger evolutionary phase compared to Class\,IIs, and are not a sub-sample seen at particular viewing angles. 
}

\maketitle


\section{Introduction} \label{Introduction}

\defcitealias{Meingast2016}{Paper~I}
\defcitealias{Furlan2016}{FFA16}
\defcitealias{Lewis2016}{LL16}
\defcitealias{Greene1994}{Greene et al.}

It is well established that star formation takes place at the coldest and densest regions of molecular clouds. With the development of infrared (IR) and millimeter facilities in recent decades, it was possible to image the early stages of the star formation process. Describing the new observables, however, is not a straightforward task, and attempts of classifying young stellar objects (YSOs) \citep[e.g.,][]{Greene1994, Evans2009} and deriving an evolutionary path from a dense core to a YSO have been plagued with uncertainties. These are mostly due to the limited sensitivity and resolution of the observations and the intrinsic complexity of the star formation process. For example, objects of similar mass can have very different observables due to the large diversity of an YSO environment and its geometry alone  \citep[e.g.,][]{Whitney2013}. In other words, it is often difficult to establish an evolutionary stage for single sources. However, one can also look at entire populations to statistically infer evolutionary properties. This is now possible with the recent deployment of several space-based and ground-based IR telescopes that observed most nearby ($<\SI{500}{pc}$) star-forming regions \citep[e.g.,][]{Evans2009, Megeath2012, Dunham2015}.

To understand and reconstruct the star formation process it is crucial to know the YSOs evolutionary stages. First attempts to classify YSOs into three evolutionary classes (I, II, III) were presented in the 80s \citep[e.g.,][]{Lada1984, Lada1987}, based on the finding that dusty envelopes and circumstellar disks cause an IR excess. These classes constitute a smooth evolutionary sequence according to the observed IR spectral energy distribution (SED), where the spectral index $\alpha$ was defined as a linear fit to the photometric near- (NIR) to mid-infrared (MIR) SED in log-space
\begin{equation}
\alpha = \frac{d \log (\lambda F_{\lambda})}{d \log \lambda},
\end{equation}
used to estimate the evolutionary Stage\footnote{Class is used for the observed SED classification, while Stage refers to the physical configuration.} \citep[e.g.,][]{Robitaille2006}. In the 90s five YSO Classes were established (0, I, flat-spectrum, II, III) \citep[e.g.,][]{Greene1994}, which are thought to be connected to the true evolutionary Stage as follows:
Class\,0 sources \citep{Andre1993} are protostars in the very early collapse phase with low blackbody temperatures ($T_\mathrm{bol}<\SI{70}{K}$), and with envelope masses that still dominate the system. They are mostly not detectable in the NIR or MIR and usually require observations at longer wavelengths.
Class\,I YSOs ($\alpha \gtrsim 0$)  are protostars (P) which are still embedded and accreting material from a surrounding envelope onto a forming circumstellar disk.
Class~II YSOs ($\alpha \lesssim 0$) are pre-main-sequence (PMS) stars surrounded by dusty circumstellar disks (D), which have dispersed their envelopes (also called T-Tauri stars). 
Finally, Class~III YSOs are likely evolved PMS stars that emerge when accretion ends and the disks dissipate by stellar radiation or winds \citep[e.g.,][]{Pillitteri2013}. They show only very little ($\alpha \lesssim -1.6$) or no IR-excess ($\alpha \lesssim -2.5$). When using selection criteria based on IR photometry, only the part of Class\,IIIs with IR-excess can be identified. 

\citet{Greene1994} introduced the flat-spectrum class (hereafter also referred to as flats), lying between Classes\,I and II with $\alpha \approx 0$. This class represents the YSOs that are not easily assignable to either protostars or disks\footnote{YSO classes are also called for simplicity: Class\,0/I - protostars (P), flat-spectrum sources - flats (F), and Class\,II/III - disks (D). Disks include Class\,IIs and the part of Class\,IIIs with IR-excess. See Tabel~\ref{tab:class2}.}, and it is not clear if they are simply a mixture of or a transitional phase between these two. Therefore, \citetalias{Greene1994} assigned them an uncertain evolutionary status. There are several reasons for this. 
Firstly, the shape of the SED can be influenced by geometric effects, like disk inclination along the line of sight to the observer \citep{Whitney2003a, Whitney2003b, Robitaille2006, Crapsi2008, Whitney2013}, or by high foreground extinction \citep{Muench2007, Forbrich2010}. For example, an evolved protostar with an almost depleted envelope or viewed pole-on, and a Class\,II source where the disk is viewed edge-on or the source is seen through high extinction, may show a similar flat-spectrum SED \citep{Whitney2003b}. On the other hand, there are several studies suggesting a younger physical stage of flats compared to Class\,IIs. \citet{Muench2007} point out, that flat-spectrum sources are considered to be protostars in a later stage of envelope dispersal or with highly flared disks. Moreover, they find that flat-spectrum sources are intrinsically more luminous than Class\,IIs, suggesting a different evolutionary stage. \citet{Greene2002}, using NIR spectroscopy, find that accretion rates of flat-spectrum sources lie in between those of Classes\,I and II (inferred from the veiling excess), suggesting a transitional evolutionary stage.
Recently, \citet{Furlan2016} find, based on SED modeling including FIR photometry, that the large majority of their studied sample of flat-spectrum sources require an envelope in their fit, indicating that these objects are still in the protostellar phase, covering different stages in their envelope evolution. At the same time, \citet{Carney2016} conclude from a molecular line study that about 30\% of previously identified Class\,I sources were more evolved Stage\,II YSOs. A similar situation was pointed out by \citet{Heiderman2015}, who find that only about 50\% of flat-spectrum sources are surrounded by envelopes. \citet{Furlan2016} point out, the differences in their findings could be due to different methods to select flats.
Indeed, different conventions do not provide easily comparable samples. The differences are driven by available photometry, the chosen spectral range to construct the spectral index, different class definitions, or even if extinction correction is applied or not \citep[e.g.,][]{Lada1987, Greene1994, Lada2006, Muench2007, Evans2009, Teixeira2012}. Until all the points above are carefully addressed for a large statistical significant sample, the nature of flat-spectrum sources will remain undetermined.

\begin{figure*}[!ht]
	\centering
	\includegraphics[ width=0.95\linewidth]{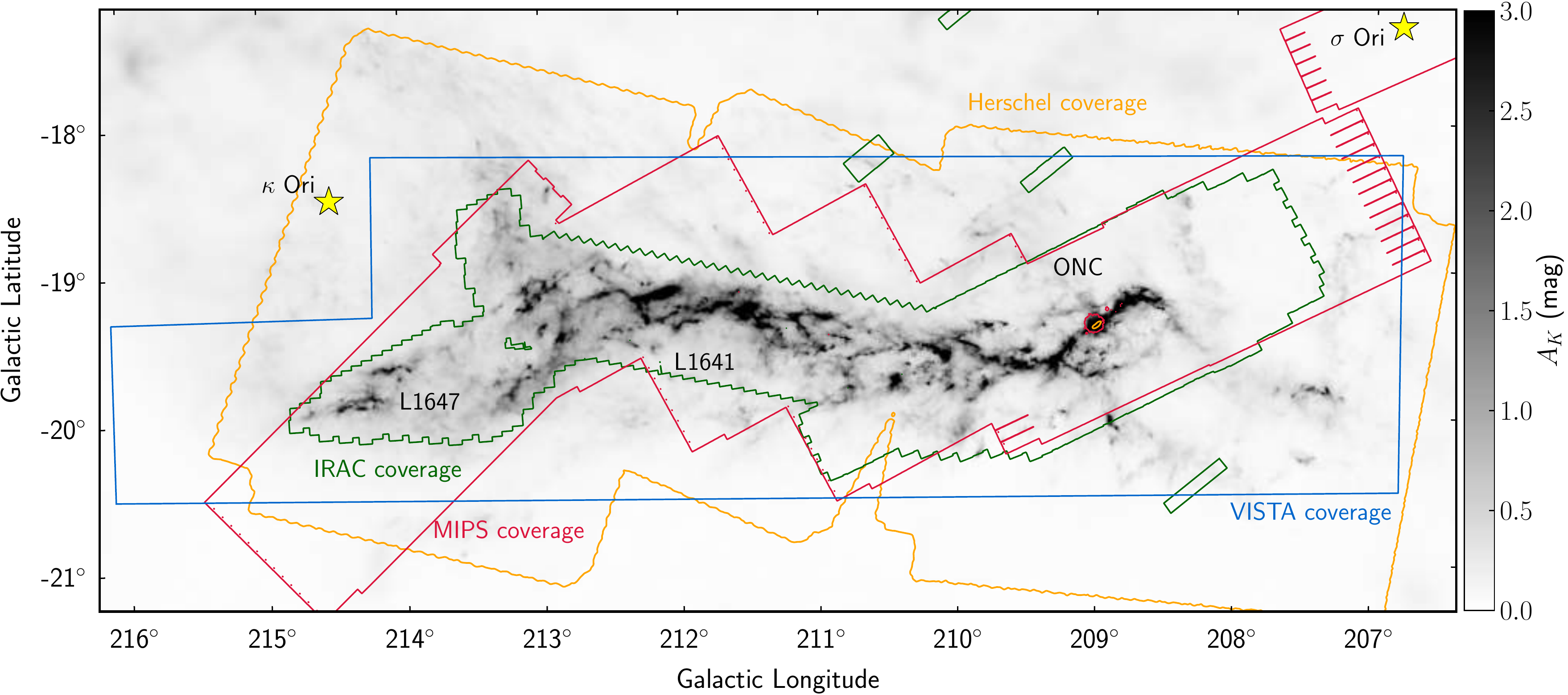} 
	\caption{Survey footprints displayed on top of the \emph{Planck-Herschel-Extinction} dust column-density map \citep{Lombardi2014}. The grayscale shows the line of sight extinction, given in $A_\mathrm{K,Herschel}$\,(mag). 
	The contours show the survey coverages for VISTA (blue), \emph{Spitzer}/IRAC1 (green), \emph{Spitzer}/MIPS1 (red), and \emph{Herschel} (orange). The small green boxes are control fields for the \emph{Spitzer}/IRAC bands, which are partially overlapping with VISTA. The red and orange circles, located at the position of the ONC, result from saturated MIPS1 and \emph{Herschel} photometry respectively. Indicated are the two main regions in Orion\,A, the Lynds dark cloud L1641 (including L1647), and the ONC region. The B-star $\kappa$-Ori and the O-star $\sigma$-Ori are marked for orientation. See also figure~1 in \citetalias{Meingast2016} for an overview of some sub-regions.}
	\label{fig:coverage}
\end{figure*}

The goal of this paper is twofold: (a) construct the most complete catalog of dusty YSOs in the Orion\,A giant molecular cloud, and (b) use it to infer on the nature of flat-spectrum sources. To achieve this we make use of the deep seeing-limited NIR VISTA photometry from the VIenna Survey In OrioN \citep[VISION,][hereafter, Paper~I]{Meingast2016} to improve on previous YSO catalogs \citep{Megeath2012,Megeath2016,Furlan2016}. Since the VISTA survey-area is larger than previously analyzed regions (Fig.~\ref{fig:coverage}), we will look for new YSO candidates in the surroundings, to improve the spatial completeness of this rich YSO sample. This we do in combination with MIR data from the Wide-field Infrared Survey Explorer \citep[\emph{WISE},][]{Wright2010} and the \emph{Spitzer} Space Telescope \citep{Werner2004}, and with FIR data from the \emph{Herschel} space observatory\footnote{\emph{Herschel} is an ESA space observatory with science instruments provided by European-led Principal Investigator consortia and with important participation from NASA.} \citep{Pilbratt2010}. Our analysis, using IR photometry, will not be sensitive to the majority of Class\,IIIs \citep[see e.g.,][]{Pillitteri2013} and we ignore PMS stars without IR-excess in this paper. Future work should consider the whole young stellar population to determine the complete star-forming history within the cloud.

The structure of the paper is as follows. In Sect.~\ref{Data} we describe the data and give a brief overview on recent Orion\,A YSO catalogs. In Sect.~\ref{Methods} we first present our methods to classify the YSOs (Sect.~\ref{classification}) and second, we discuss our methods to evaluate the contamination (false positives) of the known YSO population (Sect.~\ref{revisit}). Example images of these are presented in Appendix~\ref{cutouts}. Finally, we present our methods to select new YSO candidates (Sect.~\ref{NewYSOs}), with a detailed description of the selection methods given in Appendix~\ref{NewYSOsA}. We will classify the YSO candidates based on extinction corrected spectral indices into Class\,I, flat-spectrum, and Class\,II/III sources, with an overview of the resulting updated YSO sample presented in Sect.~\ref{Results}, and the corresponding table in Appendix~\ref{Catalogs}. In Sect.~\ref{Discussion} we discuss the issues that come with YSO classification and we infer on the meaning of the flat-spectrum sources by looking at their spatial distribution with respect to regions of high dust-column density\footnote{Dust column-density, as traced by \emph{Herschel}, is not directly tracing the dense gas. Therefore, the true volume density is unknown.}. Finally, we give a summary in Sect.~\ref{Summary}.


\section{Data} \label{Data}

We use archival IR data and the new deep NIR VISTA data to re-examine the already studied YSO population in Orion\,A \citep{Megeath2012, Megeath2016, Furlan2016, Lewis2016}, and to select new YSO candidates in a larger field covered by VISTA. In Fig.~\ref{fig:coverage} the footprints of the surveys used in this work are shown. The blue VISTA contour is the region investigated in this work ($\SI{\sim18.3}{deg^2}$), while the green \emph{Spitzer}/IRAC region ($\SI{\sim7}{deg^2}$) was investigated by \citet{Megeath2012, Megeath2016}. This improved coverage allows for a spatially more complete sample.
The background image is the \emph{Planck-Herschel-Extinction} dust column-density map from \citet{Lombardi2014} (hereafter, \emph{Herschel map}), with a resolution of \SI{36}{\arcsec}. The \emph{Herschel map} is used to estimate the total line of sight extinction at the position of the YSO candidates, to distinguish between regions of high and low dust column-density.
The dust optical depth was converted by \citet{Lombardi2014} to extinction ($A_\mathrm{K}$) using a 2MASS\footnote{2MASS - The 2 Micron All-Sky Survey \citep{Skrutskie2006}} NIR extinction map \citep{Lombardi2011}. They find a linear conversion factor of $\gamma = 2640$. Based on a recent extinction map, constructed with VISTA and \emph{Spitzer} data, we use an updated conversion factor of $\gamma = 3050$ \citep[][Paper~II]{Meingast2018}. 
Hereafter, we use the abbreviation $A_\mathrm{K,Herschel}$ when referring to extinctions extracted from the \emph{Herschel map}.


\subsection{VISTA near-infrared data} \label{NIR}

In the first paper of this series, introducing VISION \citepalias{Meingast2016}, we obtained deep NIR $J$, $H$, and $K_S$ photometry (see Table~\ref{tab:bands}), covering the entire Orion\,A cloud, using the Visible and Infrared Survey Telescope for Astronomy \citep[VISTA,][]{Emerson2006} operated by the European Southern Observatory (ESO). We gain angular resolution and sensitivity compared to previous NIR surveys (e.g., 2MASS), reaching 90\% completeness limits of 20.4, 19.9 and \SI{19.0}{mag} in $J$, $H$, and $K_S$ respectively. The survey reaches a seeing limited resolution of almost \SI{0.6}{\arcsec} (median seeing of \SI{0.72}{\arcsec}).
Compared to 2MASS, the sensitivity of VISTA goes about $\mathrm{4-5}$ magnitudes deeper, and the resolution improved by about a factor of 3. Therefore, the VISION catalog contains about a factor of ten more sources in the covered area ($\sim$800,000 point-sources). This allows for an improved YSO classification, and a better distinction of background galaxies or extended nebulous IR emission from YSO candidates. To estimate the colors and magnitudes of background and extra-galactic contamination we use the VISTA control field observed during the survey, which is shifted about $+\SI{22}{\degree}$ in Galactic longitudes ($l$) and lies at about the same Galactic latitude ($b$), covering $\SI{\sim1.8}{deg^2}$ in the sky \citepalias[10\% of the science field coverage, see figure~4 in][]{Meingast2016}.

\begin{table}[!ht]
\begin{center} 
\caption{Overview of the used photometric NIR and MIR bands.} 
\label{tab:bands}
\resizebox{\columnwidth}{!}{
\begin{tabular}{lccccc}  
\hline \hline

  \multicolumn{1}{c}{Survey} &
  \multicolumn{1}{c}{Band} &
  \multicolumn{1}{c}{$\lambda$\tablefootmark{a}} &
  \multicolumn{1}{c}{F$_{\nu0}$\tablefootmark{b}} &
  \multicolumn{1}{c}{FWHM\tablefootmark{c}} &
  \multicolumn{1}{c}{$A_{\lambda}/A_{K_S}$\tablefootmark{d}} \\
  
  \multicolumn{1}{c}{} &
  \multicolumn{1}{c}{} &
  \multicolumn{1}{c}{($\SI{}{\micro\meter})$} &
  \multicolumn{1}{c}{(Jy)} &
  \multicolumn{1}{c}{($\SI{}{\arcsec})$} &
  \multicolumn{1}{c}{} \\
  
\hline
VISTA  & $J$   & 1.25 &           1594.0\phantom{000} & \phantom{0}0.78 & 2.50   \\ 
(1)    & $H$   & 1.65 &           1024.0\phantom{000} & \phantom{0}0.75 & 1.55  \\
       & $K_S$ & 2.15 & \phantom{0}666.7\phantom{000} & \phantom{0}0.8\phantom{0}  & 1.00   \\
\hline
\emph{Spitzer}  & $I1$  & \phantom{0}3.6 & \phantom{0}280.9\phantom{000} & \phantom{0}1.66 & 0.64    \\
IRAC \& MIPS    & $I2$  & \phantom{0}4.5 & \phantom{0}179.7\phantom{000} & \phantom{0}1.72 & 0.56     \\
(2)             & $I3$  & \phantom{0}5.8 & \phantom{0}115.0\phantom{000} & \phantom{0}1.88 & 0.50    \\
                & $I4$  & \phantom{0}8.0 & \phantom{00}64.9\phantom{000} & \phantom{0}1.98 & 0.51   \\
                & $M1$  &           24.0 & \phantom{000}7.17\phantom{00} & \phantom{0}6\phantom{.00}  & 0.45   \\
\hline  
\emph{WISE}  & $W1$  & \phantom{0}3.4 & \phantom{0}309.540\phantom{0} & \phantom{0}6.1\phantom{0}  & 0.79   \\
(3)          & $W2$  & \phantom{0}4.6 & \phantom{0}171.787\phantom{0} & \phantom{0}6.4\phantom{0}  & 0.55 \\
             & $W3$  &          12.0  & \phantom{00}31.674\phantom{0} & \phantom{0}6.5\phantom{0}  & 0.61 \\
             & $W4$  &          22.0  & \phantom{000}8.363\phantom{0} &           12.0\phantom{0}  & 0.43 \\
\hline 
\end{tabular}}

    \tablefoot{
        \tablefoottext{a}{Central wavelength.}
        \tablefoottext{b}{Zero magnitude flux density.}
        \tablefoottext{c}{Mean image quality.}
        \tablefoottext{d}{The extinction laws $A_{\lambda}/A_{K_S}$ for VISTA and \emph{Spitzer} are taken from \citet{Meingast2018}, and for \emph{WISE} they are provided by S.~Meingast.}
              }
  \tablebib{
  (1) \citet{Meingast2016};
  (2) IRAC Instrument Handbook (2015) \url{http://irsa.ipac.caltech.edu/data/SPITZER/docs/irac/iracinstrumenthandbook}; 
  MIPS Instrument Handbook (2011) \url{http://irsa.ipac.caltech.edu/data/SPITZER/docs/mips/mipsinstrumenthandbook};
  (3) \citet{Cutri2013}
  }
\end{center}
\end{table}


\subsection{Mid- to far-infrared data, and existing Orion\,A YSO catalogs} 
\label{MIR}

\begin{table*}[!ht] 
\begin{center} 
\small
\caption{Comparison of the MGM and \citetalias{Furlan2016} YSO classification.}
\begin{tabular}{lccccccccc}
\hline \hline

  \multicolumn{2}{c}{} &
  \multicolumn{7}{c}{\citetalias{Furlan2016} (HOPS sources)\tablefootmark{a}} &
  \multicolumn{1}{c}{} \\
  
  \multicolumn{2}{c}{} &
  \multicolumn{1}{c}{ALL} &
  \multicolumn{1}{c}{Class\,0} &
  \multicolumn{1}{c}{Class\,I} &
  \multicolumn{1}{c}{Flats} &
  \multicolumn{1}{c}{Class\,II} &
  \multicolumn{1}{c}{Galaxies} &
  \multicolumn{1}{c}{Uncertain} &
  \multicolumn{1}{c}{} \\
\cmidrule(lr){3-9}
   &  & 309 (278)\tablefootmark{d} &  60 (60) & 103 (93) & 104 (88) & 16 (11) & 22 (22) & 4 (4) &  \\
\cmidrule(lr){3-9}
  \multicolumn{2}{c}{MGM Class\tablefootmark{c}} &
  \multicolumn{7}{c}{number of sources in both catalogs\tablefootmark{b}} &
  \multicolumn{1}{c}{not in \citetalias{Furlan2016}\tablefootmark{f}} \\
\hline
  P  & \phantom{0}330 &         235             (223) & 47 & 90 (87) & 86 (77)&  1 (1)  & 10 & 1   & 95   \\
  FP & \phantom{00}49 & \phantom{0}13 \phantom{00}(8) & 2  &  3 (1)  & 4 (2)  &  1 (0)  & 3  & --- & 36   \\
  RP & \phantom{000}6 & \phantom{00}6 \phantom{00}(5) & 2  &  1 (0)  & ---    & ---     & 1  & 2   & ---  \\
  D  &           2442 & \phantom{0}39 \phantom{0}(30) & 1  &  6 (5)  & 13 (9) & 14 (10) & 5  & --- & 2403 \\
\hline  
\multicolumn{2}{r}{not in MGM\tablefootmark{e}:} & \phantom{0}16 \phantom{0}(12) & 8 & 3 (0) & 1 (0) & --- & 3 & 1 &  \\
\hline
\end{tabular} 
\label{tab:MegFur2}

    \tablefoot{
        \tablefoottext{a}{The top row shows the \citetalias{Furlan2016} classification, including their extra-galactic and uncertain candidates.}
        \tablefoottext{b}{In the middle, the number of overlapping sources of the two catalogs is listed for each sample.}
        \tablefoottext{c}{The first column shows the MGM classification: protostar (P), faint protostar (FP), red protostar (RP), and disk candidate (D).}
        \tablefoottext{d}{The first number are all listed HOPS sources in \citetalias{Furlan2016}, and the second number in paranthesis are the sources where SED modeling was applied for sources with sufficient PACS photometry.}
        \tablefoottext{e}{The last row lists the number of sources that are only in \citetalias{Furlan2016} and not in MGM.}
        \tablefoottext{f}{The last column lists the number of sources that are only in MGM but not in \citetalias{Furlan2016}.}
              }
\end{center}
\end{table*}

\citet{Megeath2012} have carried out a comprehensive study of the dusty young stellar population in Orion\,A, presenting a sample of 2818 YSO candidates with IR-excess. The catalog was slightly updated by \citet{Megeath2016} to a new sample of 2827 candidates, which we call hereafter simply MGM sample. They obtained \emph{Spitzer} MIR photometry, using the Infra-Red Array Camera \citep[IRAC,][]{Fazio2004}, and the Multiband Imaging Photometer for \emph{Spitzer} \citep[MIPS,][]{Rieke2004} (see Table~\ref{tab:bands}). The MGM selection is based on eight band color-color and color-magnitude diagram selections (including 2MASS), also described in \citet{Megeath2009, Gutermuth2009}, and \citet{Kryukova2012}. The 2827 YSO candidates are separated into protostars (P, RP, FP, 385) and disk dominated PMS stars (D, 2442), which roughly correspond to Class\,I and Class\,II YSO candidates. They give three sub-samples for protostars; the main protostar candidates (P), red protostar candidates (RP) with only a measurement in $M1$, and faint protostar candidates (FP), while the latter is a more unreliable sample (see Table~\ref{tab:MegFur2}).

In addition, we use FIR data from the \emph{Herschel} Photoconductor Array Camera and Spectrometer \citep[PACS,][]{Poglitsch2010} at 70, 100, and $\SI{160}{\micro\meter}$. \emph{Herschel} observed Orion\,A during the \emph{Herschel Orion Protostar Survey} \citep[HOPS, see also][]{Stanke2010, Fischer2010, Ali2010, Fischer2013, Manoj2013, Stutz2013, Tobin2015}.
\citet{Furlan2016} (hereafter, \citetalias{Furlan2016}) discuss 309 of these HOPS sources in Orion\,A, with 293 (95\%) sources being a sub-sample of the MGM YSOs. Considering the \citetalias{Furlan2016} sample as an update to MGM, there are 2817 YSO candidates in Orion\,A, classified into 60 Class\,0, 234 Class\,I, 104 flat-spectrum, and 2419\,Class\,II sources. An overview and comparison of the two catalogs is shown in Table~\ref{tab:MegFur2}. \citetalias{Furlan2016} classify the sources based on the bolometric temperature and the spectral index from 4 to $\SI{24}{\micro\meter}$ ($\alpha_\mathrm{I2M}$). They perform SED modeling to determine different stellar properties, by combining PACS with \emph{Spitzer} photometry, \emph{Spitzer}/IRS spectra, and APEX 350 and $\SI{870}{\micro\meter}$ data \citep{Stutz2013}. However, modeling was only done for a sub-sample of 278 sources, due to limited PACS photometry for the rest. Out of the total 309 HOPS sources they classify 283 as YSO candidates and the remaining 26 as extra-galactic contamination or uncertain candidates (see Table~\ref{tab:MegFur2}). We further use the \emph{Herschel}/PACS point-source catalog \citep[HPPSC,][]{Marton2017} to look for matches which are not in the HOPS catalog.

To select new YSO candidates in regions beyond \emph{Spitzer}/IRAC (Sect.~\ref{NewYSOs}) we add MIR all-sky photometry from \emph{WISE} \citep[AllWISE data release,][]{Cutri2013}. \emph{WISE} observed in four bands (see Table~\ref{tab:bands}), with the sensitivity limits varying from about 17, 16, 11, to \SI{7}{mag} for $W1$-$4$\footnote{Given as \texttt{w?mpro} in the AllWISE catalog, abbreviated as $W?$ in this work. The ``?'' is used as placeholder for 1, 2, 3, or 4.}. 
The wavelength coverage is similar to \emph{Spitzer} \citep[see Table~\ref{tab:bands} and figure~1 in][]{Jarrett2011}, especially for $W1$/$I1$, $W2$/$I2$, and $W4$/$M1$. The $W3$ band covers a broader range around $\SI{12}{\micro\meter}$ and overlaps with $I4$ which is centered at $\SI{8}{\micro\meter}$. Both are influenced by PAH emission (polycyclic-aromatic-hydrocarbons), which is excited by UV radiation and emitted in the IR. Hence, typical sources of PAH emission are massive star-forming regions. This leads to higher contamination in these bands, especially near the Orion Nebula Cluster \citep[ONC, e.g.,][]{Hillenbrand1998A, Lada2000B}. Also star-forming galaxies show PAH emission, which can be erroneously identified as YSOs, which will be addressed in Sect.~\ref{revisit}. The lower resolution and sensitivity of \emph{WISE} compared to \emph{Spitzer} results in higher confusion caused by extended MIR emission. Especially the $W4$ band is significantly contaminated by extended thermal emission, amplified by its low resolution.


\subsection{Ancillary Data} \label{Aux}

Orion\,A is one of the most favorable sites to study star formation, being the closest massive star-forming region to earth \citep[$\SI{{\sim}414}{pc}$][]{Menten2007}. Hence, there is a large number of studies and data available, especially for the prominent ONC region \citepalias[see figure~1 in][]{Meingast2016}. The mentioned catalogs \citepalias[MGM,][]{Furlan2016} include members already reported in earlier smaller scale studies. To perform a more complete study, we add the following published datasets. 

Several spectroscopic and optical surveys are available for the ONC region \citep{Hillenbrand1997, Hillenbrand2000, DaRio2009, Elek2013, Pettersson2014} and the dark cloud L1641 \citep{Fang2009, Fang2013, Hsu2012, Hsu2013, DaRio2016}. Spectroscopic surveys provide information on spectral types and on extinction, and allow classification into classical and  weak-line T-Tauri stars (CTTS, WTTS). These are PMS stars showing typical emission (e.g., H$\alpha$) or absorption (e.g., Lithium, Li\,I) lines, which are indicators for youth. For example, strong H$\alpha$ emission is caused by gas accretion onto the surface of the stellar photosphere. Accordingly, it probes the gaseous component of the circumstellar disk, while IR-excess probes the dusty component.
\citet{Kim2013,Kim2016} present a study of transition disks \citep[TD,][]{Cieza2008, Cieza2010, Muzerolle2010} for Orion\,A.
These are circumstellar disks with inner dust holes filled with gas, and optically thick outer gas+dust disks \citep{Teixeira2012}. They show no or weak excess from 3 to $\SI{8}{\micro\meter}$ (probing the inner disk) but a significant excess at longer wavelengths ($\lambda \gtrsim \SI{10}{\micro\meter}$, probing the outer disk).

Moreover, we add optical data from the Sloan Digital Sky Survey DR12 \citep[SDSS,][]{Alam2015}, which does not cover all of Orion\,A, but large parts near the ONC, and the Pan-STARRS survey \citep{Flewelling2016}, which covers the whole region. The optical data allows to construct more complete SEDs. These are helpful when investigating especially critical sources, with unclear classification. 

To further confirm the young nature of stars we add X-ray observations from XMM-Newton and \textit{Chandra}. XMM-Newton data is available for the L1641 \citep{Pillitteri2013} and the L1647 region \citep[$\kappa$\,Ori,][]{Pillitteri2016}\footnote{Download from \url{https://nxsa.esac.esa.int}; the coordinates provided by \citet{Pillitteri2016} resulted in an erroneous cross-match.}, and \textit{Chandra} data is available for the ONC \citep[COUP,][]{GetmanA2005, GetmanB2005} and for regions north and south to the ONC \citep[SFINCS,][]{Getman2017}. This information is listed in the final catalog (Tabel~\ref{tab:master}) in the column ``X'', which indicates if the source was detected in X-rays.

\subsection{Combined data catalog}

We combine the different data sets to one data catalog, adopting the cross-match radius to the resolution. First, VISTA is cross-matched with the whole \emph{Spitzer} data catalog\footnote{Available at: \url{http://astro1.physics.utoledo.edu/~megeath/Orion/The_Spitzer_Orion_Survey.html}}, containing the MGM YSO sample. Second, data of the 309 HOPS \citepalias{Furlan2016} sources are added, of which most are a sub-sample of the MGM catalog. Next, AllWISE MIR data is cross-matched. Due to the lower angular resolution of \emph{WISE} ($\SI{\sim6}{\arcsec}$, \emph{Spitzer} $\SI{\sim1.7}{\arcsec}$, VISTA $\SI{\sim0.7}{\arcsec}$) multiple VISTA sources can lie inside one unresolved \emph{WISE} source, which can lead to misidentifications. This can contribute to contamination and incompleteness of the final sample in ways that are difficult to characterize. This is addressed in Sect.~\ref{Completeness} where we discuss the completeness of our final sample. Finally, all auxiliary data are added to complement the data catalog with the available information from the literature.


\section{Methods} \label{Methods}

In this Section we first present our methods of YSO classification. Next, we discuss the methods to revisit existing catalogs and clean them of possible false positives. Finally, we give an overview of our methods to add new candidates, while the detailed selection procedure is explained in Appendix~\ref{NewYSOsA}.

\begin{figure*}[!ht]
    \centering
        \includegraphics[width=0.85\linewidth]{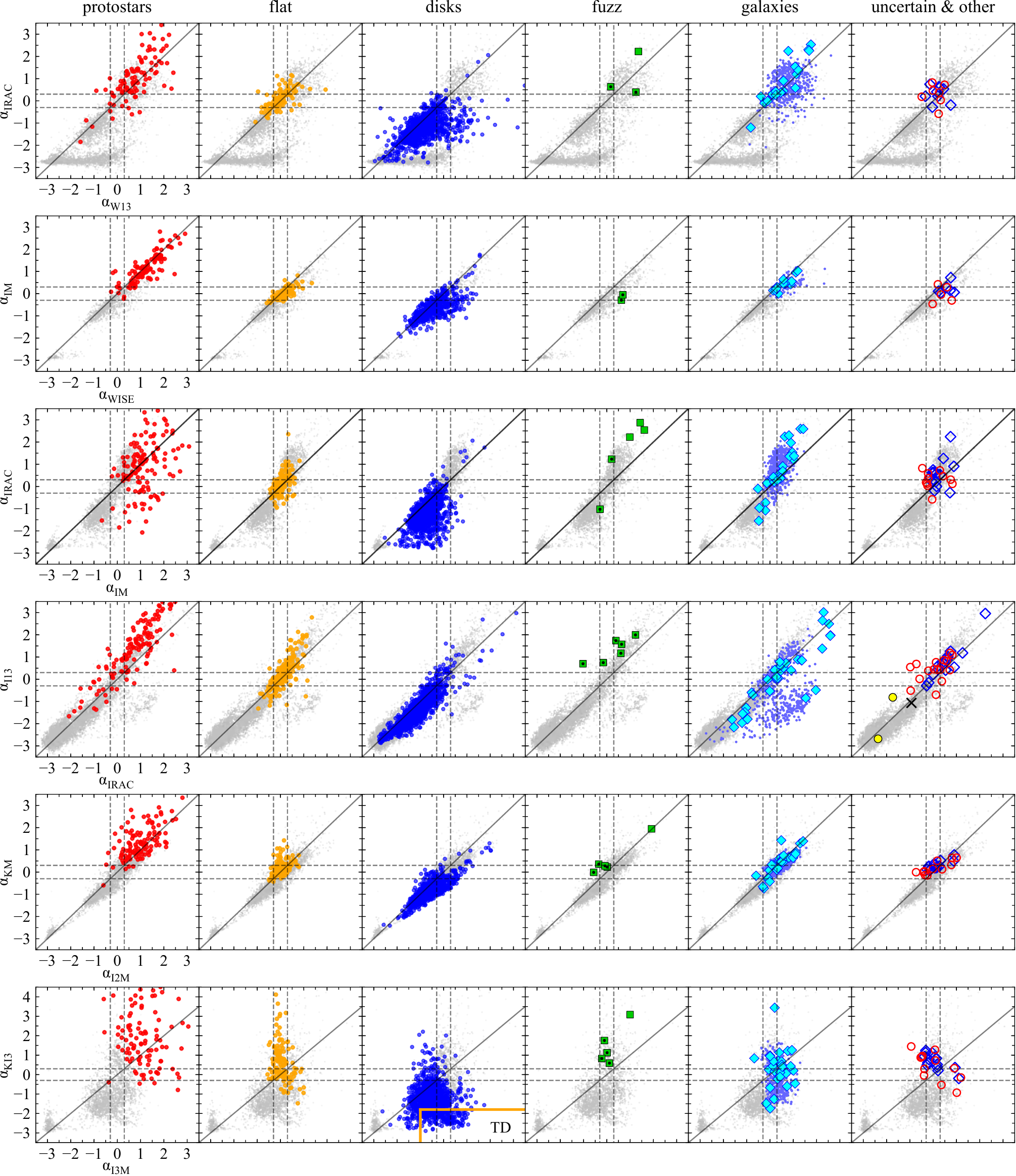}
    \caption{Comparing various observed spectral indices ($\alpha$). The used spectral indices are listed in Table~\ref{tab:alphas}. Shown are YSO candidates and false positives, with symbols like in Fig.~\ref{fig:ccds}. A slope one line is given (black solid line), highlighting where different spectral indices give the same value. The range for flat-spectrum sources is indicated by dashed lines. The two top rows show a comparison of \emph{Spitzer} and \emph{WISE} spectral indices, covering approximately the same spectral range. The protostars contain sources with declining $\alpha$ in some spectral ranges. We found that most of these are influenced by visible outflows. The solid orange outline (last row, third column) indicates a transition disk selection (see text for more explanations).
    }
    \label{fig:alpha}
\end{figure*}

\subsection{YSO classification} \label{classification}

The YSO classification in this Paper is not solely based on a classical spectral index classification, but rather a combination of investigating various spectral index ranges, of including FIR information, visual inspection, and individual SED inspection. 
\begin{table}[!ht] 
\begin{center}
\small
\caption{Adopted YSO classification based on the MIR SED.} 
\resizebox{\columnwidth}{!}{%
\begin{tabular}{ccc}
\hline \hline
Class designation & A.d.\tablefootmark{a} & Spectral Index \\
\hline
  \begin{tabular}{l}
    Class\,0/I (protostars) \\
    flat-spectrum sources (flats) \\
    Class\,II (thick disks) \\
    Class\,III (anemic or thin disks) \\
    Class\,III (disk-less PMS) / MS stars\tablefootmark{b} 
  \end{tabular}  &                
  \begin{tabular}{c}
    P        \\
    F        \\
    D        \\
    D, AD    \\
    III, MS  \\
  \end{tabular}  &                
  \begin{tabular}{r@{$\alpha$}c}
    $+0.3    <$ \, &  \\
    $-0.3 \leq$ \, & \, $\leq +0.3$   \\
    $-1.6    <$ \, & \, $< -0.3$     \\
    $-2.5    <$ \, & \, $\leq -1.6$  \\
                   & \, $\leq -2.5$  \\
  \end{tabular}  \\
\hline                              
\end{tabular}}
\label{tab:class2}
\tablefoot{ \tablefoottext{a}{Alternative designation.}
            \tablefoottext{b}{Disk-less pre-main-sequence stars or main-sequence stars.}}
\end{center}
\end{table}
As an initial estimate, we adopt the YSO classification based on the MIR spectral index similar to \citet{Greene1994}, as given in Table~\ref{tab:class2} and we refine the classification by using the methods listed above.

The lower spectral index limit for sources with IR-excess is given by \citet{Lada2006} with $\alpha_\mathrm{IRAC} > -2.56$. Below, the SED reflects the photosphere of the star. They state that effects of different spectral types have no significant influence on this value, therefore it is an upper limit for sources with no IR-excess. We adopt a value of -2.5, due to the uncertainties in the \emph{Spitzer} photometry, influencing especially sources near the ONC. 
Although the value is defined for $\alpha_\mathrm{IRAC}$, it can also be applied to other spectral index ranges as an upper limit, which is highlighted in Fig.~\ref{fig:alpha}, where we compare various spectral indices. The scatter at the main-sequence (MS) star locus at about $-3$ is caused by extincted sources. 

Class\,IIIs, by the definition of \citet{Greene1994}, include sources with weak IR-excess, due to optically thin disk remnants, also called anemic disks \citep[AD,][]{Lada2006}. The distinction between Classes\,II and III was set due to findings of \citet{Andre1994}, where they find a sharp threshold in millimeter flux density at $\alpha\approx-1.5$. 
However, we do not separate the disk bearing YSOs, meaning Class\,IIs and the part of Class\,IIIs with IR-excess, but call them collectively disks (D), when analyzing the sample in Sect.~\ref{Distribution}. Nevertheless, we label these sources separately in the final catalog with ``D'' and ``AD'', respectively. 
In a similar manner, we do not distinguish between Classes\,I and 0, and call them collectively protostars (P). Again, we label these candidates separately in the final catalog with ``I'' and ``0'', respectively, mainly based on the information from \citet{Stutz2013} and \citetalias{Furlan2016}.
Sources with flat-spectra likely correspond to YSO candidates with envelope remnants on the verge to the disk dominated PMS phase \citepalias{Furlan2016}, but they can also be an effect of disk inclination or foreground extinction, as highlighted in Sect.~\ref{Introduction}. Therefore, this class remains suspicious, and will be addressed in more detail in Sect.~\ref{flats}.

To calculate the spectral index we use all available photometry from 2 to $\SI{24}{\micro\meter}$. This range from the NIR to the MIR is often used to define the classes \citep[e.g.,][]{Dunham2015, Heiderman2015, Kim2016}. However, as mentioned above, we also compare with other ranges like given in Table~\ref{tab:alphas}, while differences are highlighted in Fig.~\ref{fig:alpha}. 
For example, a comparison of $\alpha_\mathrm{IRAC}$ and $\alpha_\mathrm{IM}$ shows (Fig.~\ref{fig:alpha}, thired row), when using only $\alpha_\mathrm{IRAC}$, some protostars would be shifted to later classes. 
Furthermore, shorter wavelength ranges probe the inner disk, while longer wavelengths from about $\SI{8}{\micro\meter}$ on-ward, probe the outer disk or envelope. This gives information on transition disk YSOs (TD, see Sect.~\ref{Aux}). These sources can be misclassified as flats or even protostars, depending on the wavelength range used. They can be selected by using for example, $\alpha_\mathrm{KI3}$ and $\alpha_\mathrm{I3M}$, highlighted at the bottom row of Fig.~\ref{fig:alpha} by the orange lines over-plotted on the disk candidate plot (third column). Again, we do not separate TDs in our statistical analysis, but we label them in the final catalog separately as ``TD''.

\begin{table}[!ht] 
\centering
\small
\caption{List of used spectral indices.} 
\label{tab:alphas}
\begin{tabular}{llcl}
\hline \hline
\multicolumn{1}{c}{Spectral} &
\multicolumn{1}{c}{band} &
\multicolumn{1}{c}{$\lambda$ range} &
\multicolumn{1}{c}{Used by, for example} \\
\multicolumn{1}{c}{indices} &
\multicolumn{1}{c}{range} &
\multicolumn{1}{c}{(\si{\micro\meter})} &
\multicolumn{1}{c}{} \\
\hline
$\alpha_\mathrm{KM}$ & $K_S$ to $M1$  & 2 - 24 & \makecell[l]{\citet{Dunham2015}\\\citet{Heiderman2015}\\\citet{Kim2016}} \\
$\alpha_\mathrm{IM}$ &   $I1$ to $M1$ & 3 - 24 &  -  \\
$\alpha_\mathrm{IRAC}$ & $I1$ to $I4$ & 3 - 8\phantom{0}  & \citet{Lada2006} \\
$\alpha_\mathrm{I2M}$  & $I2$ to $M1$ & 4 - 24 & \citetalias{Furlan2016} \\
$\alpha_\mathrm{I3M}$  & $I3$ to $M1$ & 5 - 24 & \citet{Muench2007} \\
$\alpha_\mathrm{KI3}$ & $K_S$ to $I3$ & 2 - 5\phantom{0}  & - \\
\hline
$\alpha_\mathrm{KW3}$   & $K_S$ to $W3$ & 2 - 12  & - \\
$\alpha_\mathrm{KW}$    & $K_S$ to $W4$ & 2 - 22  & - \\
$\alpha_\mathrm{W13}$   & $W1$ to $W3$ & 3 - 12  & - \\
$\alpha_\mathrm{WISE}$  & $W1$ to $W4$ & 3 - 22  & - \\
$\alpha_\mathrm{KW12M}$ & $K_S$ to $M1$ & 2 - 24  & - \\
\hline                              
\end{tabular}
\tablefoot{The top block shows spectral indices including VISTA and \emph{Spitzer} bands, and the bottom block VISTA and \emph{WISE} bands. All available bands are used between the individually given ranges, however, \emph{Spitzer} and \emph{WISE} are not mixed, except for the last spectral index, where we combine $K_S$ with $W12$ and $M1$, used for sources beyond the IRAC coverage but still inside the MIPS coverage.}
\end{table}

Not only the used spectral index but also foreground extinction can affect the classification by shifting, for example, Class\,II sources to the flat-spectrum or Class\,I regime \citep[e.g.,][]{Muench2007, Forbrich2010}, whereas longer MIR wavelength bands are less effected by extinction. \citet{Muench2007} point out that background stars can not mimic a Class\,I source when using $\alpha_\mathrm{I3M}$, even at extinctions as high as $A_\mathrm{K} \approx \SI{20}{mag}$ (see their figure~20). 

To correct for extinction we de-redden the photometry, by estimating the line-of-sight extinction toward each source individually, relative to the $K_S$ band ($A_\mathrm{K}$). We denote the de-reddened spectral index as $\alpha_0$.
When available, we use literature values for line of sight extinctions obtained via spectral surveys \citep{Hillenbrand1997, Fang2009, Kim2013, Kim2016, Furlan2016}. To convert from $A_\mathrm{V}$ to $A_\mathrm{K}$ we use $A_\mathrm{K}/A_\mathrm{V} = 0.112$ from \citet{Rieke1985}.
Else we calculate the line of sight foreground extinction with the NICER technique \citep[Near-Infrared Color Excess Revisited,][]{Lombardi2001}, using the extinction laws listed in Table~\ref{tab:bands}.
The method uses NIR JHK photometry, with the intrinsic color derived from the VISTA control field. In our case the intrinsic color corresponds to the location of M-stars.
If not all of the three NIR bands have valid measurements we use only two NIR bands for an estimate (i.e., E($J-H$), E($J-K_S$), or E($H-K_S$)).
For sources with only one or no NIR observation, we use the new PNICER technique \citep{Meingast2017}, which is a probabilistic machine learning approach, enabling the inclusion of MIR bands, to estimate extinction. The finally used method (to determine extinction) is given in Table~\ref{tab:master} in column ``{\tt AK\_method}''.
For YSOs or galaxies the calculated line-of-sight extinction might overestimate the actual foreground extinction, due to the intrinsic reddening by circum-stellar material, and galaxies show redder colors also due to dust and star formation. 
We compare the individual line-of-sight extinction  ($A_\mathrm{K,IR}$) with the total line-of-sight extinction at the position of each source as extracted from the \emph{Herschel map} ($A_\mathrm{K,Herschel}$), and find that especially for galaxies, extinctions estimated from their IR colors ($A_\mathrm{K,IR}$) are mostly larger than $A_\mathrm{K,Herschel}$. 
Therefore, we use $A_\mathrm{K,Herschel}$ if  $(A_\mathrm{K,Herschel} + \mathrm{err}\_A_\mathrm{K,Herschel}) < A_\mathrm{K,IR}$ to de-redden such sources\footnote{The error of $A_\mathrm{K,Herschel}$ is extracted from the \emph{Herschel} error map \citep{Lombardi2014}.}. 

Unfortunately, we can not use the same set of spectral indices for all sources due to different survey coverages and different sensitivities of the various bands. For sources that lie outside the IRAC covered region, thus new YSO candidates (Sect.~\ref{NewYSOs}), we use \emph{WISE} spectral indices, or a combination with VISTA or Spitzer/$M1$ (see Table~\ref{tab:alphas}). This leads to an inconsistent classification, as highlighted in Fig.~\ref{fig:alpha}. The spectral index $\alpha_\mathrm{W13}$ is significantly different to $\alpha_\mathrm{IRAC}$, where contaminated $W3$ photometry (PAH emission) produces a shift of MS stars (or Class\,IIIs) to redder colors at the bottom of the diagram. The spectral index $\alpha_\mathrm{WISE}$ shows a better correlation with $\alpha_\mathrm{IM}$, but not all $M1$ observed sources also show a significant $W4$ measurement due to the inferior sensitivity of \emph{WISE}. Nevertheless, newly selected candidates beyond the Spitzer coverage are less affected by contamination and therefore have overall more reliable WISE photometry. 

To summarize the classification process, for the final YSO classification we do not use the MIR spectral index blindly, by strictly following a simple cut using a single spectral index range. Instead, we look at different spectral indices, as listed in Table~\ref{tab:alphas}, and check if they consistently show a rising, flat, or declining slope, and compare with the de-reddened spectral indices $\alpha_0$. For border cases, especially close to the flat-spectrum range, with no clear trend for the different spectral indices, we individually check the SEDs to make a final decision, and investigate the FIR range, mostly adopting the classification by \citetalias{Furlan2016}.
Due to considering different spectral index ranges, the flat-spectrum sources in Fig.~\ref{fig:alpha} (showing the observed spectral index $\alpha$), do not fall exactly in the range of $-0.3 < \alpha < 0.3$ for all indices.

Additionally, we use visual information, as also discussed below (Sect.~\ref{revisit}). The VISTA images reveal outflows, cavities, jets, and reflection nebulae, which were taken into account as confirmation for the protostellar nature. For example, jets and outflow shocks close to the source affect the $\SI{4.5}{\micro\meter}$ range ($I2$ or $W2$) \citep{Evans2009}, and reflected light in the outflow cavity can affect NIR bands \citep{Crapsi2008}. 
Moreover, the silicate absorption feature, located at about $\SI{10}{\micro\meter}$ ($I4$ or $W3$), is caused by protostellar envelopes or edge-on disks \citep{Crapsi2008}, or by layers of high column-density in front of the source.
Therefore, many protostars do not show a rising $\alpha_\mathrm{IRAC}$, while clearly rising when including $M1$ (Fig.~\ref{fig:alpha}). The protostellar nature is also clarified with FIR data, because most protostars correspond to a PACS point sources \citepalias{Furlan2016}, and show a clear peak in the FIR, which is not reflected in the MIR spectral index.
  
Finally, contaminated photometry can produce a fake IR-excess, which is not always easy to account for. Especially high-mass star-forming regions can cause a lot of such contamination (image artifacts, saturation, nebulosities, extended emission, cloud-edges).
We exclude bands if their photometry where found to result from contamination during visual inspection.


\subsection{Revisiting the known YSO population - Methods to evaluate false positives} \label{revisit}

The \emph{Spitzer} based MGM catalog is currently the reference for the YSO population in Orion\,A, with updates from \citetalias{Furlan2016} and \citet{Lewis2016} (hereafter, LL16). In this section we present a combination of methods to evaluate the contamination (false positives) of existing YSO catalogs. These are visual inspection, position of sources in color-color and color-magnitude diagrams, effects of extinction, source morphology (extension flags) from VISTA, and information from the literature. \\

\textbf{Visual Inspection.}
All previously identified YSO candidates were visually inspected, using the VISTA images \citepalias{Meingast2016}, and images of \emph{Spitzer}/IRAC/MIPS, \emph{WISE}, 2MASS, \emph{Herschel}/PACS, DSS, and SDSS\footnote{VISTA, \emph{WISE}, 2MASS, PACS, DSS, SDSS images are available via Aladin \citep{Bonnarel2000, Boch2014}.}. This enables us to identify resolved galaxies (G), IR nebulosities (fuzz), or image artifacts like diffraction-spikes or airy-rings (C, for contamination). The abbreviations given in parenthesis are also used in the final catalog (Table~\ref{tab:master}) in column ``{\tt Class}''.
Visual inspection can further be used as a confirmation for protostars, which show visible outflows, jets, reflection nebulae, or cavities. Examples of these can be found in Appendix \ref{cutouts}.

\begin{figure*}[!ht]
    \centering
        \includegraphics[width=0.85\linewidth]{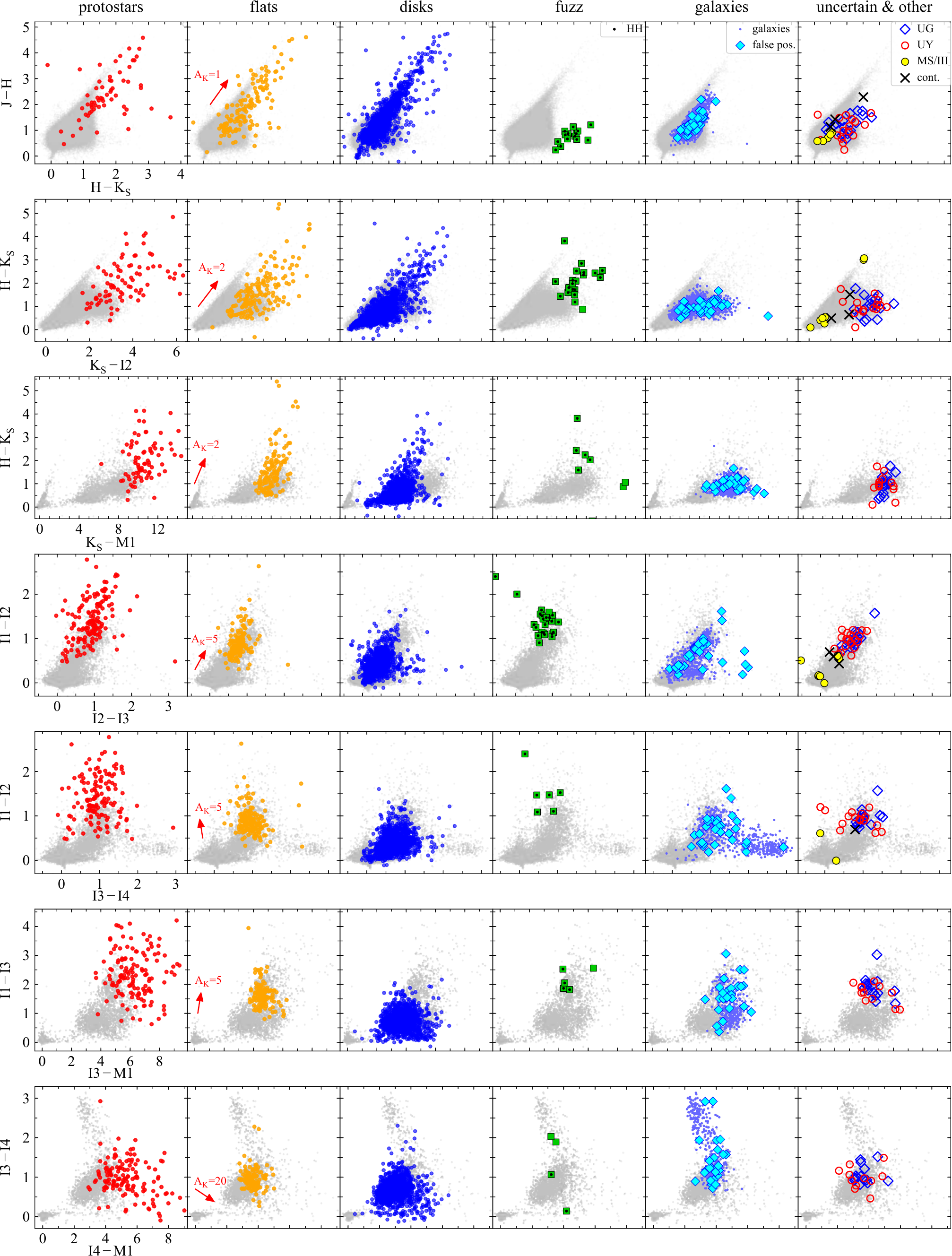}
    \caption{Seven selected color-color diagrams composed of VISTA and \emph{Spitzer} photometry, showing YSO candidates and false positives. Gray dots represent all sources toward Orion\,A with $\mathrm{errors}<\SI{0.2}{mag}$ for the given bands. From left to right we show the three YSO classes, protostars (red), flat-spectrum sources (orange), disk sources (blue); and contaminating objects: fuzzy nebulous contamination (green), galaxy contamination (cyan), and finally other contamination or uncertain objects. These are MS stars or Class\,III candidates (filled yellow circles), contamination due to image artifacst (black crosses), uncertain candidates (red open circles), and uncertain galaxy candidates (blue open diamonds). In column four showing fuzzy contamination, we highlight Herbig-Haro objects with a black dot, while sources without dot are mostly cloud-edges or other nebulous structures. In column five we show galaxies (blue dot symbols) and previous YSO candidates identified as galaxies (false positives, filled cyan diamonds). This highlights that galaxies often occupy similar color spaces as YSOs, especially the ones previously classified as YSO candidates. However, most tend to be fainter for the distance of Orion\,A (see CMDs, Fig.~\ref{fig:cmds}).}
    \label{fig:ccds}
\end{figure*}

\begin{figure*}[!ht]
    \centering
        \includegraphics[width=0.85\linewidth]{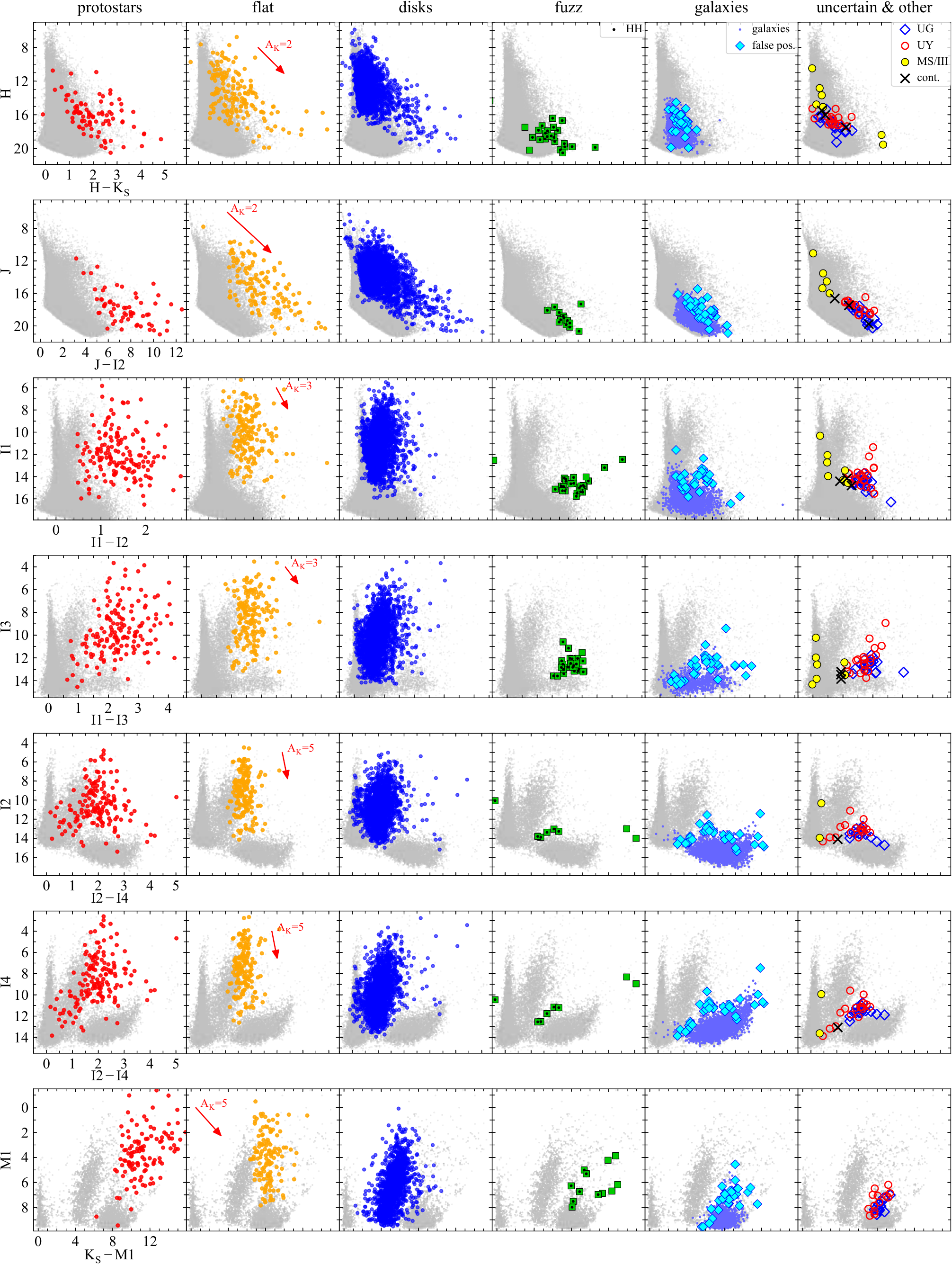}
    \caption{Seven selected color-magnitude diagrams, showing YSO candidates and false positives, as in Fig.~\ref{fig:ccds}. The CMDs highlight the brightness differences of the classes and contaminating objects. False positives (fuzz, galaxies) and uncertain sources tend to be fainter. Galaxies, previously identified as YSO candidates (false pos., cyan diamonds) are generally brighter than average galaxies (bluish dots, other identified galaxies). Looking at YSO candidates, flats tend to be overall brighter in the MIR compared to disk candidates.}
    \label{fig:cmds}
\end{figure*}

\textbf{Color-color and color-magnitude diagrams.} 
In parallel to visual inspection, various color-color (CCD) and color-magnitude diagrams (CMD) are checked, similar to those used by \citet{Gutermuth2009} and \citet{Megeath2012}. We specifically check if the color of each source is consistent with the typical color of the proposed class. The diagrams include photometry in the NIR and MIR from VISTA and \emph{Spitzer}. We show examples in Figs.~\ref{fig:ccds} and \ref{fig:cmds} separately for the three different YSO classes (protostars, flats, disks, as classified in Sect.~\ref{classification}), and for false positives (fuzz, galaxies, uncertain and other objects). In the fifth column, showing galaxies including false positives, one can see that YSOs and some types of galaxies occupy similar color spaces in most diagrams. Here, the false-positive YSO identifications are typically found toward the bright end of the galaxy locus.

\textbf{Extinction.}
As an additional indicator, the total column-density toward single sources can be used. For example, it is unlikely that protostars are associated with low extinction regions since they are still surrounded by their dust envelope. As investigated by \citet{Lada2010}, a typical star formation extinction threshold is $A_\mathrm{K}>\SI{0.8}{mag}$. We use the \emph{Herschel map} to infer the total dust column-density toward each source ($A_\mathrm{K,Herschel}$). However, this needs to be handled carefully, since we can not rule out the presence of unresolved structure beyond the resolution of \emph{Herschel} ($\SI{36}{\arcsec}$, $\SI{\sim0.07}{pc}$ @ $\SI{414}{pc}$). Therefore, if no or only little extinction is located at the position of a candidate protostar, the source is further investigated \citepalias[see also,][]{Lewis2016}, to look for other signs of youth (e.g., outflows, PACS detection). Else it is flagged as uncertain (U). On the other hand, if a source is above the adopted threshold, does not immediately confirm its YSO nature. For example, bright galaxies can be detectable through extinction as high as $A_\mathrm{K}\SI{\sim2}{mag}$. Finally, disk sources do not necessarily have to be connected to regions of higher dust column-density. During their typical age of a few million years \citep[e.g.,][]{Evans2009, Dunham2015} they could have already moved away from their birthplace, or the clouds out of which they have formed might have dissipated \citepalias{Lewis2016}.

\textbf{Source morphology.}
The VISTA source catalog provides two extension flags. \texttt{ClassSex} refers to a source's morphology as determined by the source extraction algorithm SExtractor, while \texttt{ClassCog} derives the morphology from variable aperture photometry in combination with machine learning techniques \citepalias[for details see][]{Meingast2016}. Values close to 0 indicate an extended object, values close to 1 point-like morphology. These flags, however, are not a universal discriminator between galaxies and stars, because protostars are often associated with extended emission or outflows. For this reason, we always use these flags in combination with visual inspection. 
While many galaxies are associated with extended morphology, faint extra-galactic objects can also appear point-like. These, however, are mostly identified in the various CCDs and CMDs. Special cases are active galactic nuclei (AGNs), which might be more difficult to distinguish, since they can appear more bright and point-like, while also showing similar IR colors as protostars or flat-spectrum sources. These can contribute to residual contamination in the final catalog.

\textbf{Information from the literature.}
We searched the literature (Sect.~\ref{Aux}) and the \mbox{SIMBAD} astronomical databases \citep{Wenger2000} for additional classification information. Oftentimes, young stars are already marked as emission line stars (\texttt{Em*}), flare stars (\texttt{Flare*}), variable stars (\texttt{V*}, \texttt{Orion\_V*}, \texttt{Irregular\_V*}), or T-Tauri stars (\texttt{TTau*}, WTTS, CTTS). Since this information is very heterogeneous, we generally do not use it for our classification. Only suspicious sources (faint, unresolved, untypical colors), which do not have an entry in these additional surveys, are marked as uncertain candidates (U). We include this information in the final catalog (Table~\ref{tab:master}).

To summarize, galaxies (G) are identified morphologically using visual inspection and extension flags in combination with colors, magnitudes, and information about extinction. If no clear morphological identification is possible, we flag some sources as uncertain galaxy candidates (UG), if their colors, magnitudes, and location at low extinction suggest the extra-galactic nature. Hence, they belong to the uncertain candidates (U or UY). Fuzzy contamination (fuzz) like nebulosities, cloud-edges, or Herbig-Haro objects, is generally identified visually, as well as photometric contamination like image artifacts (C).



\subsection{New YSO candidates} \label{NewYSOs}

Here, we shortly describe our methods to add new YSO candidates, while the detailed procedure is described in Appendix~\ref{NewYSOsA}. The methods are mainly based on NIR and MIR color-color and color-magnitude diagram selection criteria, and we also add few sources using PACS photometry.
To add new YSO candidates in the surroundings of the \emph{Spitzer}/IRAC surveyed region (outside IRAC regions, green contour, Fig.~\ref{fig:coverage}), we made use of the larger coverage of VISTA (blue contour). To this end, we constructed color and magnitude diagrams using VISTA combined with \emph{WISE} and partially \emph{Spitzer}/$M1$ (red contour). \emph{WISE} requires special treatment, especially concerning the two longer wavelengts bands $W3$ and $W4$, due to the low resolution and high contamination caused by extended MIR emission, already highlighted in Sect.~\ref{MIR}. The selection conditions for \emph{WISE} data were informed by previous works \citep{Jarrett2011, Rebull2011, Koenig2012, Koenig2014, Koenig2015}, however we adjusted them for our purpose (see Appendix~\ref{NewWISEA}). Moreover, we selected new YSO candidates also inside the IRAC region in combination with \emph{Spitzer} photometry, by applying different selection criteria compared to previous studies, and by including VISTA instead of 2MASS. Additionally, we used the PACS point source catalog and visual inspection of the PACS images, to add further new YSO candidates.


\section{Results} \label{Results}

In this section we firstly summarize our results for the revisited catalogs, and secondly for the new YSO candidates. Finally we give an overview of the updated YSO catalog, combining the two samples.


\subsection{Results for revisited YSO candidates} \label{Results revisited}

\begin{figure*}[!ht]
    \centering
    \begin{minipage}[t]{1\linewidth}
        \centering
        \includegraphics[width=\linewidth]{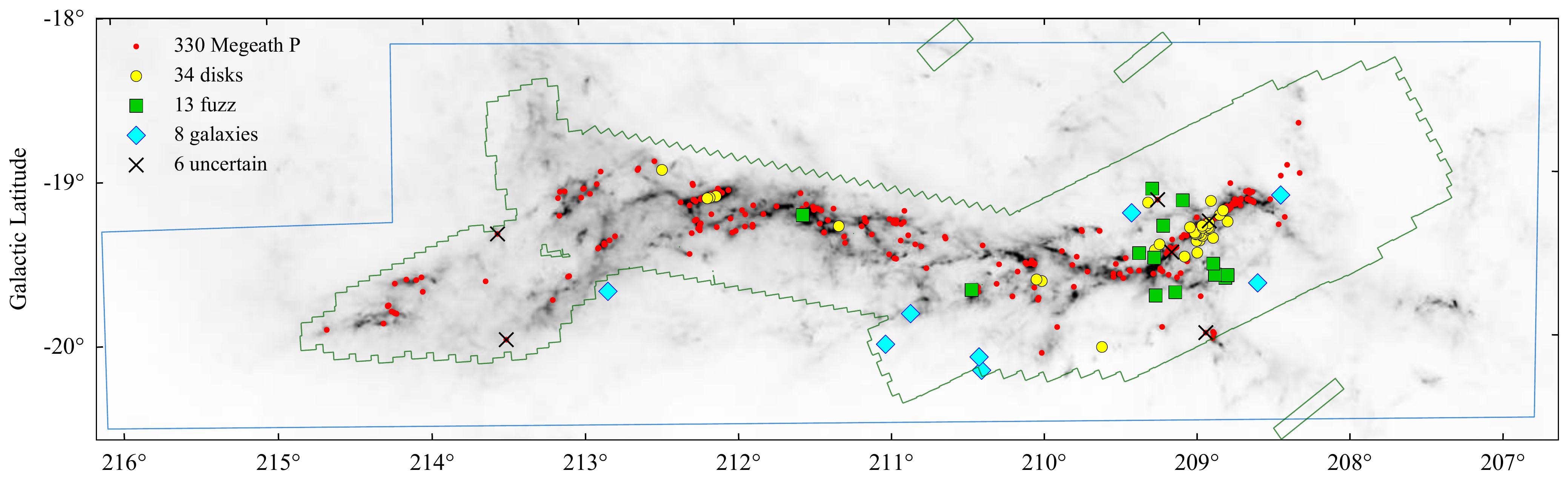}
    \end{minipage}%
    \vfill
    \begin{minipage}[t]{1\linewidth}
        \centering
        \includegraphics[width=\linewidth]{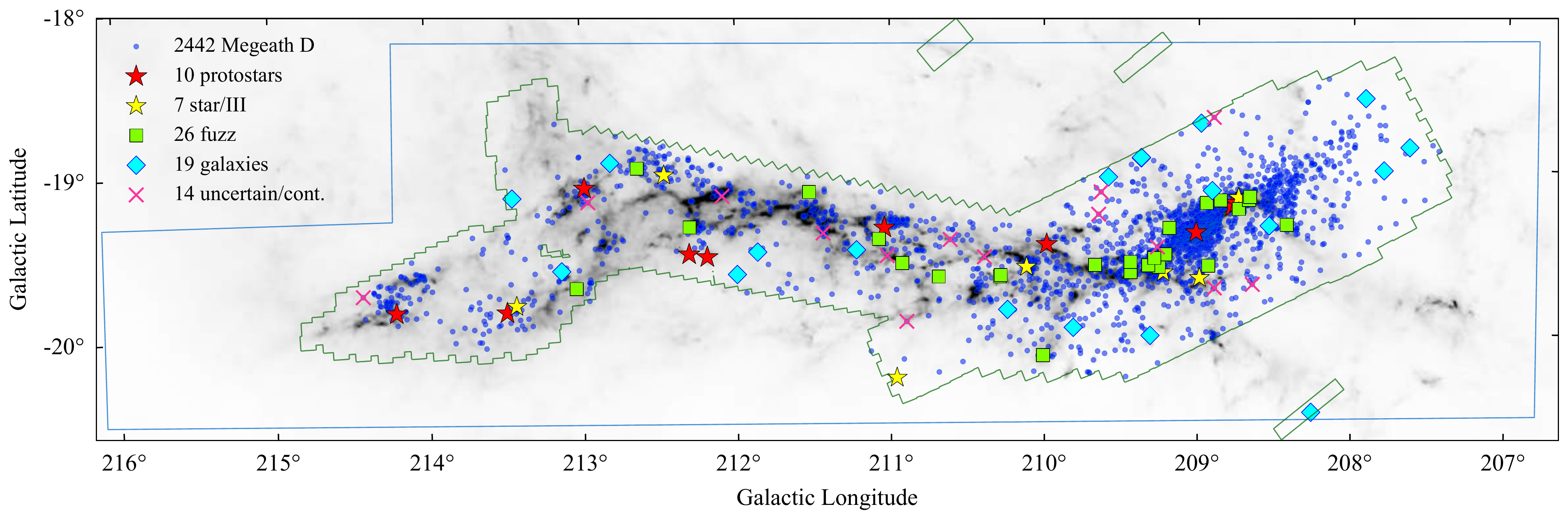}
    \end{minipage}%
    \caption{Distribution of the MGM YSO candidates displayed on the \emph{Herschel map} showing false positives and re-classification. Here, the re-classification does not consider flat-spectrum sources.
    \textit{Top:} The 330 MGM protostar candidates (P, red dots), of which 33 are re-classified as disk candidates (yellow filled circles).
    \textit{Bottom:} The 2442 MGM disk candidates (D, blue dots), of which ten are re-classified as protostar candidates (red filled stars), and seven as MS stars or Class\,III sources (yellow filled stars).
    Other symbols are false positives, like extra-galactic contamination (cyan filled diamonds), and fuzzy nebulous contamination (green filled squares) as given in the legends. The ``$\times$'' symbol marks uncertain sources and sources contaminated by image artifacts (``uncertain/cont.'').}
    \label{fig:mapfake}
\end{figure*}

With the methods described above we revisited the 2839 previously identified YSO candidates from MGM (2827) and \citetalias{Furlan2016} (283), resulting in 2706 ($\sim$95\%) YSO candidates in the updated catalog (Table~\ref{tab:master}). We re-classify them as described in Sect.~\ref{classification}, with the resulting number-counts for each class listed in Table~\ref{tab:reclassified}. Out of the 133 ($\sim$5\%) excluded candidates, there are 92 ($\sim$3\%) false positives and 41 ($\sim$2\%) uncertain sources. Most of the uncertain sources are faint objects with untypical properties (colors, magnitudes, location), for which we can not tell with our criteria and the available data if they are faint YSOs, extra-galactic (e.g., AGNs), background giants (e.g., AGBs), or even brown dwarfs. Follow up observations are needed to clarify the nature of these sources, like spectra \citep[e.g., NIR spectra,][]{Greene2002}, or looking for envelope tracers \citep[e.g., HCO$^+$,][]{Heiderman2015}. We still expect a residual degree of contamination in our final selection mainly due to AGNs or AGBs. AGNs especially influence the flat and protostar range \citep{Stern2005}, and AGBs the anemic disk (Class\,III) range \citep{Dunham2015}. The false positives include seven sources that do not show any IR-excess beside some reddening effects due to extinction. These are flagged as main-sequence star (MS) or Class\,III candidate (III; if X-ray source or emission line star, see Sect.~\ref{Aux}). Statistical overviews of the different types of contamination are listed for the MGM and \citetalias{Furlan2016} samples in Tables~\ref{tab:Meg} and \ref{tab:Furlan}, respectively. For the whole MGM sample we get a lower limit of contamination of about 3\% to 5\%, while for the \citetalias{Furlan2016} sample we get a very low contamination fraction of $<1\%$, when considering the sample with applied SED modeling (see first row Table~\ref{tab:Furlan}).

\begin{table}[!hb] 
\small
\begin{center} 
\caption{Re-classification summarized.} 
\begin{tabular}{llccc}
\hline \hline
  \multicolumn{2}{c}{} &
  \multicolumn{3}{c}{YSO Classes\tablefootmark{a}} \\
\cmidrule(lr){3-5}
  \multicolumn{1}{c}{Sample} &
  \multicolumn{1}{c}{YSOs\tablefootmark{b}} &
  \multicolumn{1}{c}{P} &
  \multicolumn{1}{c}{F} &
  \multicolumn{1}{c}{D} \\
\hline 
Revisited\tablefootmark{c} & 2706 & 182 (6.7\%) & 177 (6.5\%) & 2347 (86.8\%)  \\
\hline
New inside\tablefootmark{d} & \phantom{0}154 & \phantom{00}1 (0.7\%) & \phantom{00}3 (1.9\%) & \phantom{0}150 (97.4\%)  \\
New ouside\tablefootmark{d} & \phantom{0}120 & \phantom{00}5 (4.2\%) & \phantom{00}5 (4.2\%) & \phantom{0}110 (91.7\%)  \\
New all\tablefootmark{d}    & \phantom{0}274 & \phantom{00}6 (2.2\%) & \phantom{00}8 (2.9\%) & \phantom{0}260 (94.9\%)  \\ 
\hline
Total\tablefootmark{c}     & 2980 & 188 (6.3\%) & 185 (6.2\%) & 2607 (87.5\%)  \\ 
\hline
\end{tabular}
\label{tab:reclassified}
\tablefoot{The percentages in parenthesis are relative to the YSO counts of each sample given in Col.~2.
        \tablefoottext{a}{Classification from this work. 
        Class\,0/I protostars (P), flat-spectrum sources (F), Class\,II/III pre-main-sequences stars with disks (D).}
        \tablefoottext{b}{Total number of YSO candidates of the given samples.}
        \tablefoottext{c}{Reclassification of revisited sources for the combined MGM and \citetalias{Furlan2016} sample (Sects.~\ref{revisit}, \ref{Results revisited}).}
        \tablefoottext{d}{Classification for new sources (Sects.~\ref{NewYSOs}, \ref{results new}), separated between candidates selected in- and outside the IRAC region.}
        \tablefoottext{c}{Total = Revisited + New all}
           }           
\end{center}
\end{table}
\begin{table*}[!ht] 
\centering
\small
\caption{The MGM YSO sample revisited in numbers.} 
\begin{tabular}{lc | cccccccccccc}
\hline \hline
  \multicolumn{2}{c}{MGM} &
  \multicolumn{12}{c}{This work} \\
\cmidrule(lr){1-2}
\cmidrule(lr){3-14}
  \multicolumn{2}{c}{} &
  \multicolumn{4}{c}{YSOs} &
  \multicolumn{4}{c}{False positives\tablefootmark{b}} &
  \multicolumn{2}{c}{Uncertain\tablefootmark{c}} &
  \multicolumn{2}{c}{Total contamination\tablefootmark{d}} \\
\cmidrule(lr){3-6}
\cmidrule(lr){7-10}
\cmidrule(lr){11-12}
\cmidrule(lr){13-14}
  \multicolumn{1}{l}{Class} &
  \multicolumn{1}{c|}{Nr} &
  \multicolumn{1}{c}{All\tablefootmark{a}} &
  \multicolumn{1}{c}{P} &
  \multicolumn{1}{c}{F} &
  \multicolumn{1}{c}{D} &
  \multicolumn{1}{c}{Galaxies} &
  \multicolumn{1}{c}{Fuzz} &
  \multicolumn{1}{c}{MS/III} &
  \multicolumn{1}{c}{Artifacts} &
  \multicolumn{1}{c}{UG} &
  \multicolumn{1}{c}{UY} &
  \multicolumn{1}{c}{f.p.} &
  \multicolumn{1}{c}{f.p.+U} \\
  
\hline
All & 2827 & 2697 (95.4\%) & 176 & 175 & 2347 & 37 & 44  & 7 & 4 & 18 & 20  & 92 \phantom{0}(3.3\%) & 130 \phantom{0}(4.6\%) \\
\hline

D & 2442 & 2376 (97.3\%) &\phantom{0}10 & \phantom{0}59 & 2307 & 19 & 26 & 7 & 3  & --- &   11 & 55 \phantom{0}(2.3\%) & \phantom{0}66 \phantom{0}(2.7\%) \\

P & \phantom{0}330 & \phantom{0}303 (91.8\%) & 159 & 110 & \phantom{00}34 & \phantom{0}8 & 13  & --- & --- & \phantom{0}2 & \phantom{0}4 & 21 \phantom{0}(6.4\%) & \phantom{0}27 \phantom{0}(8.2\%) \\

FP & \phantom{00}49 & \phantom{00}15 (30.6\%) &\phantom{00}2 & \phantom{00}7 & \phantom{000}6 & 10 & \phantom{0}4 & --- & --- & 15 & \phantom{0}5 & 14 (28.6\%) & \phantom{0}34 (69.4\%) \\

RP & \phantom{000}6 & \phantom{000}3 (50.0\%) &  \phantom{00}3 & --- & --- & ---  &\phantom{0}1 & --- & 1 & \phantom{0}1 & --- & \phantom{0}2 (33.3\%) & \phantom{00}3 (50.0\%) \\
\hline

P,FP,RP & \phantom{0}385 & \phantom{0}321 (83.4\%) & 164 & 117 & \phantom{00}40  & 18 & 18 & --- & 1 & 18 & 9 & 37 \phantom{0}(9.6\%) & \phantom{0}64 (16.6\%) \\

\hline
\end{tabular}
\label{tab:Meg}

\tablefoot{
Shown is the re-classification and proposed contamination (false positives) from this work. Uncertain sources are given separately.
        \tablefoottext{a}{The total number of remaining YSO candidates is the sum of the three classes (P, F, D).}
        \tablefoottext{b}{Different types of false positives: Galaxies, Nebulosities (Fuzz), main-sequence stars or Class\,III candidates (MS/III), and contamination from image artifacts (C).}
        \tablefoottext{c}{The number of uncertain objects are given separately for uncertain galaxy  candidates (UG), and uncertain YSO candidates (UY).}
        \tablefoottext{d}{Summarized contamination, giving a lower and upper limit based on the sum of false positives (f.p.) and the sum when including uncertain candidates (f.p.+U)}
           }
\end{table*}

\begin{table*}[!ht]
\small
\begin{center} 
\caption{The \citetalias{Furlan2016} HOPS sample revisited in numbers. Similar to Table~\ref{tab:Meg}.} 
\begin{tabular}{llc | c | ccc | cccc}
\hline \hline
 \multicolumn{3}{c}{\citetalias{Furlan2016}} &
 \multicolumn{8}{c}{This work}  \\
\cmidrule(lr){1-3}
\cmidrule(lr){4-11}
 \multicolumn{3}{c}{} &
 \multicolumn{4}{c}{YSOs} &
 \multicolumn{3}{c}{False positives} &
 \multicolumn{1}{c}{Uncertain}\\
\cmidrule(lr){4-7}
\cmidrule(lr){8-10}
\cmidrule(lr){10-11}
 \multicolumn{1}{c}{Type} &
 \multicolumn{1}{c}{Class} &
 \multicolumn{1}{c|}{Nr.} &
 \multicolumn{1}{c}{All} &
 \multicolumn{1}{|c}{P} &
 \multicolumn{1}{c}{F} &
 \multicolumn{1}{c|}{D} &
 \multicolumn{1}{c}{Galaxies} &
 \multicolumn{1}{c}{Fuzz} &
 \multicolumn{1}{c}{Artifacts} &
 \multicolumn{1}{c}{UY} \\
\hline
YSOs & Modeled &           252 &           250 &          149 &            83 &           18 & 1    & 1   & --- & --- \\
\hline
YSOs & All     &           283 &           272 &          151 &            93 &           28 & 5    & 2   & --- & 4   \\

 & Class\,0  & \phantom{0}60 & \phantom{0}60 & \phantom{0}57 & \phantom{0}3 &          --- & ---  & --- & --- & --- \\
 & Class\,I  &           103 & \phantom{0}93 & \phantom{0}83 & \phantom{0}6 & \phantom{0}4 & 4    & 2   & --- & 4   \\
 & Flat      &           104 &           103 & \phantom{0}11 &           80 &           12 & 1    & --- & --- & --- \\
 & Class\,II & \phantom{0}16 & \phantom{0}16 &           --- & \phantom{0}4 &           12 & ---  & --- & --- & --- \\
\hline
Other & Galaxies  & \phantom{0}22 & \phantom{0}11 & 2 &  5  &  4   &   6   &  2   & 1   &  2   \\
      & Uncertain & \phantom{00}4 & \phantom{00}1 & 1 & --- & ---  &  ---  &  1   & 1   &  1  \\

\hline
\end{tabular}
\label{tab:Furlan}
\end{center}
\end{table*}
\begin{table*}[!ht]
\small
\begin{center} 
\caption{The 44 low--$A_K$ MGM protostar candidates revisited by \citetalias{Lewis2016}, compared to our results. Similar to Tabels~\ref{tab:Meg} and \ref{tab:Furlan}.}
\begin{tabular}{llc|cccccc}
\hline \hline
 \multicolumn{3}{c}{\citetalias{Lewis2016}} &
 \multicolumn{6}{c}{This work} \\
\cmidrule(lr){1-3}
\cmidrule(lr){4-9}
 \multicolumn{3}{c}{} &
 \multicolumn{3}{c}{YSOs}  &
 \multicolumn{2}{c}{False positives}  &
 \multicolumn{1}{c}{Uncertain}  \\
\cmidrule(lr){4-6}
\cmidrule(lr){7-8}
\cmidrule(lr){8-9}
 \multicolumn{1}{c}{Type} &
 \multicolumn{1}{c}{} &
 \multicolumn{1}{c|}{Nr.} &
 \multicolumn{1}{c}{P} &
 \multicolumn{1}{c}{F} &
 \multicolumn{1}{c}{D} &
 \multicolumn{1}{c}{Galaxies} &
 \multicolumn{1}{c}{Fuzz} &
 \multicolumn{1}{c}{UY} \\
\hline
\multicolumn{2}{l}{all MGM low--$A_K$ P} & 44  &  1  & 13  &  9  &  8  & 10  &  3   \\
\hline                    
YSOs &  Stage I   & 10  &  1  &  4  &  2  &  1  & --- &  2   \\
     &  Stage II  & 18  & --- &  9  &  6  &  3  & --- &  ---   \\
Other & Galaxies  & 4   & --- & --- & --- &  4  & --- & ---  \\
      & Fuzz      & 9   & --- & --- & --- & --- &  9  & ---  \\
      & Uncertain & 3   & --- & --- &  1  & --- &  1  &  1   \\
\hline
\end{tabular}
\label{tab:John}
\end{center}
\end{table*}
In Figure \ref{fig:mapfake} we show the distribution of the MGM YSO candidates and the proposed false positives, since this sample is the most used reference up to date. The top map shows the 330 more reliable MGM protostar candidates (P), and the bottom shows the 2442 MGM disk candidates (D). Previous MGM P candidates which are more scattered\footnote{in other words, not connected to regions of high dust column-density, or less clustered environments.} turned out to be extra-galactic contamination, uncertain sources \citepalias[see also][]{Lewis2016}, or false positives due to MIR nebulosities. The latter tend to be located close to the ONC region, as expected, due to higher contamination caused by the bright nebula. We compare our findings with the contamination estimates discussed in \citet{Megeath2012}. 
For the region inside the IRAC coverage ($\SI{\sim7}{deg^2}$) they expect about 44 false positives due to extra-galactic contamination. We find 37 galaxies (G) and 18 galaxy candidates (UG) in this region. There are further 20 uncertain YSO candidates, which might also be of extra-galactic nature. For the sub-samples D, P, and FP, \citet{Megeath2012} estimate about $\sim$11, $\sim$20, and $\sim$13 extra-galactic contaminants, respectively. Inside their given errors this corresponds roughly to the 19\,G, 8\,G (+2\,UG), and 10\,G (+15\,UG), that we found for each sub-sample. These are still lower limits, since, as already mentioned, remaining contamination by point-like extra-galactic sources can not be ruled out entirely. However, considering that our findings correspond well with the MGM contamination estimates, residual contaminants are likely a negligible fraction. Considering contamination due to nebulosities in the MIR, we get about 4\% and 1\% in the MGM P and D samples, respectively, although \citet{Megeath2012} estimated it to be a more negligible fraction. This fact highlights the unfortunate sensitivity to point-like outflow knots and cloud edges of MIR observations, influencing especially a protostar sample. However, Figure~\ref{fig:ccds} shows that for example, Herbig-Haro objects often show distinct colors in the NIR (first row, forth column) and in some other IR regimes. Hence, a careful color selection can mitigate at least some of these contaminants.

Furthermore, we compare our findings to the results in \citetalias{Lewis2016} (Table~\ref{tab:John}), who revisited a sub-sample of 44 MGM protostars (P) that are located at low dust column-density ($A_\mathrm{K}<\SI{0.8}{mag}$). These 44 sources are of interest to test the assumption that protostars (or star-formation) are connected to a certain extinction threshold \citep{Lada2010}. \citetalias{Lewis2016} concluded that ten out of the 44 low--$A_\mathrm{K}$ MGM Ps are likely Stage\,I candidates based on SED modeling \citep{Robitaille2006}, and discuss scenarios to explain the absence of significant dust at the location of these sources, including  source migration or ejection, and dust dissipation due to protostellar outflows. They use the same \emph{Herschel map} \citep{Lombardi2014} to estimate the extinction at the position of each YSO. However, based on the updated conversion factor from optical depth to $A_\mathrm{K}$ (see Sect.~\ref{Data}), the number of MGM protostars below the extinction threshold changes to 42. One of the two sources, which are now above the threshold, was classified as fuzz (MGM\,1286)  by \citetalias{Lewis2016} and the other as Stage\,I protostar (MGM\,333). The latter was classified as flat by \citetalias{Furlan2016}. However, we classify it as Class\,II candidate, due to the declining spectral index when including the K-band. Also, it lacks a PACS counterpart and is visible in the optical (Pan-STARRS $\mathit{g}=\SI{20.6}{mag}$). Moreover, it was classified as transition disk candidate in \citet{Kim2013,Kim2016}, which explains the flattish $\alpha$ in the MIR. In total, we find only one reliable protostar candidate among their rest nine Stage\,I sources (Table~\ref{tab:John}). Six more are likely more evolved YSO candidates (two disks, four flats), one is a galaxy, and two are uncertain sources, which need more investigation, to test the theory of an extinction threshold correctly. The rest 34 MGM Ps are re-classified by \citetalias{Lewis2016} into 18 Stage\,II candidates and 16 false positives or uncertain candidates. We confirm 15 of these Stage\,II sources as YSO candidates (six disks, nine flats) and we re-include one of their uncertain sources as disk candidate.

\begin{figure*}[!ht]
	\centering
        \includegraphics[width=1\linewidth]{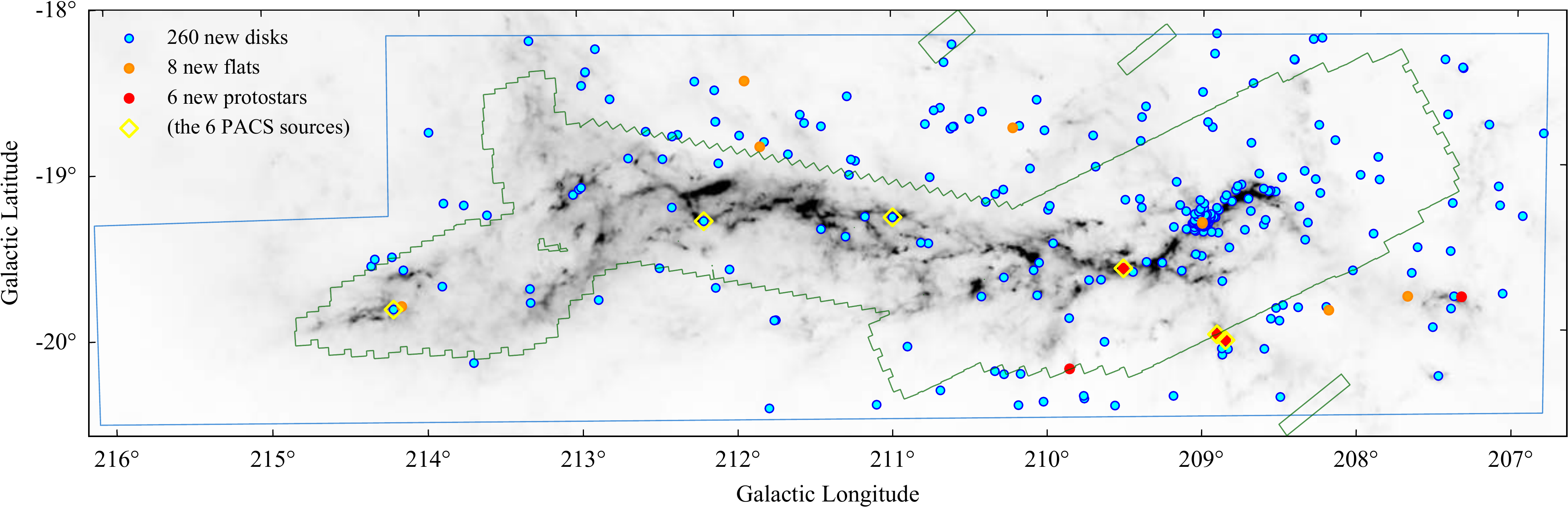}
	\caption{Distribution of the 274 new YSO candidates selected in this work. 
	Most sources were classified as new disk candidates (blue), and only few as flat-spectrum (orange) and protostar (red) candidates (see legend). The six extra PACS sources (3\,P, 3\,TD, Sect.~\ref{results new}), are additionally highlighted with yellow open diamonds. Sources beyond the green IRAC coverage are new VISTA/\emph{WISE} or VISTA/$M1$ selected YSO candidates, located in areas not included in previous studies.
	We note that there are only five red dots visible, marking the six new protostar candidates, because two of them are lying very close to each other, so the dots are blended.}
	\label{fig:mapnew_class}
\end{figure*}

Most of the \citetalias{Furlan2016} HOPS YSOs we confirm as reliable candidates ($\sim$96.1\%), especially those where SED modeling was possible ($\sim$99.2\%) (see also Table~\ref{tab:Furlan}). \citetalias{Furlan2016} list 22 galaxy candidates, of which 19 are MGM YSO candidates. We confirm 11 out of the 19 as YSO candidates. One reason for this misidentification of galaxies by \citetalias{Furlan2016} could be insufficient data quality. For example, if some bands are contaminated by artifacts or extended emission the SED might not fit to any YSO model. Also spectra in star-forming regions, which are affected by PAH emission, can be a combination of the YSO plus the nebulous surroundings, which can produce similar spectra as star-forming galaxies. Indeed, most of these sources are near regions of extended emission, and are at the same time associated with high extinction regions. This makes it unlikely that these are background galaxies, also given the fact that they are well visible in the NIR. 
Finally, there are one modeled \citetalias{Furlan2016} Class\,I candidate, and four modeled \citetalias{Lewis2016} YSO candidates (one Stage\,I, three Stage\,II), which we identify as resolved galaxies from visual inspection of the VISTA images. These findings are of interest, as they show that modeling alone is not always reliably separating YSO candidates from extra-galactic contamination, which was also pointed out by \citet{Evans2009} and \citet{Furlan2016}.


\subsection{Results for new YSO candidates} \label{results new}

With the color based NIR and MIR selection criteria (see Appendix~\ref{NewYSOsA}) we are able to add 268 new YSO candidates inside the whole VISTA coverage. Separating selections from inside (VISTA/\emph{Spitzer}) and outside (VISTA/\emph{WISE}/$M1$)\footnote{The VISTA/\emph{WISE} based selection adds 104 sources, leading to 117 in combination with the VISTA/$M1$ selection (when including the red MIPS coverage) beyond the IRAC coverage. Therefore, 43 sources are selected by both methods, meaning 13 are only selected by VISTA/$M1$.} the IRAC region, we select 151 and 117 new YSO candidates, respectively.

We add further six YSO candidates by using the PACS point source catalog (HPPSC) and PACS images. 
Two of these are new protostar candidates, located at a prominent young clustering, south-west of the ONC (Haro4-145 cluster, see also Appendix~\ref{cutouts} and Fig.~\ref{fig:cluster}), of which one is likely a new Class\,0 protostar (ID\,116363), not yet discussed in previous works.
Furthermore, we add another new protostar candidate (ID\,213612), detected during visual inspection, located inside the IRAC region right next to a known Class\,0 source (MGM\,1121, separation $\SI{\sim5}{\arcsec}$). 
This new candidate shows a prominent outflow cavity in the NIR. Both, the Class\,0 and the new candidate, lie on top of an elongated PACS source, and are also highlighted by \citet{Tobin2017-TALK} as protostar binary candidate. 
Finally, we add three transition disk candidates (ID 377204, 459841, 522530). These sources show no NIR or MIR excess, but a clear PACS excess, indicating an outer disk. Visually they seem to be surrounded by reflection nebulae in the NIR. See also Appendix~\ref{cutouts}, Fig.~\ref{fig:special}.

In total we add 274 new YSO candidates to the Orion\,A catalog inside the VISTA coverage, with 155 selected inside and 119 selected outside the IRAC region. The sources are classified with the methods discussed in Sect~\ref{classification} into six new protostar candidates, eight new flat-spectrum candidates, and 260 new disk candidates (Table~\ref{tab:reclassified} and Fig.~\ref{fig:mapnew_class}). Sources inside the IRAC coverage might have been missed previously due to different selection criteria, and by adding VISTA we gain sensitivity in the NIR. YSOs selected near the ONC often lack longer wavelength measurements ($\lambda \ngtr \SI{5}{\micro\meter}$), which can lead to erroneous classification of these sources. There are 67 such new disk candidates (333 disks total) with the longest measured wavelength at $\SI{4.5}{\micro\meter}$ ($I2$), mostly near the ONC. This lack of longer MIR observations leads to less reliable classifications. Therefore, some of these disks can still be flat-spectrum or protostar candidates. They can also be influenced by contamination near the ONC, that is not reflected in the photometry error, and for the same reason the extinction correction can be erroneous. However, visual inspection of these sources does not show signs of deep embeddedness or outflows, therefore, the Class\,II status is more likely than an earlier class. 

In Figure \ref{fig:mapnew_class} the new YSO candidates are shown on top of the \emph{Herschel map} including VISTA and \emph{Spitzer} survey contours. The new candidates in the surroundings are often located near the IRAC coverage, especially near the ONC region. Beyond the L1641 region to the Galactic south-east we find almost no new YSOs, whereas to the Galactic north of L1641 and to the Galactic south-west we find some scattering of new candidates. 
Overall, the distribution of the new candidates highlights the influence of the massive ONC, by showing a larger scatter near this region. A more detailed analysis of the (2D) distribution of our final sample will be discussed in Sect.~\ref{Distribution}.


\subsection{Final YSO sample and YSO re-classification} \label{Results all}

\begin{figure*}[!ht]
    \centering
    \begin{minipage}{1\linewidth}
        \centering
        \includegraphics[width=\linewidth]{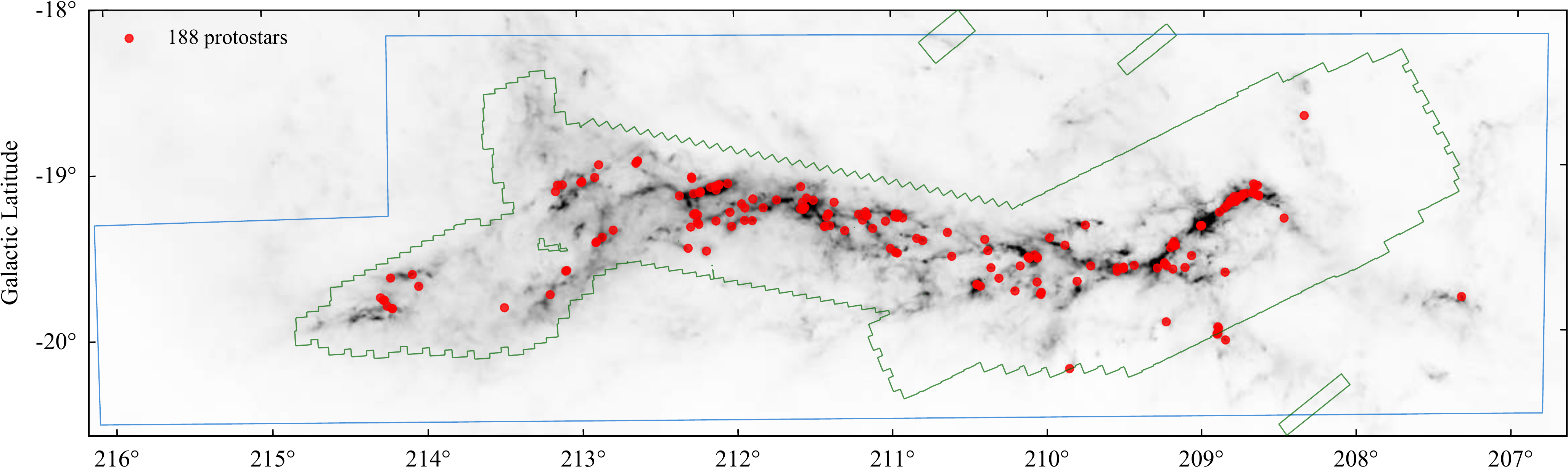}
    \end{minipage}%
    \vfill
    \begin{minipage}{1\linewidth}
        \centering
        \includegraphics[width=\linewidth]{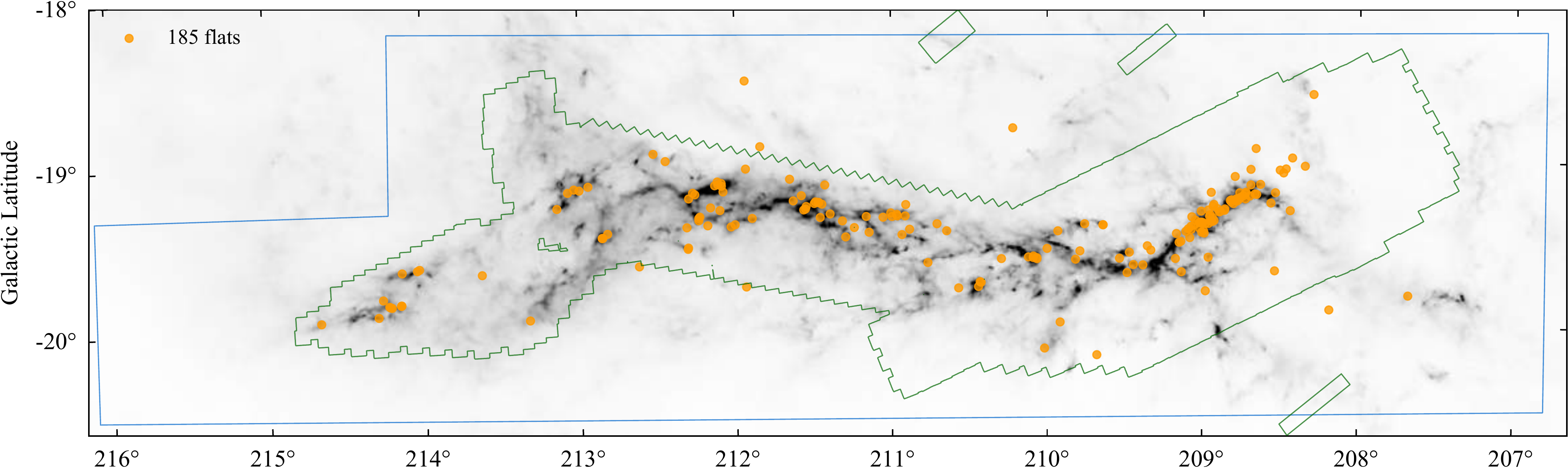}
    \end{minipage}%
    \vfill
    \begin{minipage}{1\linewidth}
        \centering
        \includegraphics[width=\linewidth]{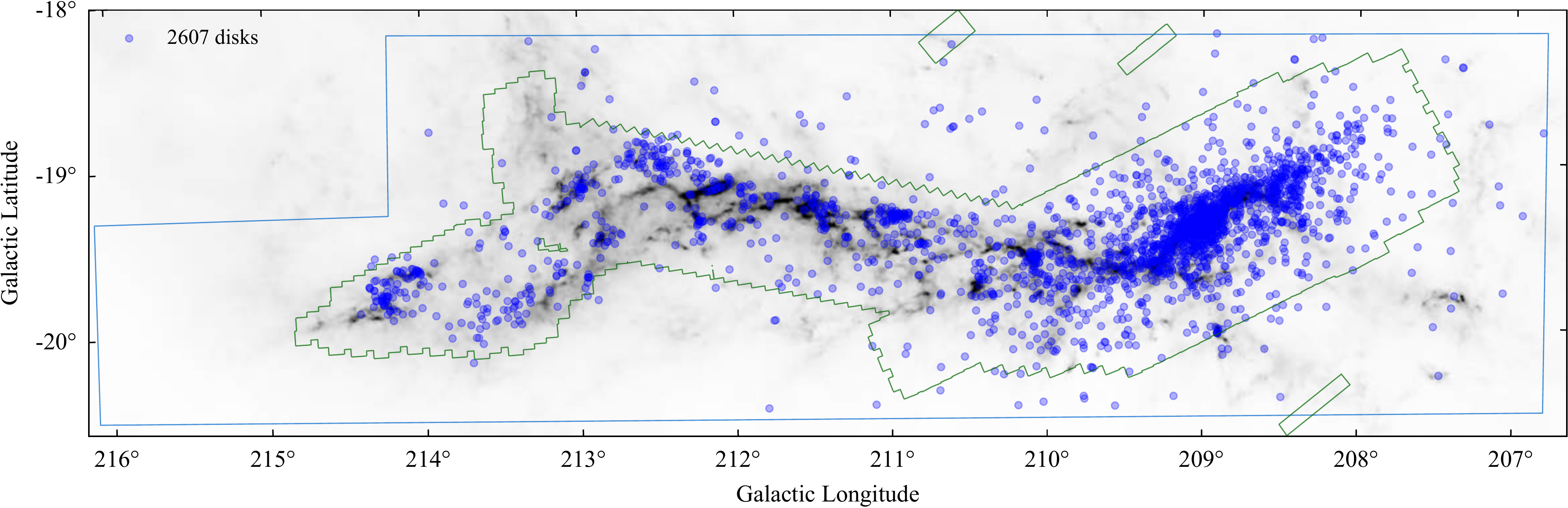}
    \end{minipage}%
    \caption{Distribution of all 2980 YSO candidates as selected and classified in this work. The classes are shown separately from top to bottom: protostar (Class\,0/I, top, red), flat-spectrum (middle, orange), and disk candidates (Class\,II/III, bottom, blue), displayed on the \emph{Herschel map}. The \emph{Spitzer}/IRAC and VISTA coverage contours are shown in green and blue, respectively. This highlights the location of new sources in the surroundings and the limitations of the VISTA survey coverage.}
    \label{fig:mapall}
\end{figure*}

The updated YSO catalog for Orion\,A contains 2980 YSO candidates with IR-excess, located inside the VISTA coverage. 
Included are the revisited 2706 YSO candidates (2839 minus 133)\footnote{133 = 92 rejected plus 41 uncertain}, plus the newly selected 274 sources. The final catalog is presented in Appendix~\ref{Catalogs}, which contains a column {\tt Class\_flag} for revisited (1), new (2), rejected (3), and uncertain candidates (4). The 2980 YSO candidates are classified into 2607 disk (Class\,II/III), 185 flat-spectrum, and 188 protostar (Class\,0/I) candidates (Table~\ref{tab:reclassified}). The flat-spectrum sources are composed of 59 previously identified MGM disks (32\%), 117 MGM protostars (63\%), and nine newly selected flats (5\%). 

We reclassify about 10\% of the MGM protostar candidates as disk candidates (34 disks out of 330 Ps, see also Fig.~\ref{fig:mapfake}).
These reclassified sources are mostly near the ONC. Reasons for the different classification are mainly due to extinction or contamination effects. When correcting for extinction, some sources do not show significant IR-excess to be classified as protostars by our methods. In addition, the different classification method compared to \citet{Megeath2012} can lead to different results, because some candidates do not show significant excess even without dereddening. Unfortunately, sources in the ONC region often lack longer wavelength detections due to saturation (e.g., missing IRAC3,4 or MIPS1) as already mentioned above. Using solely NIR colors can be ambiguous, therefore, we used visual inspection and a more detailed SED examination for a final decision (see Sect.~\ref{revisit}). For example, if the VISTA image shows a bluish NIR source and if the source has an optical counterpart it is very unlikely to be an embedded protostar. 
There are also rather exotic protostar candidates, showing typical protostellar-like red NIR to MIR colors but without a MIPS1 or PACS counterpart (in surroundings where these bands are not yet saturated). Other sources have a cataloged photometry entry in $M1$ while the images show only extended fuzzy counterparts in MIPS from the surrounding cloud structure, therefore, these are contaminated by extended emission. Overall, regions with extended emission (mainly near the ONC) are very critical areas, and the YSO classification in such regions is likely more prone to errors than in other regions.

A FIR measurement is another indicator for youth, since more evolved YSOs are too week in the FIR to be detected. 
We check especially the protostars and flats samples if they show a corresponding \emph{Herschel}/PACS counterpart by (a) using HOPS information from \citetalias{Furlan2016}, (b) using the \emph{Herschel} PACS point source catalog, and (c) visually inspecting the PACS images. 
Considering the 188 Class\,0/I candidates, there are 168 sources (89\%) with a clear PACS counterpart. Out of the remaining 20 sources there are six with no counterpart and for 14 we can not tell 
due to extended emission, crowded regions, or saturated regions near the ONC. This makes these 20 sources suspicious or more uncertain protostar candidates.
Out of the 185 flat-spectrum sources, 102 (55\%) coincide with a PACS point-source, suggesting that these flats might still be associated with envelopes. For the rest, there are 35 (19\%) without PACS, and for 48 (26\%) we can not be sure, due to mentioned contamination issues. The flats with PACS are overall brighter than those without (see also Sect.~\ref{flats}). 
We did not check all disk candidates visually for PACS counterparts but looked for cross-matches with HOPS or the HPPSC. Out of the 2607 disk candidates, 249 ($\sim$10\%) are clearly associated with a PACS point source. 

The resulting spatial distribution of the three YSO classes is presented in Fig.~\ref{fig:mapall}. By eliminating false positives, the distribution of protostars now appears to be less scattered and more confined to regions of high dust column-density. Moreover, protostars and flats show a similar distribution and are almost equal in sample sizes. Both seem to be connected to or located near regions of high dust column-density, whereas the disk sources are already more dispersed, while also larger in number. Hence, we quantify this behavior in Sect.~\ref{Distribution}.


\section{Discussion} \label{Discussion}

In this section we firstly discuss the completeness of the YSO sample, and secondly the issues that come with YSO classification, especially concerning the flat-spectrum class. Finally we discuss the distribution of the three YSO classes with respect to regions of high dust column-density. This is done with a statistical approach, to rule out that flat spectrum sources are solely a mixture of protostars and disks.


\subsection{Completeness} \label{Completeness}

Estimating the completeness of our selection, or any similar selection, is complicated. We will partly refer to \citet{Megeath2016}, who estimated the completeness of the MGM sample in two ways. First, they estimated the nebular background and source confusion, using the route median square deviations (RMEDSQ) of the IRAC pixels surrounding each YSO candidate. This gives an estimate of the incompleteness due to local MIR background emission, which is spatially varying, and increasing with stellar density. 
Second, they used COUP data at the ONC, to estimate the incompleteness in the crowded ONC region, which is affected by very bright IR nebulosity, and high extinction. They do this by carefully comparing the number of COUP sources with and without IR counterparts to their known \emph{Spitzer} YSOs (MGM sample). 
With this approach they correct the number of YSO candidates with IR-excess in Orion\,A from 2821\footnote{This number does not include the six red protostar candidates (RP), since the completeness was estimated for IRAC.} to 3191, using the COUP correction, and finally to 4199, using the correction due to local MIR background emission. This means an incompleteness of about 49\% for the Orion\,A sample inside the IRAC coverage.

The YSO sample in this paper, inside the IRAC coverage, includes the revisited 2694 MGM sources\footnote{2821 MGM sample minus 127 (90 false positives and 37 uncertain)} plus the new 151 candidates added inside the IRAC region, leading to 2845 YSO candidates. 
The final number is similar to the original MGM sample size, therefore, we adopt their completeness estimate of about 49\% as an upper limit. 

We now focus on the COUP covered region containing 630 MGM sources. \citet{Megeath2016} estimate 370 extra sources after applying the COUP correction, meaning there should be about 1000 YSOs with IR-excess in the relatively small coverage. We added 73 new candidates in this region (Sect.~\ref{NewYSOs}), of which 56 are X-ray detected COUP sources, meaning that we were only able to add about 15\% of the estimated missing sources toward the ONC, or about 20\% including the 17 sources without an X-ray counterpart. Assuming MGM completeness, we are still missing about 30\% of YSOs with IR-excess toward the ONC. 
Also of note in this context, about 75\% of the newly identified YSOs with IR-excess have an X-ray counterpart within the COUP coverage. This provides an independent support for these new candidates. Considering the whole YSO sample (revisited + new), there are about 81\% IR YSOs with an X-ray counterpart within the COUP coverage.

The \emph{WISE} completeness is not directly comparable to the \emph{Spitzer} completeness. The inferior resolution and sensitivity of \emph{WISE} misses faint sources and sources in crowded regions.
To test the VISTA/\emph{WISE} selection presented in this work (Appendix~\ref{NewWISEA}), we check how many sources can be recovered inside the IRAC region, restricting this analysis to L1641 ($l > \SI{210}{\degree}$). This is a fair comparison for regions not as complicated as the ONC in the MIR (\emph{WISE} saturates toward the ONC). We are able to recover about 59\% of previously known YSO candidates in L1641. 
This shows what can be achieved with \emph{WISE} in combination with deep NIR data in low-mass star-forming regions.
Including the ONC region we recover only about 38\%, highlighting the influence of massive-star-forming regions on low resolution MIR data.

\begin{figure}[!ht]
    \centering
    \includegraphics[width=0.7\linewidth]{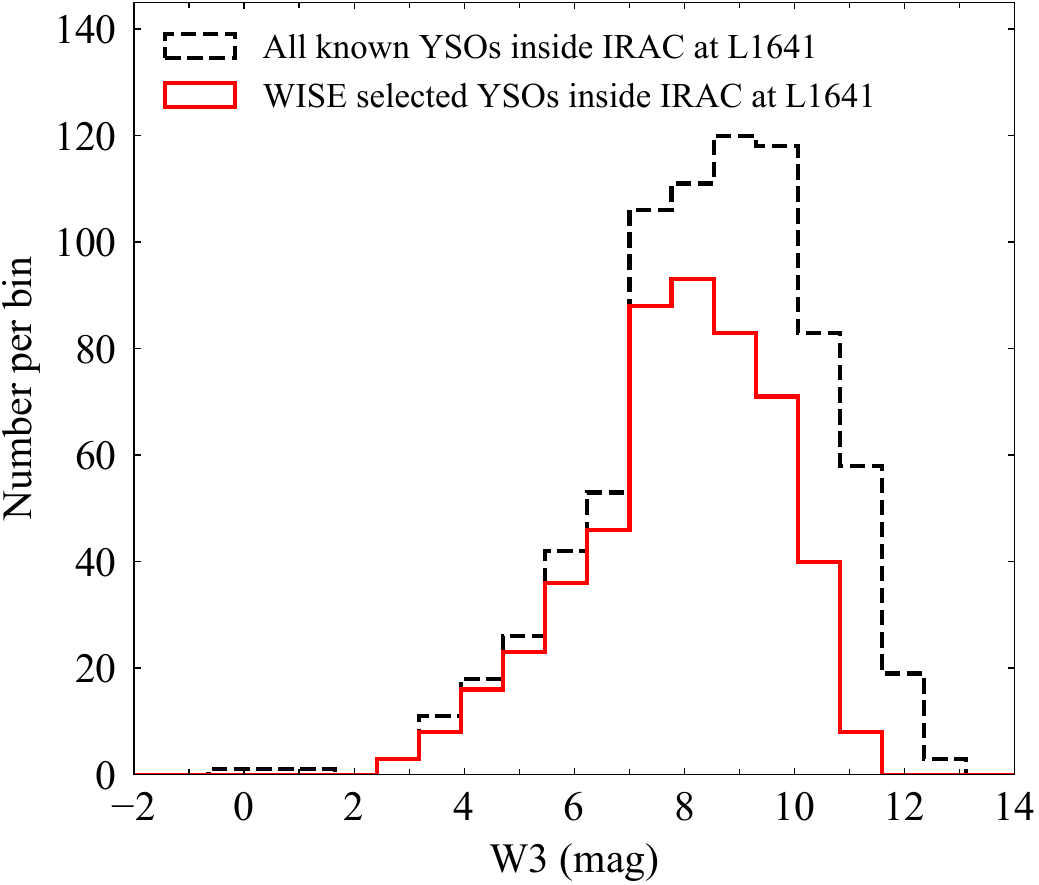}
    \caption{Histogram of the WISE $W3$ magnitude for all known YSO candidates inside the IRAC region. Included are only sources located at L1641 ($l>\SI{210}{\degree}$) with \texttt{w3sigmpro<0.5} (black dashed line). The red solid line represents YSOs selected by the VISTA/\emph{WISE} selection in the same region.}
    \label{fig:W3_hist}
\end{figure}

To test the effect of crowding on \emph{WISE} based selections, we redo the recovery test by comparing to only those MGM YSOs in L1641 with no other \emph{Spitzer} source closer than \SI{6}{\arcsec} as nearest neighbor. Surprisingly, we do not find a significant difference, and get again a recovery rate of about 59\% when comparing only to non-crowded \emph{Spitzer} YSOs\footnote{ The recovery rate is about 53\% when comparing the non-crowded VISTA/\emph{WISE} selection to all \emph{Spitzer} observed YSOs in L1641.}. This suggests that \emph{WISE} is mainly limited by sensitivity issues, since we are losing mostly faint YSO candidates, due to our error and magnitude cuts and various steps to clean the \emph{WISE} data of extended emission. 
This is highlighted in Fig.~\ref{fig:W3_hist}, comparing the $W3$ magnitude of all known YSO candidates in the IRAC L1641 region to those selected by VISTA/\emph{WISE}. The \emph{WISE} selection is especially incomplete for sources fainter $W3 \gtrsim \SI{7}{mag}$.

We can use the 59\% recovery rate to estimate the completeness of our VISTA/\emph{WISE} selection outside IRAC, where we added 104 new YSO candidates with this method. If the YSO density beyond the IRAC coverage is similar to a low-mass star-forming region like L1641, we would expect about 72 additional YSO candidates in the surroundings. Adding this to the 104, a \emph{Spitzer} based selection would have selected 176 candidates. Now we can add the \citet{Megeath2016} completeness estimate for the  \emph{Spitzer} YSOs of 49\%. With this we get an upper limit of new YSO candidates beyond the IRAC region of 359 sources. 
The combined VISTA/\emph{WISE} and VISTA/$M1$ selections give 117 new YSO candidates outside IRAC, meaning we selected only about 1/3 of possible new candidates. 
However, the completeness was estimated by \citet{Megeath2016} based on bright MIR nebulosity near the \emph{Spitzer} YSOs in the whole Orion\,A region, including the ONC. Since regions outside IRAC are less influenced by background MIR emission, the 359 are indeed an upper limit, as it is  likely that we are missing less sources toward these regions. Moreover, the YSO density decreases beyond the IRAC coverage, meaning less crowded sources, which also suggests that, locally, we likely miss less than two thirds of the YSO candidates.

Comparing the $\sim$59\% recovery rate with other \emph{WISE} based selections from the literature, we find that we recover slightly more than \citet{Koenig2015}, with a 50\% recovery rate. They compare the \citet{Koenig2014} \emph{WISE} YSO selection scheme with \emph{Spitzer} selected YSO candidates in various regions. 
Another previous \emph{WISE} based study covering the whole Orion\,A region \citep{Marton2016}, using machine learning based selection criteria, recovers about 20\% of YSO candidates at Orion\,A, with a small fraction of contamination ($\sim$3\%). 
Compared to our selection, we recover almost twice as much, considering the 38\% recovery rate when including the ONC.


\subsection{Inferring on the meaning of flat-spectrum sources} \label{flats}

In this work we perform YSO classification based on the MIR spectral index, defined by the observed IR excess. This is sometimes just a rough estimate of the true evolutionary Stage, however, for low-mass stars the method is a well established tool \citep[e.g.,][]{Lada2006}. Using a grid of SED models would give more detailed results, by taking into account inclination and/or extinction effects \citep[e.g.,][]{Whitney2003a, Robitaille2006, Crapsi2008, Forbrich2010, Furlan2016}, although to accurately model an entire YSO population would require complete reliable photometric observations, covering wide wavelength ranges, ideally reaching from the optical to the FIR and mm-range, as shown by \citet{Furlan2016}. Additionally, obtaining spectra (e.g., \emph{Spitzer}/IRS) gives further useful insights into emission and absorption lines (H$_\alpha$, H$_2$O, Si, PAH). However, this would exceed the scope of this paper, where we focus on a statistically significant sample, while a detailed analysis using SED modeling is generally only possible for smaller sub-samples of a YSO population.
Therefore, we like to review the various uncertainties influencing the reliability of YSO classification. 

Particularly uncertain are the flat-spectrum sources, since they are at the border between Classes\,I and II, spanning a narrow spectral index range. The physical significance of defining flat-spectrum sources between $-0.3 < \alpha < 0.3$, as suggested by \citet{Greene1994}, is highly debatable. For example, \citet{Teixeira2012} use $-0.5 < \alpha < 0.5$. Also, if there would be a physically meaningful separate class between Classes I and II, one would expect three distinct over-densities in the various CCDs and CMDs, which is not observed. 
To make things even more complicated, more massive stars disperse their disks faster \citep{Lada2006}, so different SED shapes are expected just as a consequence of the mass distribution of the YSOs, even if all stars had the same age. 
Moreover, \citet{Whitney2004} show that also the luminosity (or mass) of the central YSO (not only of the disk) influences the SED shape. For example, the emission of the stellar photosphere of a low mass star peaks in the NIR, while for more massive stars it peaks in the optical. This means the latter contribute less to the NIR part of the SED, which leads to a slightly more rising observed IR SED, even if the disk mass and extend is the same as that of the lower mass star. This suggests that classification is a function of luminosity, which can actually be seen in some color-magnitude diagrams (see Fig.~\ref{fig:cmds}), or when plotting $\alpha$ versus a magnitude. 
Therefore, the traditional definition of flat-spectrum sources seems to introduce a bias toward brighter sources.
Moreover, \citet{Heiderman2015} looked for envelope tracers in a significant sample of Class\,0/I and flat-spectrum sources, and find that about 50\% of flats are true Stage\,0/I sources. Therefore, they conclude, that nothing distinctive occurs within the flat-spectrum category, suggesting that this category has no physical significance.

On top of that, classification can be influenced by different geometric effects, like disk inclination or extinction effects, as highlighted in the introduction. \citet{Crapsi2008} find that seeing the disk near edge-on, can be responsible for most observed flats, also suggested by \citet{Chiang1999} and \citet{Whitney2003a, Whitney2003b, Whitney2004}. 
To test the effect of disk inclination, we use the SED models of \citet{Robitaille2006} for Class\,II YSOs. For sources with more than $\SI{2}{M_{\sun}}$ and inclinations greater than $\SI{75}{\degree}$ (disk is seen near edge-on) we find that less than 1\% are misidentified as flat-spectrum sources. However, the majority of YSOs are low-mass stars ($M<\SI{2}{M_\sun}$), for which we estimate that about 3.6\% would show flat spectra due to disk inclination effects, which corresponds to about half of the observed flat-spectrum sources in our sample.  
This is similar to the $\sim$50\% Stage\,II sources, that \citet{Heiderman2015} find in their flats sample. 
However, \citet{Muench2007} find that flats tend to be overall more luminous than disks. At the same time they point out that edge-on disks tend to be sub-luminous, due to obscuration by the disk. If flats were caused largely by inclination effects, this would contradict the first statement.
We checked the luminosity of the YSOs by calculating the bolometric luminosity (L$_\mathrm{bol}$) with the method from \citet{MyersLadd1993}.
We get a median of $(0.2 \pm 0.1) \mathrm{L}_{\odot}$, $(0.7 \pm 0.6) \mathrm{L}_{\odot}$, and $(1.4 \pm 1.1) \mathrm{L}_{\odot}$, for disks, flats, and protostars, respectively. For the dereddened photometry we get $(0.3 \pm 0.3) \mathrm{L}_{\odot}$, $(1.6 \pm 1.4) \mathrm{L}_{\odot}$, and $(2.0 \pm 1.7) \mathrm{L}_{\odot}$. Indeed, the flats are overall more luminous than the disks, also, they lie in-between the disks and protostars. 
This gives the impression that flat-spectrum sources can be interpreted as a transitional evolutionary class.

Moreover, \citet{Furlan2016} showed that most of their investigated sample of flat-spectrum sources in Orion\,A show signs of envelopes when applying SED modeling. They point out, that this sample likely represents protostars at different stages in their envelope evolution.
\citet{Megeath2012}, who investigated the whole dusty YSO population, only presented a simple color based separation into disk dominated PMS stars (D) and protostars (P) which are similar to Classes\,II and I, respectively.
The flat-spectrum sources in this paper are composed of 32\% previously classified MGM Ds and 63\% MGM Ps, which would also suggest at first guess that these sources are likely younger compared to the average Class\,II, and not simply disk inclination effects. 
Moreover, \citetalias{Lewis2016} point out, that 15\footnote{Actually they find 18 Stage\,II sources, however, we excluded three false positives.} out of their investigated 44 low-$A_\mathrm{K}$ MGM protostars are modeled as Stage\,II YSOs, which show overall a rather flat SED. They suggest that these are still very young, likely at the beginning of the disk dominated PMS phase, and therefore were misclassified as protostars previously.
Also pointed out by \citetalias{Lewis2016}, the median spectral index for disks between 2 and $\SI{8}{\micro\meter}$ ($\alpha_\mathrm{KI}$) is about $-1.33$, which is the expected value for a spatially flat accretion (or reprocessing) disk. We get a similar median for this spectral index of $-1.42$ to $-1.28$ (de-reddened and observed).
This suggests that the majority of the disk sources are not highly flared. This might be explained by sufficient dust settling onto the circumstellar disk during the Class\,II evolution \citep{DAlessio1999, Lewis2016}.

We can contribute to this discussion by looking at the spatial distribution of the various YSO classes with respect to regions of high dust column-density.
Figure~\ref{fig:mapall} suggests a stronger connection of protostars and also flats to these regions, while disk sources are more dispersed.
The stronger connection to denser cloud regions of these two classes was also pointed out by \citet{Heiderman2010} and \citet{Heiderman2015}.
However, if flats were a result of disk inclination effects, they should be more evenly distributed, similar to confirmed disks. Unfortunately, the sample sizes are not directly comparable. There are about a factor of ten more disks than flats or protostars. To make sure that we are not dealing with small number statistics, we will quantify the distribution in the next section.


\subsection{Distribution of YSOs with respect to regions of high dust column-density} \label{Distribution}

\begin{figure*}[!ht]
    \centering
    \includegraphics[width=1\linewidth]{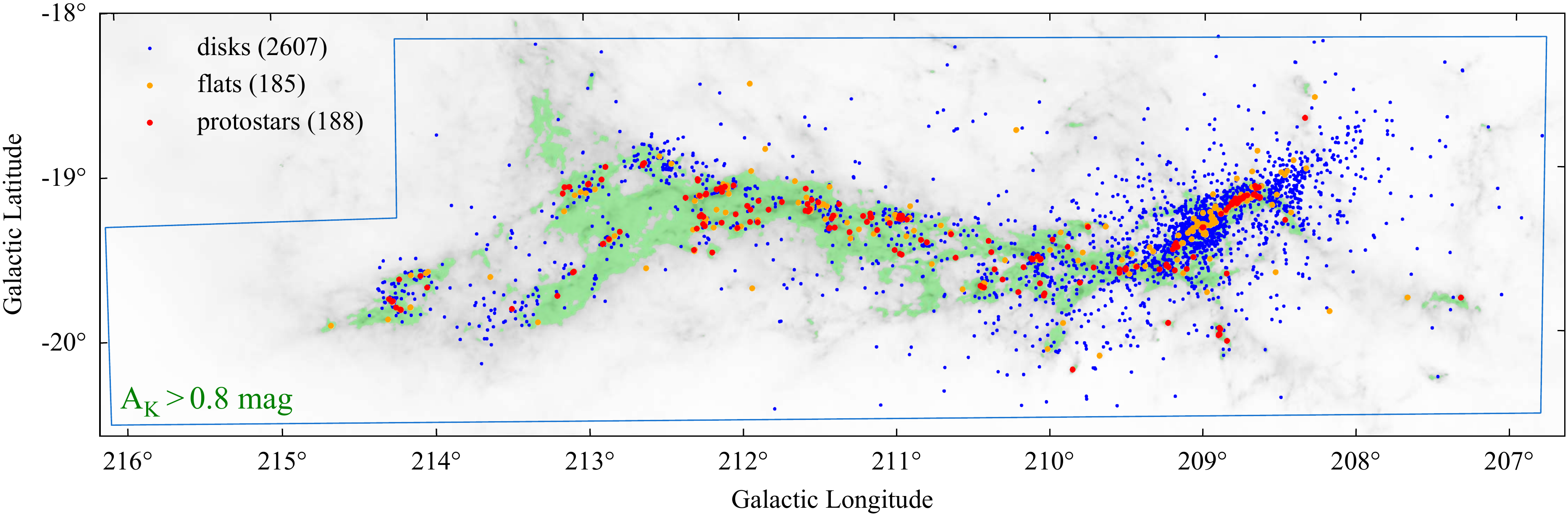}
    \caption{The green shaded area indicates the extinction threshold of $A_\mathrm{K,Herschel}>\SI{0.8}{mag}$. Superimposed are the YSO candidates (see legend). For each source we calculate the projected distance to the closest pixel in the \emph{Herschel map} (green) above the extinction threshold. The resulting normalized cumulative distribution is shown in Fig.~\ref{fig:NND}.}
    \label{fig:NNDmap}
\end{figure*}

\begin{figure}[!ht]
    \centering
    \includegraphics[width=1\columnwidth]{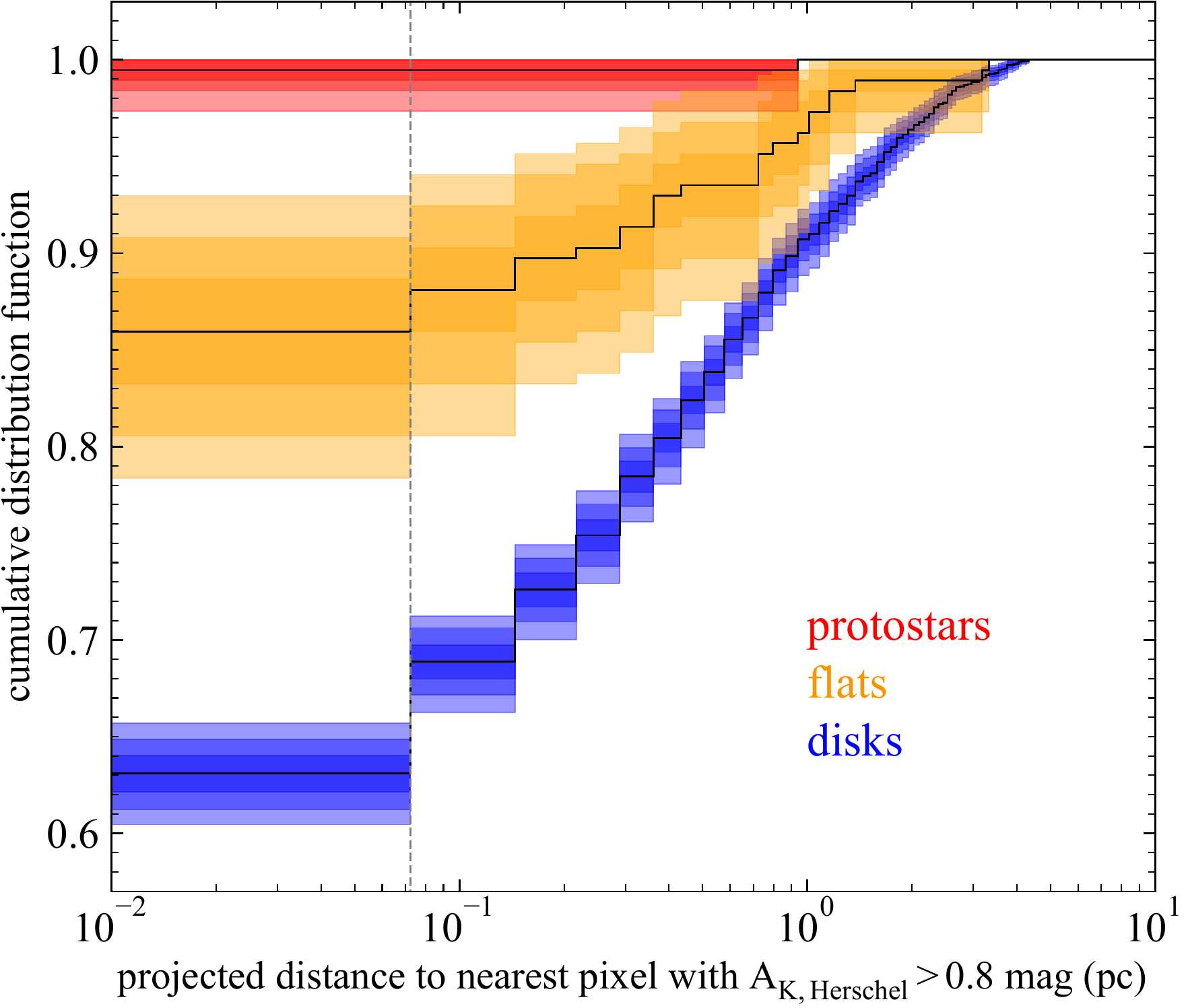}
    \caption{Normalized cumulative distribution of the projected distances of YSOs to the nearest \emph{Herschel map} pixel with $A_\mathrm{K,Herschel}>\SI{0.8}{mag}$ (green pixels, Fig.~\ref{fig:NNDmap}). Separated in colors are the three YSO classes: Class\,I (protostars, red), flat-spectrum sources (flats, orange), and Class\,II/III (disks, blue). The observed distributions are shown as black lines, and the color shaded areas show the confidence interval contours at 68.3\%, 95\%, and 99.7\% ($\SI{1}{\sigma}$, $\SI{2}{\sigma}$, and $\SI{3}{\sigma}$). The first bin, indicated by the vertical gray dashed line, gives the resolution of \emph{Herschel} (\SI{36}{\arcsec}, \SI{0.07}{pc} @ \SI{414}{pc}). Hence, sources in the first bin are projected directly on top of regions of high dust column-density.}
        \label{fig:NND}
\end{figure}

The spatial distribution of YSOs in Orion\,A was investigated by \citet{Gutermuth2011}, \citet{Pillitteri2013} and \citet{Megeath2016},
and a connection of protostars with high column-density was pointed out by \citet{Megeath2012}. Such a behavior was also highlighted, for example, by \citet{Muench2007, Jorgensen2007, Lada2010}; or \citet{Hacar2016}. Moreover, \citet{Heiderman2015} indicated the same for flats. Furthermore, \citet{Teixeira2012} show that sources with thick disks are stronger connected to high-extinction regions as compared to more evolved anemic disks in NGC\,2264. 
Recently, \citet{Hacar2017} found a strong correlation of protostars (Class~0/I) with the dense gas structure in NGC\,1333 (not only high column-density), by observing N$_2$H$^+$ line emission, as dense gas tracer. However, they do not find a significant connection of flat spectrum sources with dense gas. Unfortunately, we do not know (yet) the distribution of dense gas (volume-density) in the whole Orion\,A region. However, we can use the dust column-density from \emph{Herschel} to infer on the connection of YSOs to a certain column-density threshold. For this we use a star formation threshold of $A_\mathrm{K}>\SI{0.8}{mag}$ as suggested by \citet{Lada2010}. This is now possible for a larger field, since most of the above listed previous studies were limited by the available survey coverages, while the majority of YSOs connected to Orion\,A should be present within the VISTA coverage. 

To quantify the distribution of the three YSO classes with respect to regions of high dust column-density, we evaluate the closest distance\footnote{astropy.coordinates.match\_coordinates\_sky \\(\url{http://www.astropy.org)}} to the next \emph{Herschel map} pixel above the threshold (Fig.~\ref{fig:NNDmap}). 
The resulting normalized cumulative distribution function of the distances (given in pc) is presented in Fig.~\ref{fig:NND}, with the bin-size corresponding to \emph{Herschel} resolution.
The displayed confidence intervals are obtained with bootstrapping. To this end, we draw random values out of each sample with replacement with 2000 iterations, while the sub-samples have the same size as the original sample size of each class. 
The resulting distributions are significantly different from each other within $3\sigma$.
Essentially all protostars are seen in projection of regions of high dust column-density ($99.5\%\pm0.5$\%), while flats also show a stronger connection ($86.0\%\pm2.7$\%) compared to disks ($63.1\%\pm1.0$\%)\footnote{The percentages in parenthesis give the fraction of sources in the first bin in Fig.~\ref{fig:NND}, with the standard deviation as uncertainty, corresponding to a 1$\sigma$ uncertainty.}. 
To get a measure for the background we check the distribution of all 800,000 VISTA sources, and find that only about 7.7\% of these sources are projected on regions above the extinction threshold.

Looking at the original MGM YSO catalog, there are $90.0\%\pm1.7\%$ Ps and $64.8\%\pm1.0\%$ Ds projected above the threshold. Compared to our results, we see that the MGM protostars show a less clear connection to regions of high dust column-density, while the disk samples are similar within the errors. Differences are due to the exclusion of false positives and YSO reclassification by including flat-spectrum sources.

To check the influence of the chosen flat-spectrum range of $-0.3<\alpha<0.3$ on the spatial distribution result in Fig.~\ref{fig:NND}, we re-did the above test with a larger range of $-0.5<\alpha<0.5$ \citep[e.g.,][]{Teixeira2012}. We find that the resulting distributions still show a significant difference between the classes, and the fraction of sources projected on top of high column-density stays essentially the same for each class.

Next, we investigate the possibility that the flat-spectrum sources presented in this paper are a simple mix of disks and protostars. For this we created a random mix of these two classes following the ratio of protostars to disks (0.072) to create a population of ``synthetic flat sources''. In this case, it is clear that the distribution of the synthetic flat sources ($65.7\%\pm3.5$\%) is substantially different from the observed distribution of flats, being actually very similar to the distribution of disks, and can be ruled out. 
As a second more stringent test, we did the same experiment, but only for disks that would be observed as flats due to inclination effects (estimated to be about 3.6\% of the total sample of disks, see Sect.~\ref{flats}). This leads to an almost even number of Ps (91) and Ds (94) to be drawn randomly from these samples. 
In this case we find that the spatial distribution of synthetic flat sources is similar within $3\sigma$ to the observed one, while being marginally different from each other within the the $1\sigma$ range, with $80.9\%\pm2.6$\% projected on regions of high column-density. Nevertheless, we would expect that in the latter scenario the flat sources would be on average fainter because of the obscuration of the edge-on disk, which is the opposite of what is observed (see Sect.~\ref{flats}). This argument was also made in \cite{Muench2007}. These simple experiments suggest that a flat-sources population created solely as a mix of protostars and disks is unlikely, or in other words, that most flats tend to be younger and therefore closer to the protostellar phase.

Overall, the spatial distribution of protostars in Orion\,A supports previous findings, for example by \citet{Hacar2017} in NGC\,1333, but, we also find a stronger correlation of flats to high column-density regions as compared to Class\,IIs. This is not supported by \citet{Hacar2017}, who does not find a significant correlation of flat-spectrum sources with dense gas (as traced by N$_2$H$^+$). One reason could be that dust column-density does not only probe the real dense gas; by using an extinction threshold of $A_\mathrm{K} > \SI{0.8}{mag}$, we also include lower density and more diffuse regions, which are not included when specifically using dense gas tracers. 
However, by investigating a much larger region in Orion\,A compared to NGC\,1333\footnote{The investigated Orion\,A region is about a factor of 300 larger than that of NGC\,1333.} and by having a larger statistically significant sample of YSOs, we can use the column-density as indicator for regions of dense cloud material. This suggests that flat-spectrum sources are indeed a younger evolutionary stage, and not simply disk inclination effects. Even if the flat sources are not anymore directly connected to dense gas, like in NGC 1333, they are still located near regions of higher column-density. This indicates that they did not have enough time to disperse sufficiently to show the same distribution as the more evolved Class\,IIs. 

To make a stronger statement about the distribution of classes, follow up observations are needed to confirm the YSO nature of uncertain sources, and also of the scattered flat-spectrum sources. 
As shown for example by \citet{Heiderman2015}, many of previously classified protostars and flat-spectrum sources that were found in regions of low dust column-density turned out to be background contamination. This can also be the case for some of the sources in the updated Orion\,A YSO catalog.


\section{Summary} \label{Summary}

We have revisited and validated previous YSO catalogs \citep{Megeath2012, Megeath2016, Furlan2016, Lewis2016} of the Orion\,A star-forming region using deep NIR ESO-VISTA data \citepalias[VISION,][]{Meingast2016}, and added new YSO candidates in the larger field covered by VISTA, in combination with \emph{Spitzer}, \emph{WISE}, and \emph{Herschel}/PACS.
We summarize our results as follows:

\begin{enumerate}

\item   We identified 274 new YSO candidates (six protostars, eight flat-spectrum sources, and 260 PMS stars with disks) of which 268 were selected by combining VISTA, \emph{Spitzer}, and \emph{WISE} based selection criteria, and six candidates were selected by including \emph{Herschel}/PACS photometry (Sects.~\ref{NewYSOs} \& \ref{results new}). A total of 119 candidates were found in regions beyond the previously analyzed \emph{Spitzer}/IRAC survey. The rest are selected in regions covered by \emph{Spitzer}/IRAC and were likely missed in previous works due to (a) different selection criteria, (b) sensitivity and saturation issues at longer wavelengths, and (c) crowding and nebula contamination, especially in regions near the ONC.

\item Among the previously known 330 protostars and 2442 disk sources from \citet{Megeath2012, Megeath2016}, contamination levels are at least 6.4\% and 2.3\% respectively (Sect.~\ref{Results revisited}), mostly due to background galaxies or unresolved nebulosities, which were identified visually (Sect.~\ref{revisit}). These numbers are lower limits, because we can not rule out a residual degree of contamination, mainly due to unresolved galaxies, or background giants.
With this we conclude that previous surveys of Orion\,A are largely reliable, especially concerning the more evolved Class\,IIs, but they are incomplete due to limited survey areas and sensitivity issues near regions of bright nebula. The latter is also still an issue for this updated catalog, although we were able to slightly reduce the incompleteness near the ONC region for sources with IR-excess, and we extended the spatial completeness by using the larger field observed by VISTA.

\item The new catalog contains 2980 YSO candidates, including the 274 new and the 2706 revisited YSO candidates (Sect.~\ref{Results all}). They are classified as 188 (6.3\%) protostar, 185 (6.2\%) flat, and 2607 (87.5\%) disk candidates, using extinction corrected spectral indices. 

\item Within the {\it Chandra} observed COUP field, 81\% of the IR YSOs, and 75\% of the newly identified IR YSOs (Sect.~\ref{Completeness}) appear to be associated with X-ray YSOs from \citet{GetmanA2005}. This provides independent support for the validity of these new IR selected YSOs. Considering the whole YSO sample from this work, about 38\% of all IR YSOs have an X-ray counterpart, while large areas of the VISTA observed region are not covered by X-ray surveys (Sect.~\ref{Aux}).

\item We estimate that a search for YSOs in Orion\,A using \emph{WISE} and VISTA recovers about 59\% of the known YSOs in the region of L1641 ($l<\SI{210}{\degree}$, excluding the ONC) inside the \emph{Spitzer}/IRAC coverage (Sects.~\ref{Completeness} \& \ref{app:recovery}). This shows what can be achieved by the all-sky \emph{WISE} survey in combination with deep NIR data in regions not contaminated by massive star formation.

\item The spatial distribution of protostars follows essentially the regions of high dust column-density ($A_\mathrm{K}>\SI{0.8}{mag}$) as traced by \emph{Herschel}. The distribution of flat-spectrum sources relative to regions of high dust column-density shows also a significantly stronger correlation with these regions compared to Class\,IIs (Sect.~\ref{Distribution}). 
This is a strong indication that they correspond to a younger evolutionary phase, and are not simply affected by disk or envelope projection effects or foreground extinction. 
With this we confirm earlier works \citep[e.g.,][]{Furlan2016} that show that flats are likely protostars at different stages in their envelope evolution. 
Other studies find \citep[e.g.,][]{Heiderman2015} that only about 50\% of flat-spectrum sources show signs of envelopes when using dense gas tracers. However, with the result from this work we can add to this discussion; maybe not all flats are still associated with envelopes and dense gas, but most are likely still very young, either at the end of the embedded protostellar phase or at the beginning of the disk-dominated PMS phase, since they had not enough time to disperse similarly as the Class\,II YSO population. 
To establish a final conclusion in the future, detailed SED modeling and follow-up observations are needed for all uncertain sources. 

\end{enumerate}

\begin{acknowledgements}
Author Gro\ss schedl greatfully acknowledges funding by the Austrian Science Fund (FWF) under project number P 26718-N27.
This work is based on observations made with ESO Telescopes at the La Silla Paranal Observatory under program ID 090.C-0797(A).
This work is part of the research program VENI with project number 639.041.644, which is (partly) financed by the Netherlands Organisation for Scientific Research (NWO). AH thanks the Spanish MINECO for support under grant AYA2016-79006-P.
We thank John~A.~Lewis for the very useful discussions on the validation of the YSO classification. 
This work is based on data obtained from 
(1) the Wide-field Infrared Survey Explorer, which is a joint project of the University of California, Los Angeles, and the Jet Propulsion Laboratory/California Institute of Technology, funded by the National Aeronautics and Space Administration; 
(2) the Spitzer Space Telescope, which is operated by the Jet Propulsion Laboratory, California Institute of Technology under a contract with NASA;
(3) Herschel, which is an ESA space observatory with science instruments provided by European-led Principal Investigator consortia and with important participation from NASA.
This work has made use of 
(1) TOPCAT \url{http://www.starlink.ac.uk/topcat}; 
(2) ``Aladin sky atlas'' developed at CDS, Strasbourg Observatory, France \citep{Bonnarel2000, Boch2014}; 
(3) the SIMBAD database and VizieR catalog access tool, operated at CDS, Strasbourg, France; 
(4) Python, \url{https://www.python.org}; and 
(5) Astropy, a community-developed core Python package for Astronomy \citep{Astropy2013};
(6) the Sloan Digital Sky Survey (SDSS), which is a joint project of The University of Chicago, Fermilab, the Institute for Advanced Study, the Japan Participation Group, The Johns Hopkins University, the Max-Planck-Institute for Astronomy, Princeton University, the United States Naval Observatory, and the University of Washington. Apache Point Observatory, site of the SDSS, is operated by the Astrophysical Research Consortium. Funding for the project has been provided by the Alfred P. Sloan Foundation, the SDSS member institutions, the National Aeronautics and Space Administration, the National Science Foundation, the U.S. Department of Energy, and Monbusho, \url{https://www.sdss.org}; 
(7) the Digitized Sky Surveys DSS, which is an optical all-sky survey produced at the Space Telescope Science Institute under U.S. Government grant NAG W-2166.
\end{acknowledgements}

\begin{flushleft}
\bibliographystyle{aa}
\bibliography{biblio} 
\end{flushleft}

\begin{appendix}

\section{Example images} \label{cutouts}

In this section we present example image cut-outs of selected new YSO candidates, revisited interesting sources, and contaminating objects. The latter are objects that were erroneously identified as YSO candidates previously (false positives).

\begin{figure*}[!ht]
	\centering
	\begin{minipage}[c]{0.79\linewidth}
	    \centering
	    \includegraphics[width=1\linewidth]{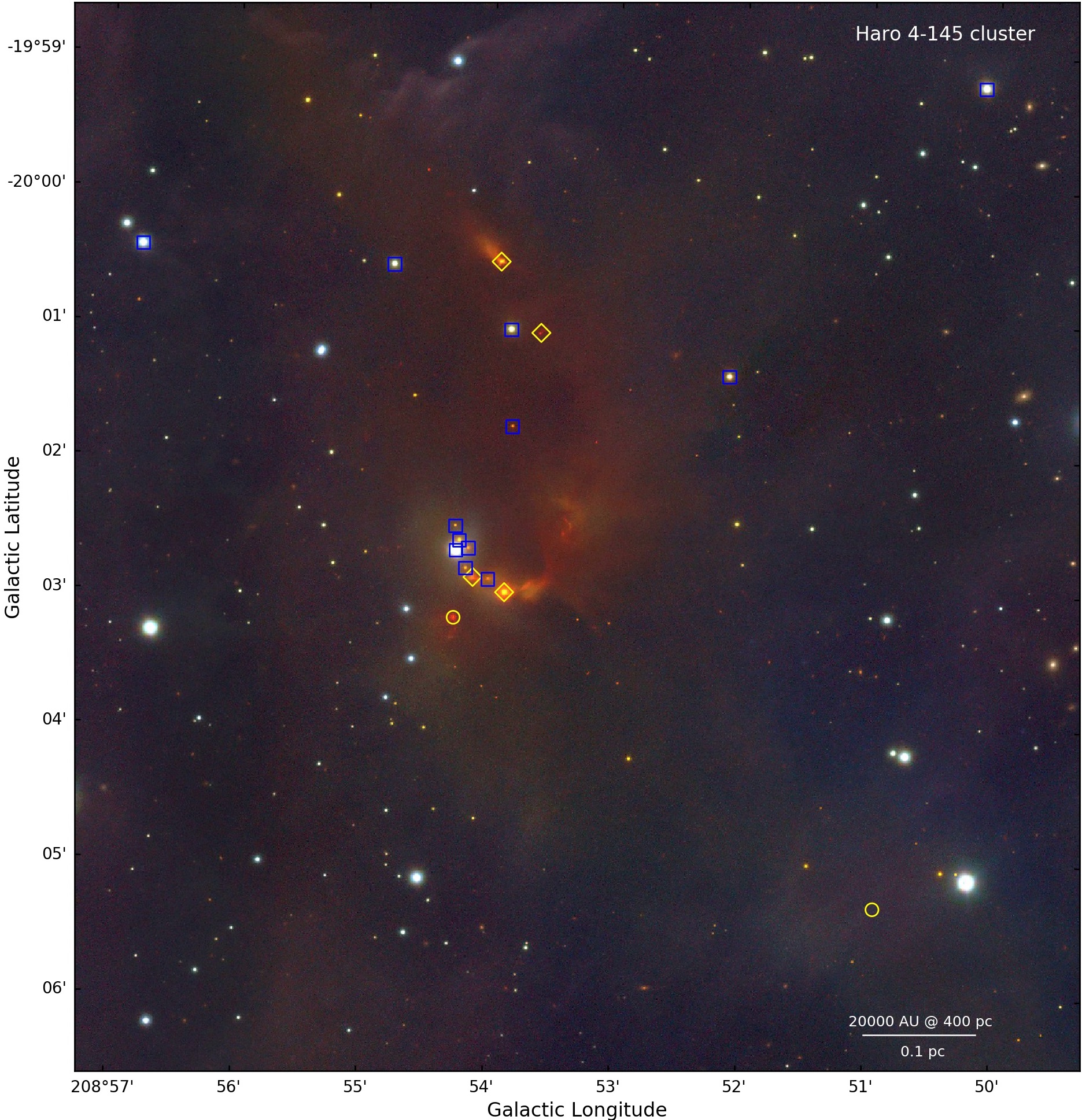}
	\end{minipage}\hfill%
	\begin{minipage}[c]{0.205\linewidth}
	    \centering
	    \includegraphics[width=1.\linewidth]{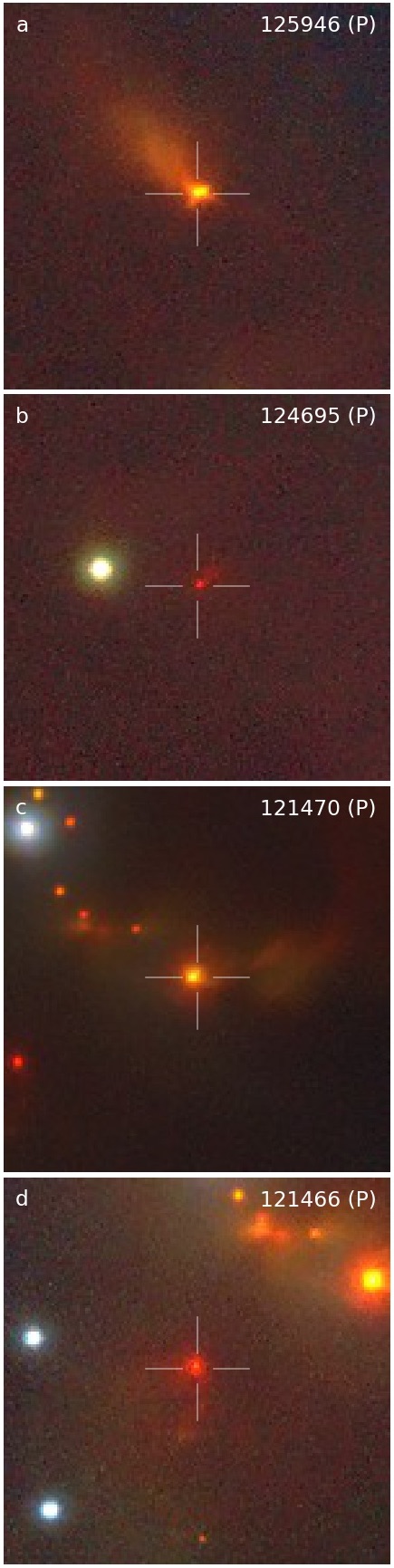}
	\end{minipage}%
	\caption{Clustering of YSOs south-west of the ONC (Haro4-145 cluster) shown by a $7.5\arcmin \times 7.9\arcmin$ VISTA 3-color composite, in Galactic coordinates. The cluster harbours at least six protostars (yellow symbols) and nine disk candidates (blue squares).  There are three more disk candidates visible in the image a bit more scattered. The two new extra protostar candidates from Sect.~\ref{NewYSOs} are the two yellow circles, while previously known protostar candidates are shown by diamond symbols. The source at the bottom right is the new Class\,0 candidate, only detected as a point-source in \emph{Herschel}/PACS. The cloud structure (region of higher dust column-density) gets visible in this NIR VISTA image. The slightly reddish illuminated cloud, containing the YSOs, seems to be separate from the background cloud structure. Moreover, background sources like galaxies are visible in this image. The four cutouts ($\SI{50}{\arcsec} \times \SI{50}{\arcsec}$) on the right show zoom-ins on four protostar candidates in this region (see text).}
	\label{fig:cluster}
\end{figure*}

\subsection{New candidates and interesting objects} \label{special}
\begin{figure*}[!ht]
    \centering
        \includegraphics[width=1\linewidth]{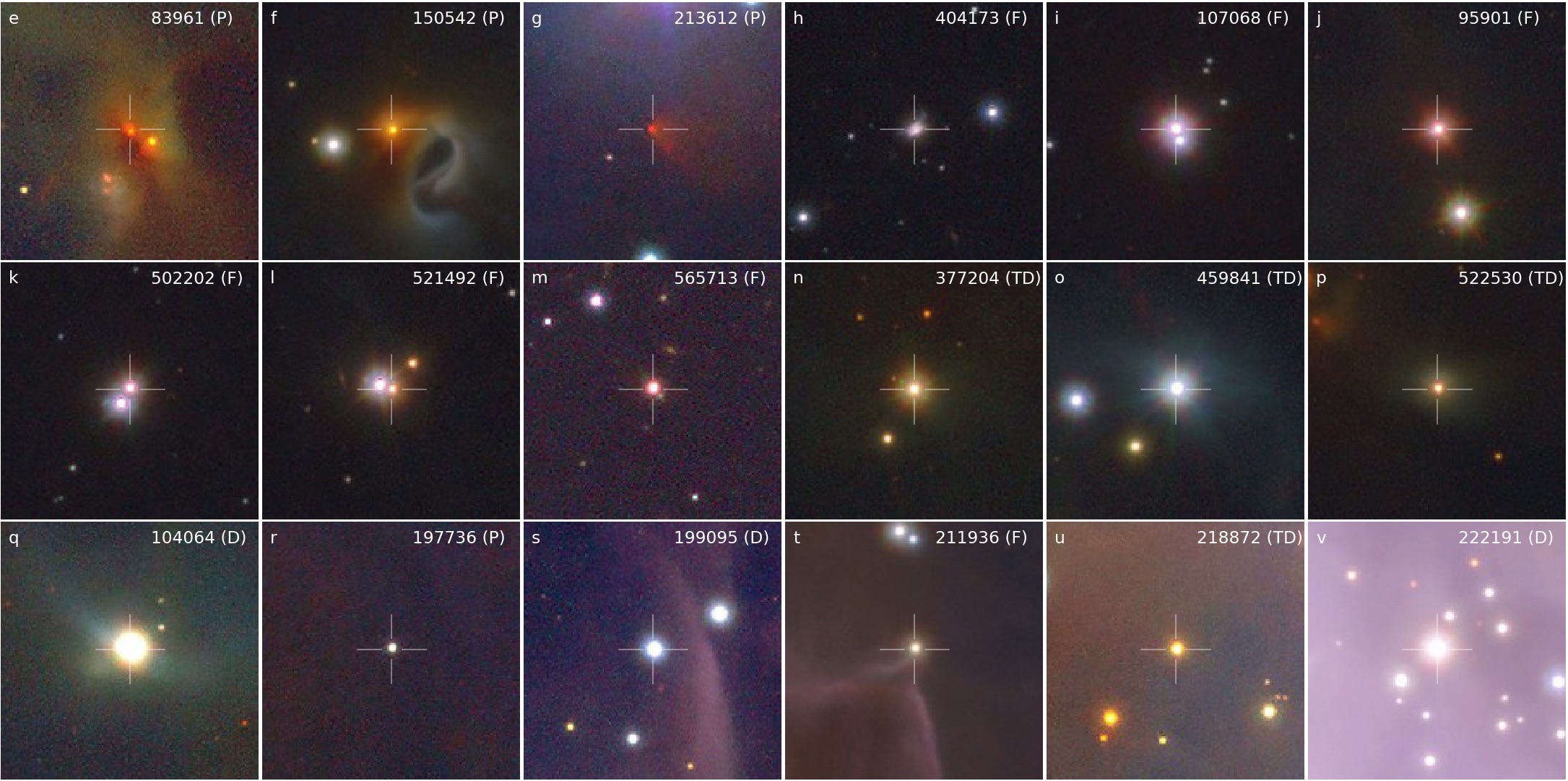}
    \caption{VISTA \SI{50}{\arcsec} $\times$ \SI{50}{\arcsec} cutouts of selected 18 YSO candidates. Sources e-p are new candidates (two top rows), and sources q-v (bottom row) are previously known objects. See text for more explanations.}
    \label{fig:special}
\end{figure*}

First we want to highlight some interesting objects, including revisited and new YSO candidates, shown in Figs.~\ref{fig:cluster} and \ref{fig:special}. The images show $\SI{50}{\arcsec} \times \SI{50}{\arcsec}$ VISTA JHK three color composites centered on the objects, oriented in Galactic coordinates. The sources are addressed in more detail below, with the listed small letters pointing to the corresponding VISTA cut-outs.

\textbf{a,b,c,d) Haro\,4-145 clustering.} This clustering (see also Sect.~\ref{NewYSOs}) was already discussed by \citet{Wang2002}, who listed six YSO candidates (IRS\,1 to IRS\,6), calling it Orion\,A-W star-forming region. We call it ``Haro\,4-145 cluster'', named after the bright Class\,II YSO at its center. A visualization of this region using a VISTA three color composite is shown in Fig.~\ref{fig:cluster}.
The clustering is located at the \emph{Spitzer}/IRAC survey border, therefore some sources were missed by MGM. 
It contains at least 15 YSO candidates (6 P, 9 D), of which nine are in the MGM sample (3 P, 6 D), and six are selected in this work.
One is a new VISTA/\emph{WISE} selected protostar candidate, which is a known \mbox{H$_2$O} maser \citep[TFT95b H2O 053014.409-053750.890,][]{Meehan1998}. 
During visual inspection we found two new protostar candidates, which also have an entry in the PACS point source catalog. One candidate (ID 121466) is very close to the water maser, and was missed by our VISTA/\emph{WISE} selection, likely because of the very crowded region and overlapping \emph{WISE} PSFs. This source is already listed in \citet{Wang2002} as IRS\,2.
The second new candidate is more isolated and lies farther to the Galactic south ($\SI{\sim3.6}{\arcmin}$). It only gets visible as a point-source in \emph{Herschel}/PACS and has no NIR or MIR counterpart, located in a region of high dust column-density ($A_\mathrm{K,Herschel}\approx\SI{4.6}{mag}$). It is likely a new Class\,0 candidate (ID 116363).
Four of the six protostar candidates located in the Haro\,4-145 clustering are shown in zoom-ins on the right side in Fig.~\ref{fig:cluster}. The two at the top (\textbf{a, b}) were selected already by MGM (125946, 124695), while the two at the bottom (\textbf{c, d}) are selected in this work (121417, 121466) and were also highlighted in  \citetalias{Meingast2016}. Source \textbf{c} is the known water maser, and source \textbf{d} is one of the new protostar candidates in this region. The source at the bottom of the overview image, which is the new Class\,0 candidate, is not highlighted separately, because is is not visible in the NIR.

\textbf{e) New protostar candidate.} One of the new protostar candidates (ID 83961), selected by VISTA/\emph{WISE}, is located in the Galactic west of the ONC at a small visible clump of higher dust column-density, visible in the \emph{Herschel map} at $(l,b)\sim(207.3,-19.8)$. VISTA resolves two sources, while \emph{WISE} shows only one point-source covering both. The second VISTA source (the right one) could also be a possible new YSO candidate. The projected spatial distance between the two sources is about \SI{4.5}{\arcsec} ($\SI{\sim1800}{AU}$ @ $\SI{400}{pc}$). The sources were already highlighted in \citetalias{Meingast2016}.

\textbf{f) Protostar HH\,83\,IRS.} The source HH\,83\,IRS (or IRAS 05311-0631, ID 150542) lies at the border of the IRAC coverage, but is observed by $I2$ and $I4$. It was already highlighted in \citetalias{Meingast2016} and discussed in \citet{Reipurth1989, Ogura1991, Moneti1995}, and \citet{Davis2011}. The source coincides with an optical jet \citep{Reipurth1989} and a reflection nebula \citep[Re 17,][]{Rolph1990}, well visible in VISTA. The protostar seems to be an isolated star-forming event in the Galactic south of L1641-North, at the location of a clump of higher dust column-density ($A_\mathrm{K,Herschel}\approx\SI{2.5}{mag}$), well visible in the \emph{Herschel map} at $(l,b) \sim (209.85,-20.27)$. The source is listed in the HPPSC, but is not included in the HOPS catalog. The distance to the object is given with $\SI{450}{pc}$ \citep{Reipurth1989}.

\textbf{g) New protostellar binary in L1641.} Next to a known Class\,0 (MGM\,1121) we select a second protostar candidate visually  (ID\,213612, see also Sect.~\ref{NewYSOs}), using PACS and VISTA images ($\SI{\sim5}{\arcsec}$ separation, \SI{2000}{AU} @ \SI{400}{pc}). The new candidate shows a prominent outflow cavity. The known Class\,0 is the faint reddish dot, to the bottom right from the new source, overshadowed by the outflow of the new candidate. Both sources correspond to an elongated PACS point-source. The system was also highlighted by \citet{Tobin2017-TALK} as protostellar binary candidate.

\textbf{h) New edge-on disk.} With VISTA/\emph{WISE} we select a new interesting object (ID\,404173) off of the cloud to the Galactic north of L1641 at $(l,b) \sim (210.22,-18.81)$. We classify the source as flat-spectrum candidate based on its spectral index, but the VISTA image shows an extended object, likely an edge-on disk candidate. The PACS images show a clear point source, as well as the \emph{Herschel map}, with a peak extinction value of $A_\mathrm{K,Herschel}\approx\SI{0.33}{mag}$, while the immediate surrounding pixels are of lower extinction ($A_\mathrm{K,Herschel} \lesssim \SI{0.1}{mag}$). This indicates a massive dusty disk or envelope. It even seems that the source lies in the middle of an almost circular excavated area (see \emph{Herschel map}) with an extension of about $\SI{30}{\arcmin}$. Its flattish edge-on appearance is similar to Gomez's Hamburger \citep{Bujarrabal2008}, which is suggested to be an A-type PMS star with a massive circumstellar disk, that is not associated with any inter-stellar molecular cloud. The new edge-on candidate needs more investigation to determine its true nature, and to rule out that it is not an exotic extra-galactic source, or a post AGB star. 

\textbf{i) New flat-spectrum candidate UY\,Ori.} One new flat-spectrum candidate selected by VISTA/\emph{WISE} (ID 107068) is the known Herbig Ae-Be (HAeBe) star UY\,Ori \citep[B9 III,][]{Vieira2003}.The \emph{WISE} point source actually covers two resolved VISTA sources. The second source is maybe a companion, with a projected distance of $\SI{\sim2.4}{\arcsec}$ ($\SI{{\sim}960}{AU}$ @ \SI{400}{pc}). We classify the source as a flat-spectrum source, even though the WISE spectral index is slightly above 0.3 ($\alpha_\mathrm{WISE}$ = 0.43). Though the local low extinction ($A_\mathrm{K,Herschel}\approx\SI{0.1}{mag}$), its location far form high extinction regions ($\SI{\sim30}{\arcmin}$), and the brightness and colors in the NIR and optical suggest that it is not a protostar. Moreover, as discussed for example by \citet{Whitney2004}, the higher luminosity and temperature of high mass stars require a different interpretation of the observed spectral index, and the classification does not follow the same standards as for low mass T-Tauri stars.

\textbf{j) New flat-spectrum candidate ID\,95901.} This new flat-spectrum candidate \citepalias[also highlighted in][]{Meingast2016}, selected by VISTA/\emph{WISE}, is a prominent bright YSO. It is located near the region of the already mentioned new protostar candidate ID\,83961 (source \textbf{e}, distance $\SI{\sim20}{\arcmin}$), at the Galactic west of the ONC. It is located at an isolated clump of higher dust column-density ($A_\mathrm{K,Herschel}\approx\SI{1}{mag}$) and is associated with a prominent PACS point-source.

\textbf{k) New flat-spectrum candidate ID\,502202.} This new flat-spectrum candidate, selected by VISTA/\emph{WISE}, appears to be a double star in VISTA, similar to UY\,Ori. It is associated with a PACS point-source, and interestingly, it also coincides with a point-like peak in the \emph{Herschel map} ($A_\mathrm{K,Herschel}\approx\SI{0.36}{mag}$).

\textbf{l) New flat-spectrum candidate ID\,521492.} This is the only new flat-spectrum candidate selected inside IRAC (by VISTA/$I24$, Appendix~\ref{HKI24}). It appears to be a double star, and was missed previously maybe due to source confusion, or different selection criteria.

\textbf{m) New flat-spectrum candidate or AGN?} The last new flat-spectrum candidate (ID\,565713) discussed here is a rather suspicious YSO candidate. It could also be a bright AGN (Blazar), since it is located a bit off from the cloud ($\SI{\sim30}{\arcmin}$) to the Galactic north of L1641 at $(l,b) \sim (211.94,-18.51)$, in a region of low extinction ($A_\mathrm{K,Herschel}\approx\SI{0.15}{mag}$). It is the only of the six new flat-spectrum sources that dose not have a PACS counterpart. Furthermore, the surroundings show a loose clustering of background galaxies. 

\textbf{add h-m)} These six new flat candidates are rather untypical flats (prominent edge-on disk, early spectral-type, luminous, double-star, or suspicious AGN candidate). Moreover, four of them (except ID 95901 and 521492) are located at regions of low dust column-density ($A_\mathrm{K,Herschel}<\SI{0.4}{mag}$).

\textbf{n,o,p) Three new PACS transition disks.} Three of the new YSO candidates show only an excess in PACS, but not in the NIR or MIR, and we classify them as transition disk candidates (ID 377204, 459841, 522530). All three seem to be surrounded by some nebulous haze, especially source 459841 (\textbf{o}) shows a prominent reflection nebula. 

\textbf{q) New disk.} Source 104064 is a prominent new disk candidate \citepalias[also highlighted in][]{Meingast2016}, selected by VISTA/\emph{WISE}, to the Galactic south-west of the cloud at $(l,b)\sim(208.6,-20.1)$. It is located at an isolated small clump of higher dust column-density ($A_\mathrm{K,Herschel}\approx\SI{1}{mag}$), and is also associated with a PACS point-source. The VISTA image reveals a prominent reflection nebula and a spiral-like structure.

\textbf{r) Suspicious protostar 2MASS\,J05344694-0544512.} This source is a suspicious protostar candidate (ID\,197736, MGM\,1238, HOPS\,24), by showing a whitish point-source in VISTA, with no signs of reddening or outflows in the NIR. Moreover, it has no clear PACS counterpart. Still, we kept it listed as protostar candidate due to its rising MIR SED.

\textbf{s) V1314\,Ori} was previously identified as protostar candidate (MGM\,1503, HOPS\,49) and listed as FUOri-type star in Simbad. The source is quite bright in the optical and NIR (Pan-STARRS $\mathit{g}=\SI{16.0}{mag}$, VISTA $K_S=\SI{12.5}{mag}$), with no significant signs of reddening in this wavelength ranges. We reclassify the source as disk candidate, even though the colors and spectral indices point to a protostar candidate, maybe due to high variability. It is also listed as CTTS in \citet{Elek2013}. Still, there is the possibility that it is a pole-on protostar. 

\textbf{t) V2168 Ori} is an emission line M-star \citep{Hillenbrand2013}, listed as galaxy in \citetalias{Furlan2016}. It coincides with some outflow or line-of-sight nebulosity. The VISTA image even gives the impression, that the source was just floating out of the cloud. 

\textbf{u) V2275\,Ori} is listed as extra-galactic (EG) in the  membership list of the COUP catalog by \citet{GetmanB2005}. However, its color and brightness suggest a Class\,II YSO, also listed as such previously (MGM\,1559, HOPS\,51). 

\textbf{v) TU\,Ori} is a known prominent YSO and was selected as a new disk candidate by VISTA/$I24$. It is given as an F or G spectral type in the literature \citep{Hillenbrand2013}, located near the ONC ($\SI{\sim2.4}{\arcmin}$ distance to Trapezium). It is not listed in MGM, despite its brightness. It is at the border to be a flat-spectrum source, and shows quite inconsistent spectral indices, maybe due to contamination effects in this region. 

\subsection{Contaminating objects} \label{cont}

In this section we show examples of contaminating ojects in Figs.~\ref{fig:cutout Meg D+P JosefaG} to \ref{fig:cutout MegD JosefaH}. The images are constructed as follows. The first column shows a VISTA JHK three color composite ($\SI{50}{\arcsec} \times \SI{50}{\arcsec}$) centered on the object and oriented in Galactic coordinates. The second to fourth columns are \emph{Spitzer} $I2$, $I4$, and $M1$ cutouts respectively, showing the same region as VISTA. In the upper left corner of the VISTA cutouts we give the running ID from this work with the classification in brackets. If present, the MGM index and classification are given as M\#(class), the HOPS index and classification as H\#(class), and/or the classification from \citetalias{Lewis2016} as LL(class). At the bottom of the VISTA cutouts we give some selected information on each source: the K$_S$ magnitude (given as K), the spectral indices $\alpha_\mathrm{KM}$ and $\alpha_\mathrm{I2M}$, the line-of-sight extinction from the \emph{Herschel map} at the position of the source (given here as A$_\mathrm{K}$), and the extension flag \texttt{ClassSex}. The last columns show six selected color-color and color-magnitude diagrams with the source highlighted by a red dot. These diagrams demonstrate the color spaces the source occupies. Some contaminating objects, especially nebulosities, do not have measurements in all bands, and therefore the diagrams do not include the source. We still keep the diagrams for consistency. Besides, the fact that fuzzy nebulous sources are not detected in all bands is another clue to identify them.

Figure~\ref{fig:cutout Meg D+P JosefaG} shows four examples of galaxies that were erroneously classified as YSO candidates in MGM (false positives). Some even show bright spiral galaxies, for example MGM\,1261 and MGM\,1517. In the latter case, one can see two bright star-forming regions in one spiral arm, that were both classified as disk candidates previously. Concerning faint and less prominent extra-galactic objects, the VISTA extension flags are used additionally to identify these as extended sources. In combination with the position at low extinction ($A_\mathrm{K,Herschel} \lesssim \SI{1}{mag}$), we can rule out that these sources are embedded protostars. Moreover, visual inspection at various wavelengths helps to rule out that the extension is not caused by outflows, surrounding a protostellar candidate. The color and magnitude diagrams (on the right) further help to confirm the nature of some sources. Galaxies tend to have red colors, similar to YSOs, but are also mostly fainter and are located close to the galaxy clump in the CMDs (last column, bottom right over-densities). 

In Figs.~\ref{fig:cutout MegP JosefaH} and \ref{fig:cutout MegD JosefaH} we show examples of VISTA identified fuzzy objects, like nebulosities, cloud edges, and Herbig-Haro objects, that were erroneously classified as YSO candidates previously. Here, the \emph{Spitzer} images are important, to rule out that the source is not an embedded source that only gets point-like at longer wavelengths. Some sources indeed almost show a point-like structure in the MIR, for example, source M\,2415 in the $M1$ cutout. Though, it is likely caused by heated material, maybe due to a shocked outflow, appearing in blue in the VISTA image. The extinction at the cite of this source is on the order of $A_\mathrm{K,Herschel}\SI{\sim0.24}{mag}$, which makes it unlikely that it is a deeply embedded source, because then it should also show up in $I4$ where no point-source can be identified. The CCDs and CMDs are shown for completeness, however, fuzzy nebulous detections which seem point-like at some wavelengths and resolutions are mostly not detected in all bands which construct the here presented diagrams.

\onecolumn
\begin{landscape}

\begin{figure}[!ht]
	\centering
	\includegraphics[width=0.9\linewidth]{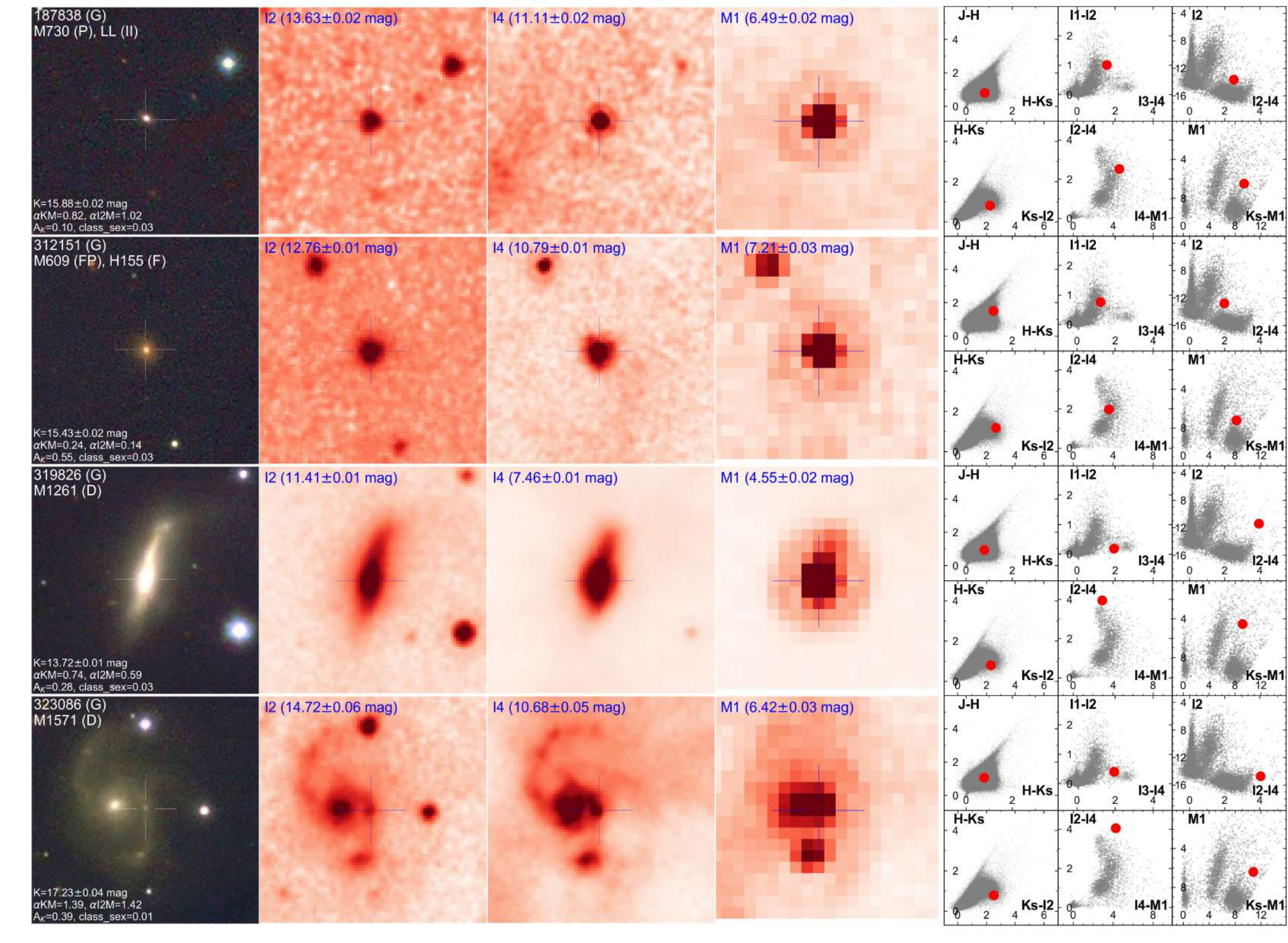}
	\caption{Examples of galaxies which were classified as YSO candidates in MGM. From left to right: VISTA three color composite, \emph{Spitzer} $I2$, $I4$, and $M1$ cutouts (\SI{50}{\arcsec} $\times$ \SI{50}{\arcsec}), oriented in Galactic coordinates, and examples of six color-color and color-magnitude diagrams, showing the displayed source as red dot. See text for more explanations.}
	\label{fig:cutout Meg D+P JosefaG}
\end{figure}

\begin{figure}[!ht]
	\centering
	\includegraphics[width=0.9\linewidth]{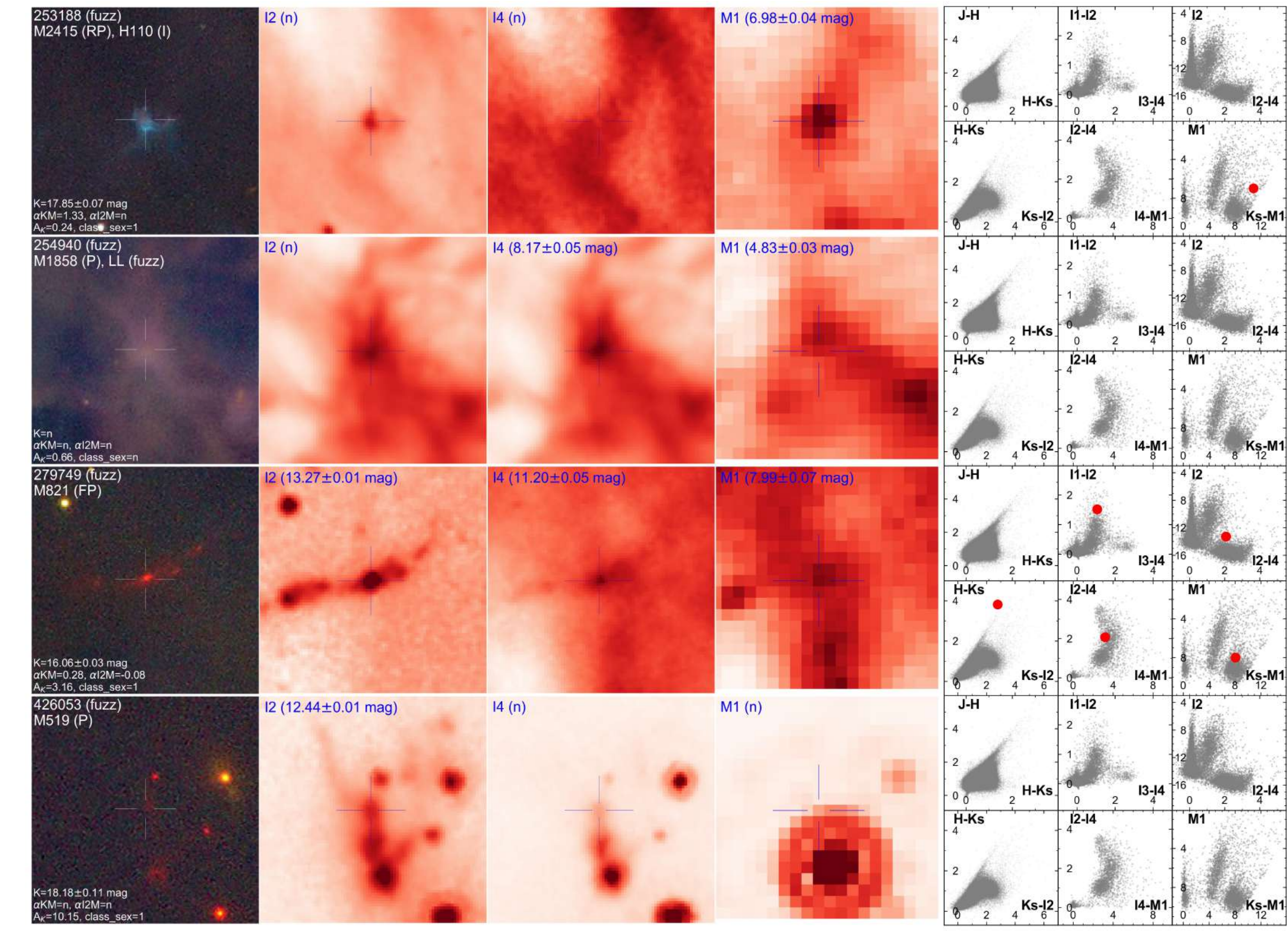}
	\caption{Examples of fuzzy objects (nebulosities, cloud edges, or parts of Herbig Haro objects) which were classified as YSO candidates in MGM. See Fig.~\ref{fig:cutout Meg D+P JosefaG} and text for more explanations.}
	\label{fig:cutout MegP JosefaH}
\end{figure}

\begin{figure}[!ht]
	\centering
	\includegraphics[width=0.9\linewidth]{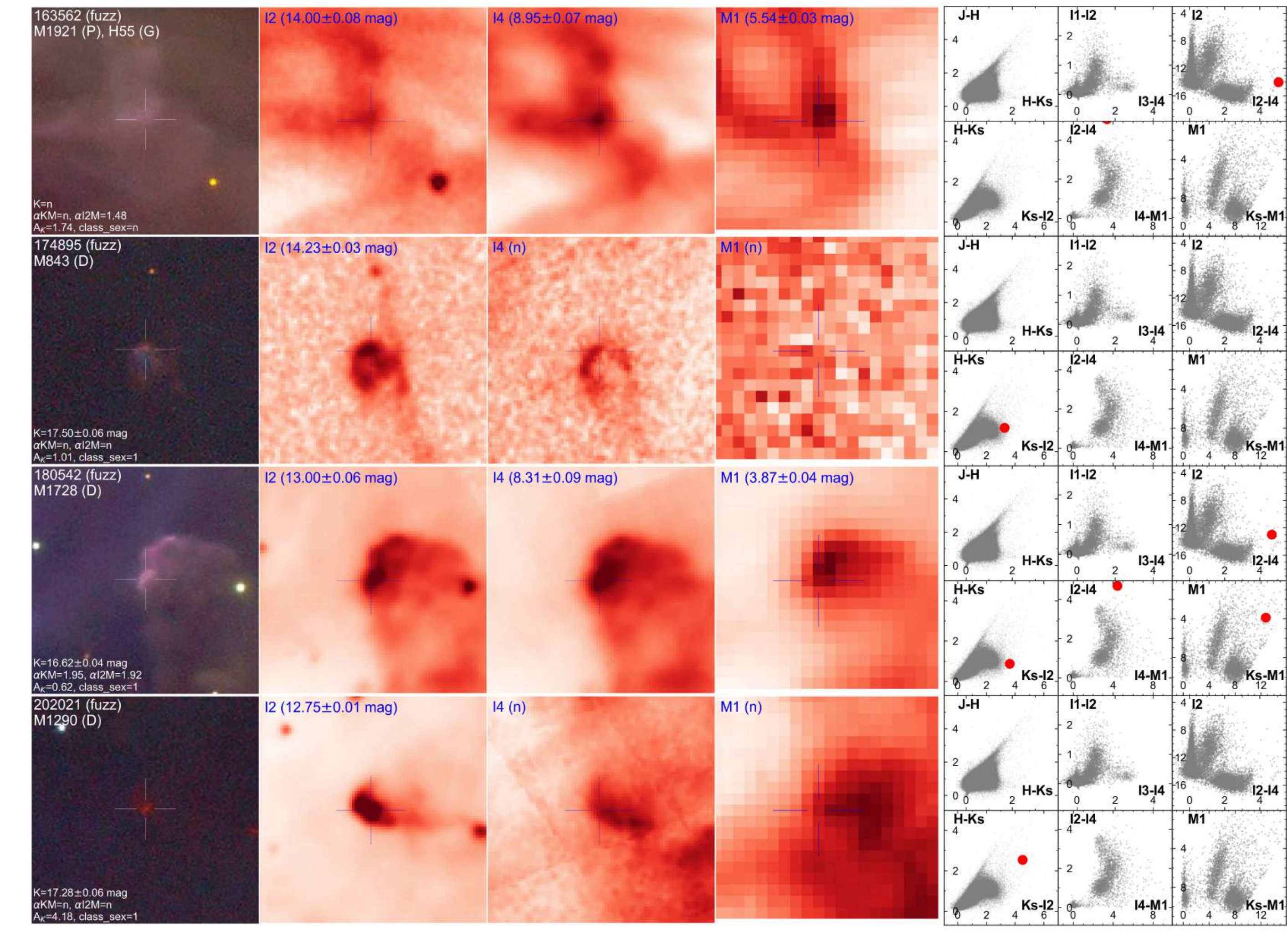}
	\caption{Same as Fig.~\ref{fig:cutout MegP JosefaH}}
	\label{fig:cutout MegD JosefaH}
\end{figure}

\end{landscape}
\twocolumn


\section{Selection conditions for new YSO candidates} \label{NewYSOsA}

Here we present the detailed description of the YSO selection methods for the new YSO candidates (see Sect.~\ref{NewYSOs}). 
We use six individual selection criteria with different band combinations to exploit various color-color and color-magnitude spaces, and the different sensitivities of the available bands. Three criteria are used to select sources inside the IRAC observed region, and the other three for outside IRAC.
To summarize the basic procedure for each selection, we 
(1) apply individual error-cuts to get rid of inferior photometry, 
(2) use a color-magnitude diagram to exclude faint unresolved star-forming galaxies and the majority of active galactic nuclei (AGN-cut),
(3) use a color-color diagram to exclude MS-stars and reddened source due to extinction, and finally 
(4) cut further color regions that are confused with contaminants. 
Such contaminants can be shocked blobs of gas outflows around young stars (influencing $I2$ or $W2$), or bright extended nebula emission, especially influencing e.g., $W2-W3$ or $W2-W4$ colors.

We decide the selection conditions by comparing with control field plots, when available, and by checking the color spaces of known YSOs and contaminating objects, found in Sect.~\ref{revisit} (see Figs.~\ref{fig:ccds} and \ref{fig:cmds}). With this we try to get a high recovery rate and at the same time avoid contamination as good as possible. The following figures display in gray all sources left after applying basic error-cuts. Diagrams for selections inside the IRAC region show the recovered protostars and disks in red and blue, respectively\footnote{For simplicity we do not include a flat-spectrum classification in these figures.}, and new candidates are highlighted with yellow diamonds. New candidates selected outside IRAC in combination with \emph{WISE} are shown separately as blue open squares. Selected sources that turned out to be false positives, or that are uncertain candidates, are marked with a black cross.   
 
\subsection{Selections using VISTA and \emph{Spitzer}} \label{NewSpitzerA}

Here we present selections based on VISTA and \emph{Spitzer} photometry, to add new YSO candidates inside and also outside the IRAC region (including $M1$). Sources might have been missed previously, due to coverage, resolution and/or sensitivity issues in the NIR, or due to different selection criteria.
To estimate the background contamination we use the control fields of IRAC that are overlapping with VISTA (see the small rectangular IRAC fields in Fig.~\ref{fig:coverage}).
These contain in total much less sources compared to the IRAC science field ($\sim$4\%), therefore we also use the knowledge of the color spaces of known objects to decide the selection conditions.


\subsubsection{VISTA and IRAC1-2} \label{HKI12}

\begin{figure}[!ht]
    \centering
    \begin{minipage}{1\linewidth}
        \centering
        \includegraphics[width=\linewidth]{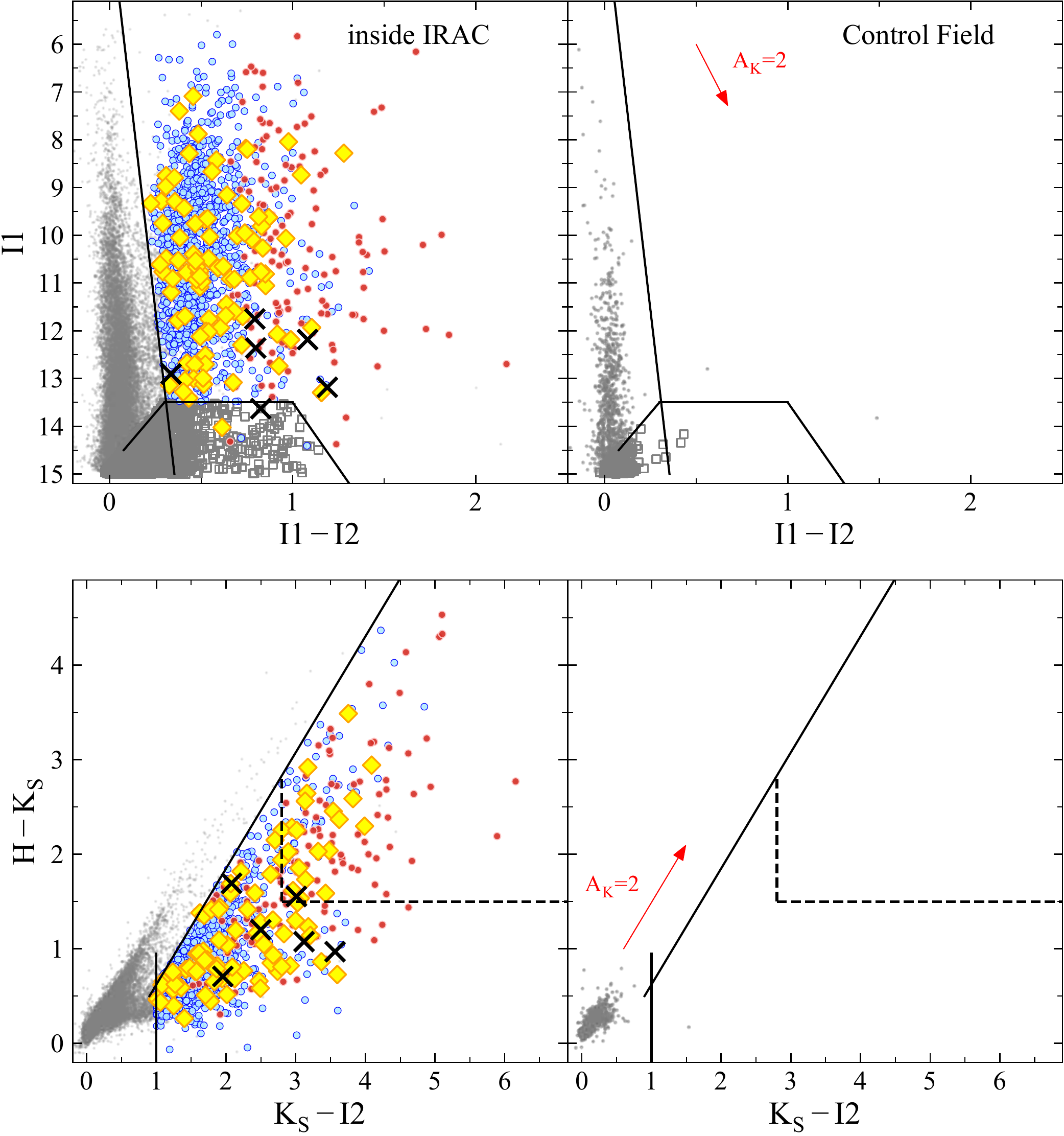}
    \end{minipage}%
    \vfill
    \begin{minipage}{0.98\linewidth}
        \centering
        \includegraphics[width=\linewidth]{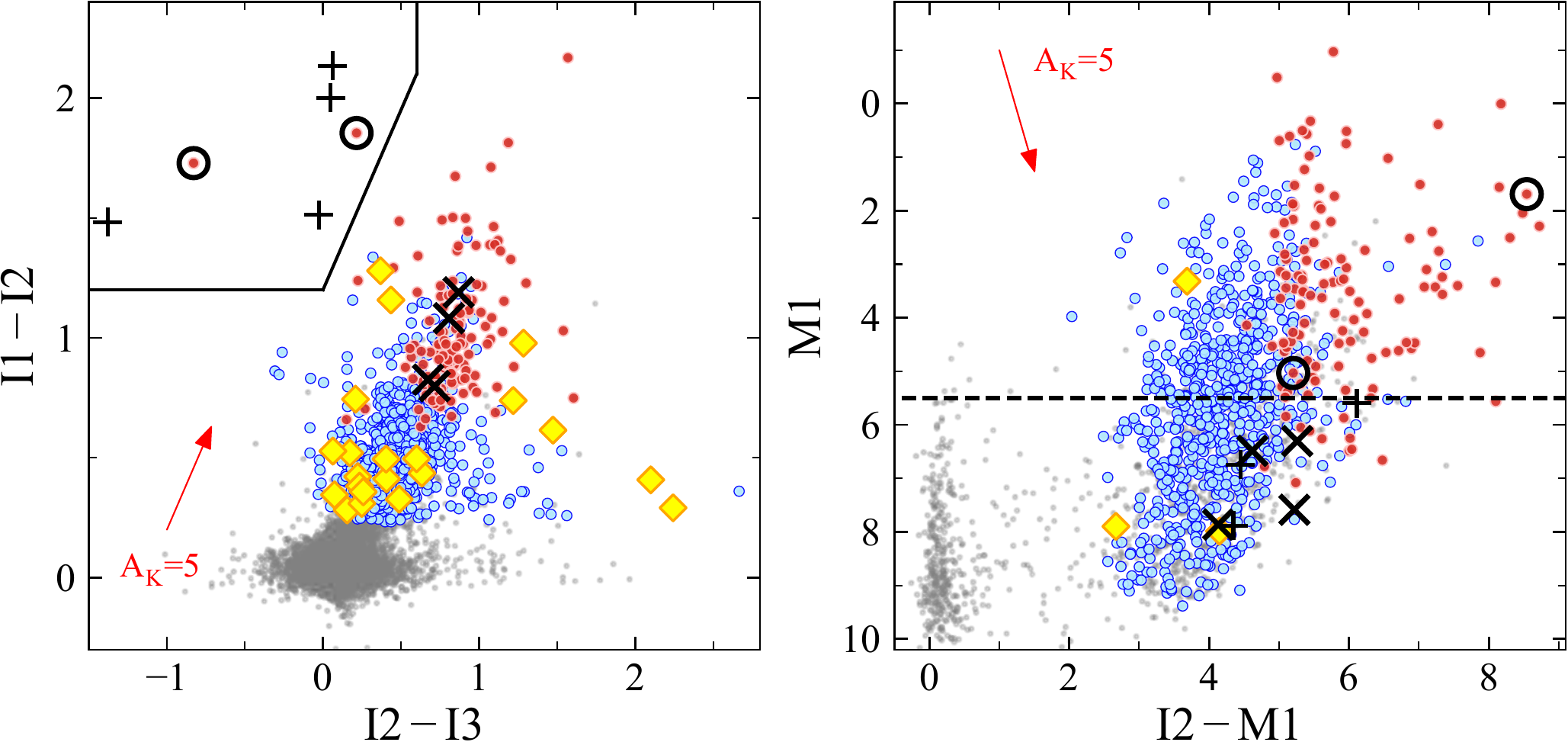}
    \end{minipage}%
    \caption{CCDs and CMDs showing the HKI12-selection. Top: $I1$ vs $I1-I2$ color-magnitude diagram for the science and control field. Middle: $H-K_s$ vs $K_s-I2$ color-color diagram for the science and control field. The science fields (left) show the YSO selection inside the IRAC coverage (yellow filled diamonds), while the recovered sources are highlighted as well (disks - blue, protostars - red). Black crosses mark contamination or uncertain sources. The control fields (right) are used to exclude background contamination. 
    The extinction vector is shown in red with a length of $A_\mathrm{K}=\SI{2}{mag}$.
    Bottom: Shock emission exclusion for sources with valid $I3$ measurements for the HKI12-selection. Symbols are the same as above in the science field plots. Additionally we show black plus symbols, marking sources that are caused by shock emission, and therefore are excluded. The open black circles mark sources which fall in the color region of shock emission, but which are likely YSO candidates, by showing a significant $M1$ measurement.}
    \label{fig:HKI12}
\end{figure}

First, we use the higher sensitivity of VISTA and \mbox{$I$1-2}, compared to 2MASS and the other \emph{Spitzer} bands. These bands are less affected by saturation and contamination near the ONC. 
However, YSO candidates are not as well separated when using this band combination compared to longer MIR bands. Sources with too little IR-excess will be missed, or are excluded by a reddening cut (parallel to extinction vector above the MS). 
We use the CMD $I1-I2$ vs $I1$ and the CCD $K_S-I2$ vs $H-K_S$ (HKI12-selection, Fig.~\ref{fig:HKI12}), and apply the following error cuts:
\begin{equation}
\begin{aligned}
\mathit{Herr}, \, K_{S}\mathit{err}, \, I1\mathit{err}, \, I2\mathit{err} &< \SI{0.1}{mag}   \\
I1, \, I2 &< \SI{14.5}{mag} \\
\mathtt{ClassSex} &> 0.1  
\end{aligned}
\end{equation}
The latter condition excludes extended objects, mostly galaxies but also sources surrounded by prominent outflows. This means we also exclude some protostar candidates with this criteria. However, we do not expect that there are still such sources left undiscovered, since protostars with prominent outflows are already selected by previous works. Furthermore, such prominent objects are eye-catchers, and are often found when visually inspecting the images.
After applying basic error cuts there are 21,641 and 953 sources left in the science and control field, respectively. Since the control field only contains about 4\% compared to the science field, we also infer on the color spaces of known YSOs to decide the selection conditions. 
To exclude further galaxies and faint unclear candidates we apply the following conditions (AGN-cut, see CMD in Fig.~\ref{fig:HKI12} top row, lower borders). The sources beyond this region (gray boxes) are excluded in the following from the CCD.\footnote{The symbols $\land$ and $\lor$ stand for the logical \texttt{AND} and \texttt{OR}, respectively.} 
\begin{equation}
\begin{aligned}
I1 &\leq 13.5 \ \lor \\
I1 &\leq -4.5 \times (I1-I2 - 0.4) + 13.5 \ \lor \\
I1 &\leq 5.5 \times (I1-I2 - 1) + 13.5  
\end{aligned}
\label{equ:HKI12-galaxies}
\end{equation}
To exclude MS stars we apply a cut in the CMD, and also in the CCD, where we cut parallel to the extinction vector.
\begin{equation}
\begin{aligned}
I1    &\leq 33 \times (I1-I2 + 0.1) \ \land \\ 
H-K_S &\leq 0.546/0.445 \times (K_S-I2 - 0.9) + 0.5 \ \land \\
K-I2 &\geq 1
\end{aligned}
\end{equation}
The selection, until this point, contains contaminating sources due to shock emission, which can be identified including $I3$ in the CCD $I1-I2$ vs $I2-I3$, similar to \citet{Gutermuth2009}. Shock emission is located in the upper left corner of this diagram (see bottom row in Fig.~\ref{fig:HKI12}). We apply the following conditions, but only for sources with $I3$ not \texttt{NULL}, and which are not in $M1$ or with $M1>\SI{5.5}{mag}$, to keep YSO candidates included, that show a significant $M1$ measurement (bottom right plot in Fig.~\ref{fig:HKI12}). 
\begin{equation}
\begin{aligned}
I1-I2 &\leq 1.5 \times (I1-I3) + 1.2 \ \lor \\
I1-I2 &\leq 1.2 \ \lor \\
I2-I3 &\geq 0.6
\end{aligned}
\end{equation}
Finally, we add sources in the top right region of the HKI12 CCD (Fig.~\ref{fig:HKI12} middle row, black dashed lines). This is a color region, that does not suffer from contamination. Therefore, we do not apply the AGN-cut here (Equ.~\ref{equ:HKI12-galaxies}).
This additional cut adds faint YSO candidates, that are located in the galaxy region of the CMD.
\begin{equation}
\begin{aligned}
H-K_S &\geq 1.5 \ \land \\
K_S-I2 &\geq 2.8 
\end{aligned}
\end{equation}

With the HKI12-selection we select 1,365 sources inside the IRAC region of which 1,270 are known YSO candidates from previous works, and 89 are new candidates (see map, yellow filled diamonds in Fig.~\ref{fig:mapnew2}). The rest six are contaminating or uncertain sources. The HKI12-selection recovers $\sim$47\% of the known YSO population.


\subsubsection{VISTA and IRAC2-4} \label{HKI24}

\begin{figure}[!ht]
    \centering
        \includegraphics[width=1\linewidth]{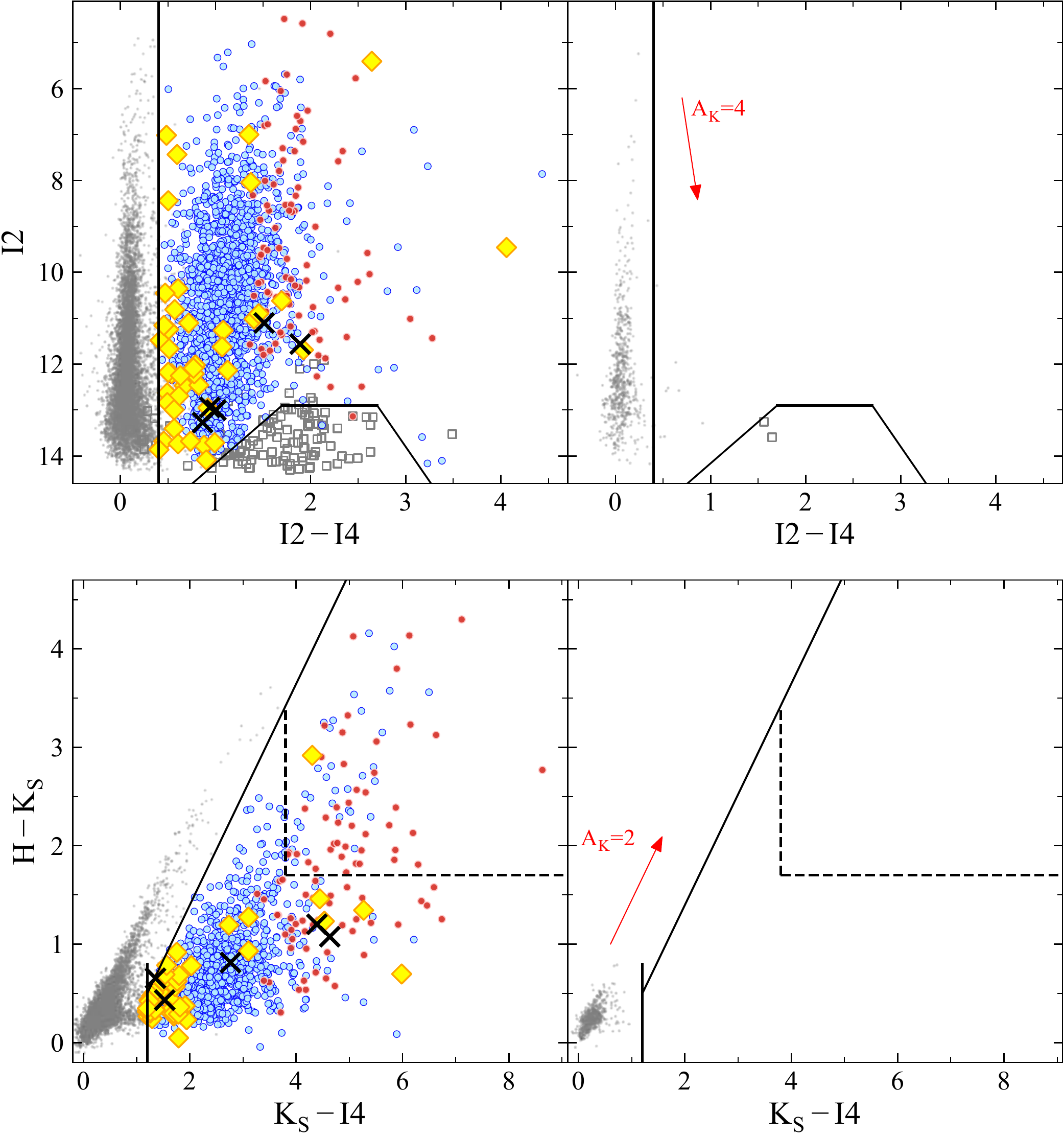}
    \caption{CCDs and CMDs showing the HKI14-selection. Top: $I2$ vs $I2-I4$ color-magnitude diagram for science and control field. Bottom: $H-K_s$ vs $K_s-I4$ color-color diagram for science and control field. Symbols as in Fig.~\ref{fig:HKI12}.}
    \label{fig:HKI24}
\end{figure}

Next we combine VISTA and \emph{Spitzer}/$I2$-4, to primarily look for possible missed disk candidates, by using different color spaces compared to previous works.
We use the CMD $I2-I4$ vs $I2$ and the CCD $K_s-I4$ vs $H-K_S$ (HKI24-selection, Fig.~\ref{fig:HKI24}) with a similar procedure as for to the HKI12-selection above. 
We apply the following basic error cuts:
\begin{equation}
\begin{aligned}
H\mathit{err}, \, K_{S}\mathit{err}, \, I2\mathit{err}, \, I4\mathit{err} &< \SI{0.1}{mag}   \\
I1, \, I2 &< \SI{14.3}{mag} \\
\mathtt{ClassSex} &> 0.1 \\
\mathtt{ClassCog} &= 1 
\end{aligned}
\end{equation}
With this there are 9316 sources left in the science field and only 539 in the control field ($\sim$6\%). 
To exclude galaxies and faint unclear candidates we apply the following conditions (AGN-cut, CMD in Fig.~\ref{fig:HKI24} top row, lower trapezoidal borders). 
\begin{equation}
\label{equ:HKI24-galaxies}
\begin{aligned}
K_S &< 15.5 \ \land \\
(I2 &\leq 12.9 \ \lor \\
I2 &\leq -1.8 \times (I2-I4 - 1.7) + 12.9 \ \lor \\
I2 &\leq 3 \times (I2-I4 - 2.7) + 12.9 ) 
\end{aligned}
\end{equation}
The sources beyond these borders (gray boxes) are excluded from the CCD. 
To exclude MS stars we apply cuts in the CMD and in the CCD, where we cut parallel to the extinction vector.
\begin{equation}
\begin{aligned}
&I2-I4 > 0.4 \ \land \ K_S-I4 \geq 1.2 \ \land \\
&H-K_s \leq 0.546/0.486 \times (K_S-I4 - 1.2) + 0.5 
\end{aligned}
\end{equation}
Again, we add sources in the top right region of the HKI24 CCD (Fig.~\ref{fig:HKI24} bottom row, black dashed lines), by not applying the AGN-cut (Equ.~\ref{equ:HKI24-galaxies}).
\begin{equation}
H-K_S \geq 1.7 \ \land \ K_S-I4 \geq 3.8 
\end{equation}

We select 1,675 sources, of which 1,626 are known YSO candidates from previous works, and 44 are new candidates (see map, yellow filled triangles in Fig.~\ref{fig:mapnew2}). The rest five were found to be false positives or are uncertain candidates. Out of the 44 only nine were already selected by HKI12. The 35 additional candidates add up to 124 new candidates inside IRAC until this point (HKI12-HKI24-selection). The HKI24-selection alone recovers $\sim$60\% of the known YSO population.


\subsubsection{VISTA and $M1$} \label{HKM}

\begin{figure}[!ht]
    \centering
        \includegraphics[width=1\linewidth]{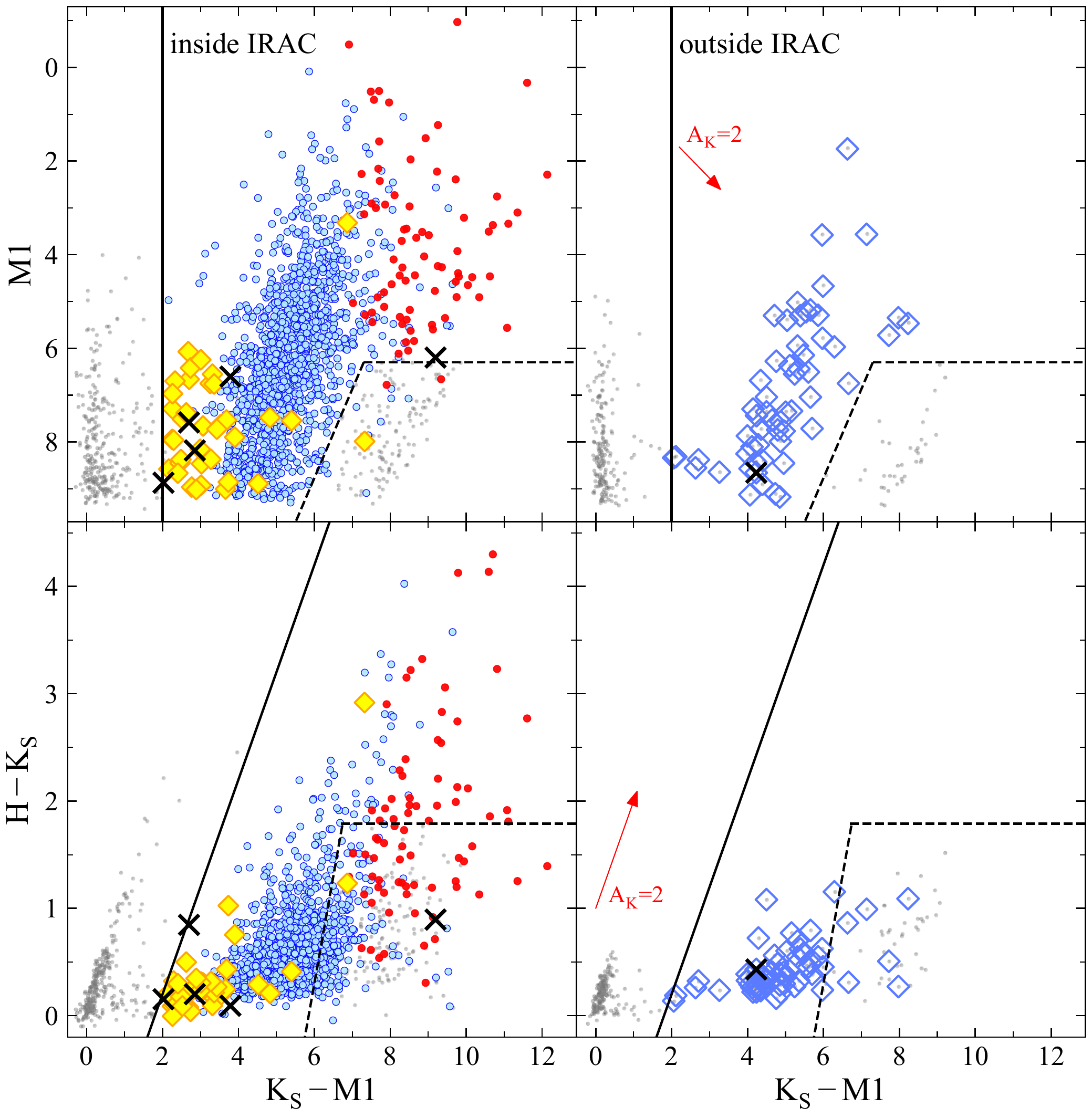}
    \caption{Top: CMD $M1$ vs $K_S-M1$. Bottom: CCD $H-K_S$ vs $K_S-M1$. Left: Selection of YSOs inside the IRAC coverage, showing the recovered sources (disks - blue, protostars - red). Yellow filled diamonds are new YSO candidates and the black crosses are contaminates or uncertain sources. Right: Selected new YSO candidates outside the IRAC region (open blue diamonds). The extinction vector is shown for both diagrams in red with a length of $A_\mathrm{K}=\SI{2}{mag}$. The black solid lines represent cuts to exclude MS stars, and the black dashed lines exclude extra-galactic sources.}
    \label{fig:HKM}
\end{figure}

We finally combine VISTA and \emph{Spitzer}/$M1$ to select new YSO candidates inside and also outside the IRAC region. This band combination is sensitive to anemic and transition disks, which often show only an IR-excess long-ward of about 10 to $\SI{20}{\micro\meter}$. 
We use the CMD $K_S-M1$ vs $M1$ and the CCD $K_S-M1$ vs $H-K_S$ (HKM-selection). There is no control field for this band combination, therefore, we decide the selection conditions solely based on known objects. We apply the following basic error cuts:
\begin{equation}
\begin{aligned}
    K_S\mathit{err}, H\mathit{err} &<  \SI{0.1}{mag}  \\  
    M1\mathit{err} &< \SI{0.15}{mag} \\
    K_S &< \SI{17}{mag} \\
    \mathtt{ClassSex} &> 0.1  \\
    \mathtt{ClassCog} &= 1 
\end{aligned}
\end{equation}
This leaves 1,793 sources inside and 271 outside the IRAC region. 
To exclude MS stars we apply rather conservative cuts in both the CMD and CCD (black solid lines in Fig.~\ref{fig:HKM}), to exclude possible fake $M1$ detections that give some sources a false IR-excess. Also AGBs are particularly influencing this region.
\begin{equation} \label{equ:HKM-MS}
\begin{aligned}
    K_S-M1 &> 2 \ \land  \\
    H-K_S &\leq 0.55/0.48 \times (K_S-M1 - 1.3) -0.5 
\end{aligned}
\end{equation}
Next we cut extra-galactic contamination in the lower right corner of the CMD, shown by the black dashed lines in the upper panels of Fig.~\ref{fig:HKM}. 
\begin{equation}
\begin{aligned}
M1 &\leq 6.25 \ \lor\\
M1 &\leq -1.9 \times (K_S-M1 - 7.3) + 6.3
\end{aligned}
\end{equation}
The remaining sources are selected as YSO candidates. 
We add further candidates in the CCD but exclude sources right to the dashed lines in the lower panels of Fig.~\ref{fig:HKM}, which is again contaminated by extra-galactic sources. 
\begin{equation}
\begin{aligned}
H-K_S &> 1.79 \ \lor \\
H-K_S &\leq 2 \times (K_S-M1 - 5.6) -0.5 
\end{aligned}
\end{equation}

We select 1,387 sources inside the IRAC region, of which 1,346 are known YSO candidates from previous works, and 36 are new candidates (see map, cyan open diamonds in Fig.~\ref{fig:mapnew2}). The rest five were found to be false positives or uncertain candidates. 
Out of the 36 new disk candidates eight are overlaps with HKI12-HKI24. The 28 extra sources were not picked up previously, likely because they show no or very little IR-excess in the shorter wavelength bands. Out of the 36 there are 31 anemic disk candidates. The combined HKI12-HKI24-HKM-selections give a total of 152 new YSO candidates inside the IRAC region.
Outside the IRAC region we select 56 new YSO candidates using HKM1 (see map, cyan filled diamonds in Fig.~\ref{fig:mapnew2}), including eight transition disks and five anemic disks.
There is only one uncertain source selected outside IRAC (black cross), which lies at the border of the MIPS1 coverage. 
The HKM-selection recovers about 50\% of the known YSO population.


\subsection{Selections using VISTA and WISE} \label{NewWISEA}

\begin{figure}[!ht]
	\centering
	\includegraphics[width=0.98\linewidth]{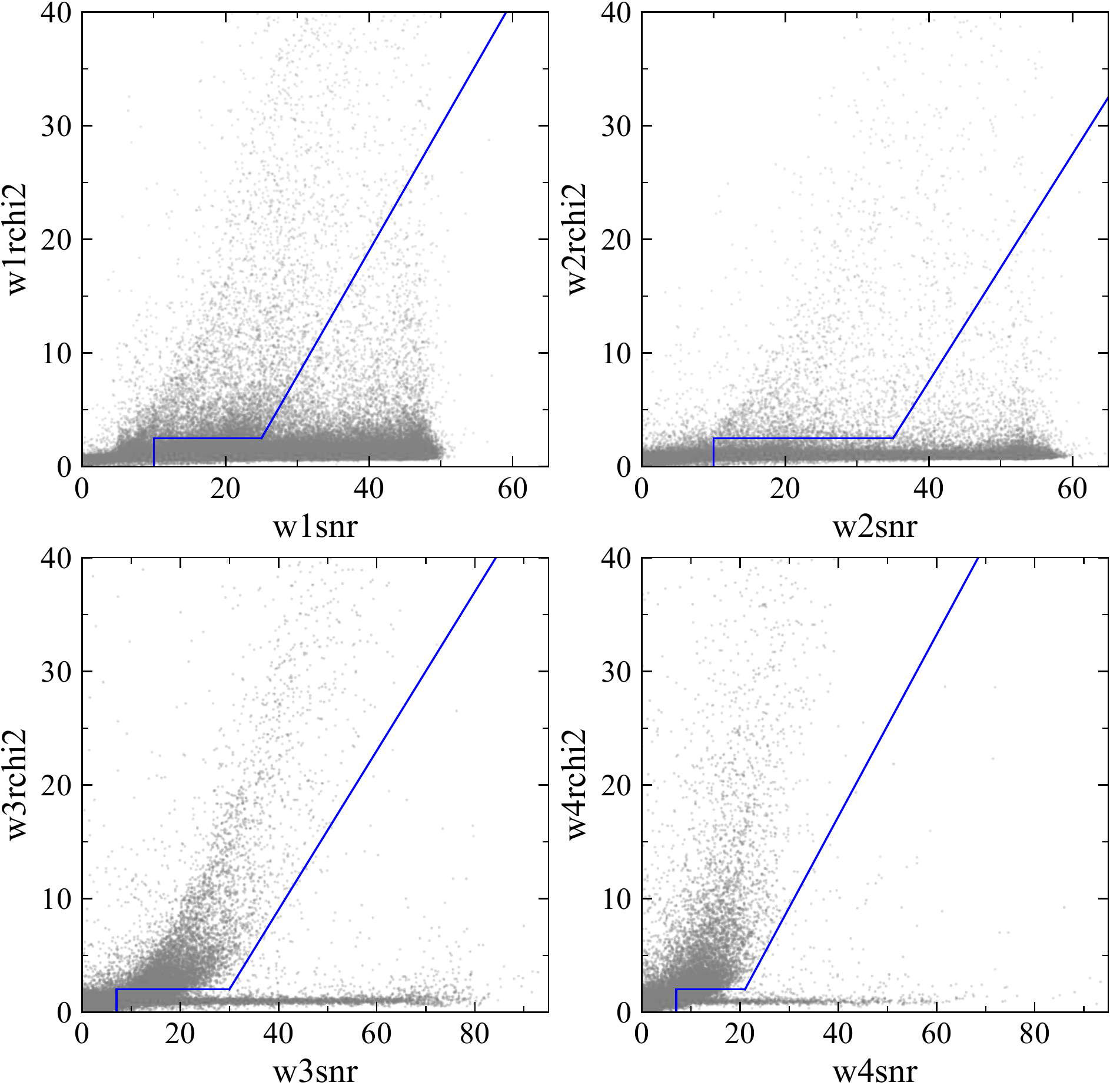}
	\caption{S/N versus reduced chi square for each \emph{WISE} band. The blue lines show the cuts (S/N-RCHI2-cut) to reduce fake point sources \citep[similar to][]{Koenig2014}. Sources to the left of the blue borders are excluded, as given in Equs.~\ref{equ:snrchi1} to \ref{equ:snrchi4}.}
	\label{fig:snrchi}
\end{figure}

We use \emph{WISE} MIR data to add additional YSO candidates outside the IRAC region, keeping the MIPS region included. To estimate the background we use the VISTA control field cross-matched with \emph{WISE} data. 
\citet{Koenig2014} presented a detailed analysis of different possible cleaning processes for AllWISE photometry in the Galactic plane, to mitigate the high contamination due to fake point sources. They find that a combination of signal to noise ratio (S/N) and reduced chi square (RCHI2) gives the best separation between fake and true point sources. We adopt their approach but modify it to be less conservative, and to recover more YSO candidates. 
The following conditions are applied for \emph{WISE} bands if they are used in one of the selections presented below. Fig.~\ref{fig:snrchi} shows the S/N-RCHI2-cut as blue solid lines, while sources to the left of these borders are excluded.
\begin{equation} \label{equ:snrchi1}
\begin{aligned}
 \mathtt{w1snr} > 10 \ \land \ (&\mathtt{w1rchi2}<2.5 \ \lor \\ 
 &\mathtt{w1rchi2} < 1.1 \times (\mathtt{w1snr}-25) + 2.5) 
\end{aligned}
\end{equation}
\begin{equation} \label{equ:snrchi2}
\begin{aligned}
 \mathtt{w2snr} > 10 \ \land \ (&\mathtt{w2rchi2}<2.5 \ \lor \\ 
 &\mathtt{w2rchi2} < 1 \times (\mathtt{w2snr}-35) + 2.5)
\end{aligned}
\end{equation}
\begin{equation} \label{equ:snrchi3}
\begin{aligned}
 \mathtt{w3snr} > 7 \ \land \ (&\mathtt{w3rchi2}<2 \ \lor \\ 
 &\mathtt{w3rchi2} < 0.5 \times (\mathtt{w3snr}+28) - 2)
\end{aligned}
\end{equation}
\begin{equation} \label{equ:snrchi4}
\begin{aligned}
 \mathtt{w4snr} > 7 \ \land \ (&\mathtt{w4rchi2}<2 \ \lor \\ 
 &\mathtt{w4rchi2} < 0.8 \times (\mathtt{w4snr}+21)-2)
\end{aligned}
\end{equation}
The S/N is directly correlated to the magnitude error (\texttt{w?sigmpro}), hence, an additional error-cut with the latter is not necessary.
Still, we require that these errors should not be \texttt{NULL} values.
\begin{equation} \label{equ:wsig}
\mathtt{w?sigmpro} \ \mathrm{not} \ \mathtt{NULL}
\end{equation}

\begin{figure}[!ht]
	\centering
	\includegraphics[width=0.98\linewidth]{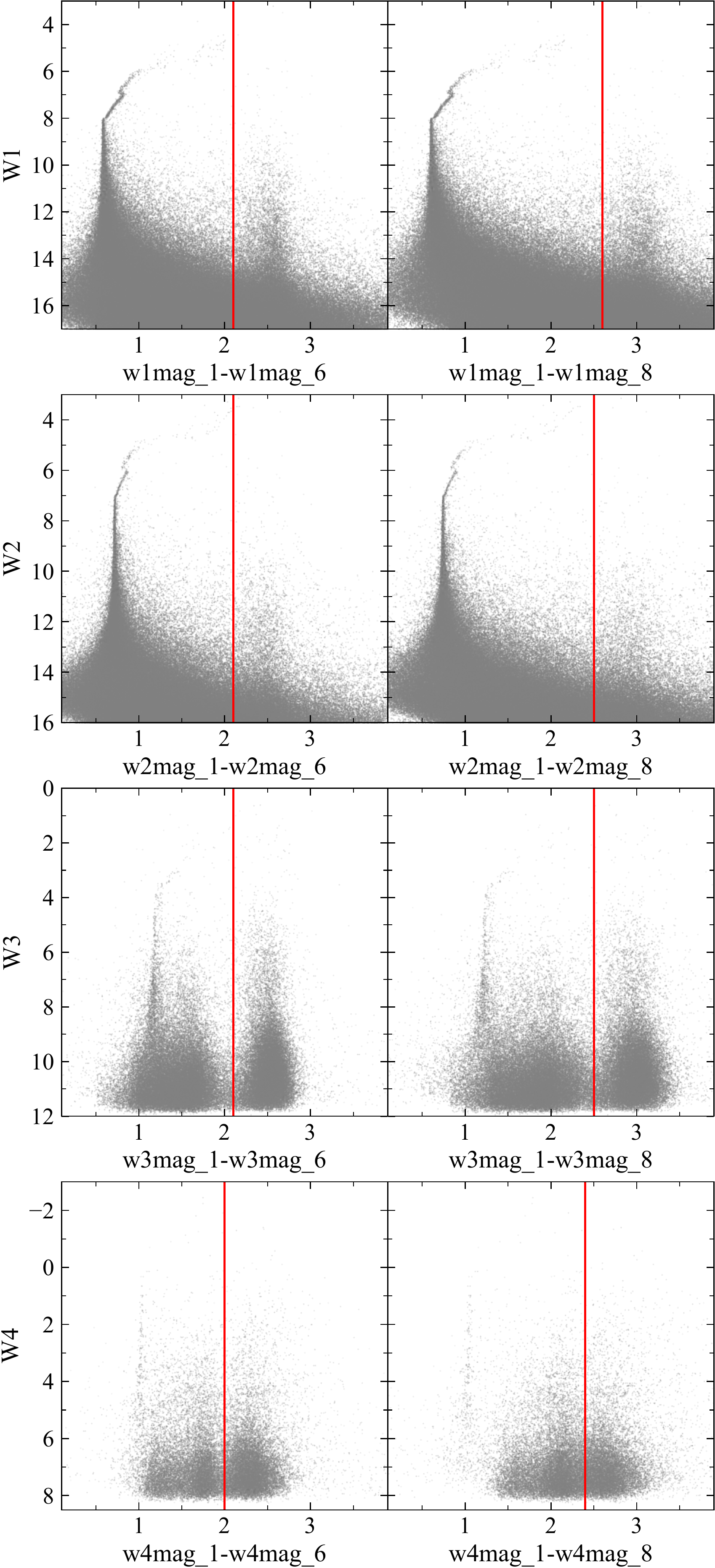}
	\caption{\emph{WISE} aperture photometry cuts to reduce extended source contamination. Gray dots are all WISE sources after applying basic error-cuts: $\mathtt{w?snr}>5$ and $\mathtt{w?sig} < 0.5$. The red solid lines show the criteria to separate point sources (left) from extended sources (right), given in Equs.~\ref{equ:apw1} to \ref{equ:apw4}.}
	\label{fig:apertures}
\end{figure}

To mitigate further contamination due to extended emission we use \emph{WISE} aperture photometry, which is provided by the AllWISE catalog for eight different aperture sizes \citep{Cutri2013}. Extended emission in the MIR is mainly caused by nebulosities, that can give an erroneous IR-excess. Especially the $W3$ band shows high contamination due to PAH emission. 
The aperture cuts were chosen using information from Figure~2 in \citet{Meisner2014}, where they show the curve-of-growth for the first six apertures of the $W3$ band. They use test sources of known type (extended or point like) and get a separation at about two when using $\mathtt{w3mag\_1}-\mathtt{w3mag\_6}$. 
We use the 6th and also the 8th (the largest) aperture, while the latter is not used by \citet{Meisner2014}.
To decide the final separation we plot $\mathtt{w?mag\_1}-\mathtt{w?mag\_6}$ and $\mathtt{w?mag\_1}-\mathtt{w?mag\_8}$ versus the magnitude of each band, shown in Fig.~\ref{fig:apertures}. 
In most of these diagrams, one can see an over-density of brighter sources, beyond $\sim$2, best visible for the $W3$ band, caused by bright nebulous structures. The $W1$ and $W2$ bands show a less prominent tip.  
Sources that do not satisfy the following conditions will be excluded if the \emph{WISE} band is used in a selection. The conditions and are shown as red solid lines in Fig.~\ref{fig:apertures}. 
\begin{equation} \label{equ:apw1}
\begin{aligned}
    \mathtt{w1mag\_1}-\mathtt{w1mag\_6} &< 2.1 \ \land \\ 
    \mathtt{w1mag\_1}-\mathtt{w1mag\_8} &< 2.6
\end{aligned}
\end{equation}
\begin{equation} \label{equ:apw2}
\begin{aligned}
    \mathtt{w2mag\_1}-\mathtt{w2mag\_6} &< 2.1 \ \land \\ 
    \mathtt{w2mag\_1}-\mathtt{w2mag\_8} &< 2.5 
\end{aligned}
\end{equation}
\begin{equation} \label{equ:apw3}
\begin{aligned}
    \mathtt{w3mag\_1}-\mathtt{w3mag\_6} &< 2.1 \ \land \\ 
    \mathtt{w3mag\_1}-\mathtt{w3mag\_8} &< 2.5
\end{aligned}
\end{equation}
\begin{equation} \label{equ:apw4}
\begin{aligned}
    \mathtt{w4mag\_1}-\mathtt{w4mag\_6} &< 2.0 \ \land \\ 
    \mathtt{w4mag\_1}-\mathtt{w4mag\_8} &< 2.4  
\end{aligned}
\end{equation}

\begin{figure*}[!ht]
	\centering
	\includegraphics[width=0.85\linewidth]{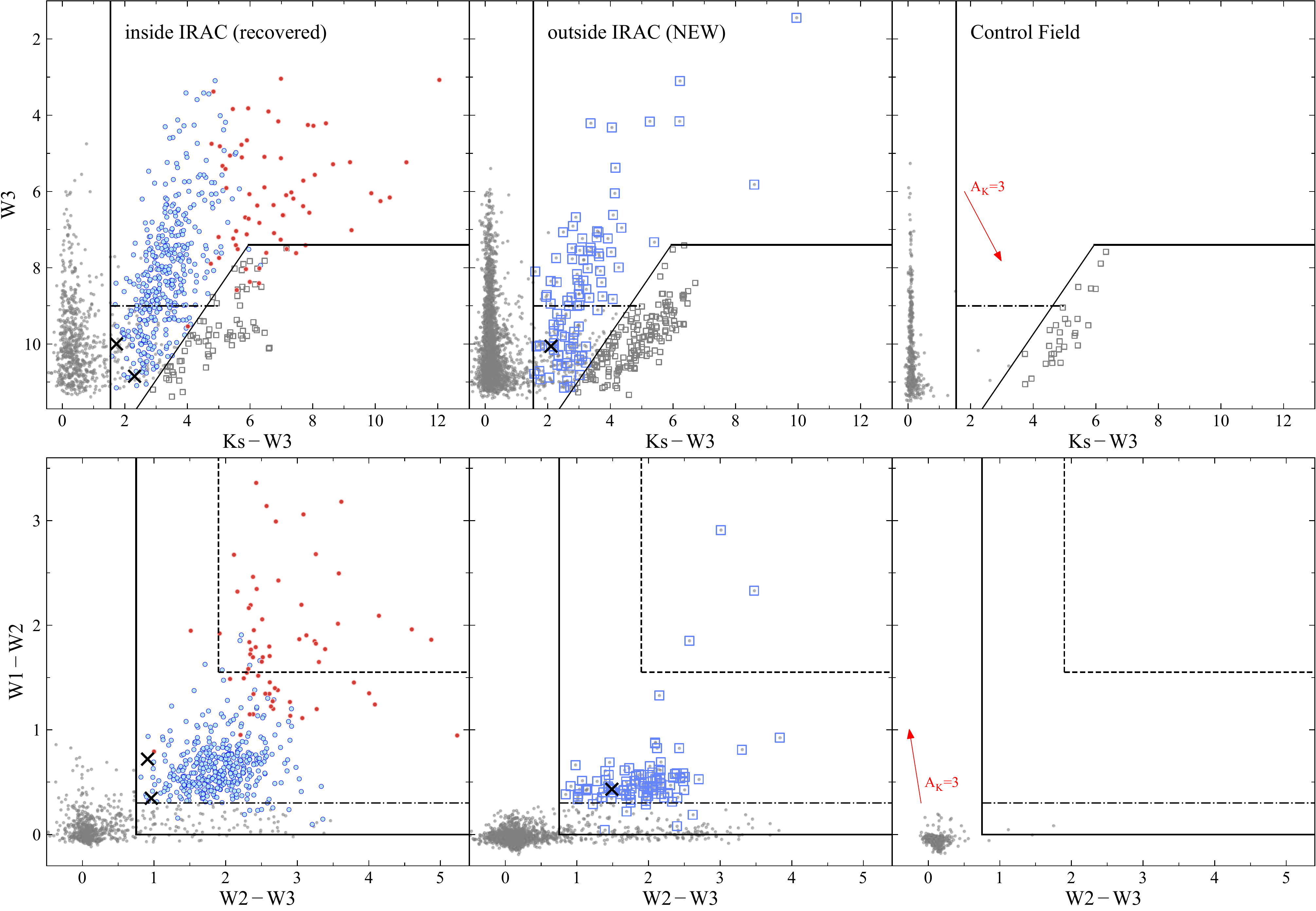}
	\caption{W123-selection. 
	Left: Orion\,A science field inside the IRAC coverage for L1641, showing recovered YSO candidates (blue - disks, red - protostars). 
	Center: Orion\,A science field outside the IRAC coverage showing new YSO candidates (blue open squares). 
	Right: Control field showing background and MS star contamination.  
	Top: $K_S-W3$ vs $W3$ CMD. To clean the sample of extra-galactic contamination we exclude sources below and right to the black lines (AGN-cut, gray open boxes). The left border reduces contaminated $W3$ photometry and eliminates MS stars.  
	Bottom: $W2-W3$ vs $W1-W2$ CCD. The black lines show the selection conditions. Sources from the AGN-cut are already removed. The slope on the left separates MS stars. The cut at the bottom reduces contaminated photometry.  
	The dashed black lines in the upper right corner show an additional selection of sources where the AGN-cut was not applied due to the fact that this region is free of any contamination. The horizontal black dashed lines in both diagrams show the exclusion condition for faint sources ($W3>\SI{9}{mag}$) which at the same time fall below $W1-W2<0.4$. 
    }
	\label{fig:w123}
\end{figure*}

Furthermore, we check if the sources are reliable detections by using the AllWISE cataloged number of individual exposures for a given band, where a profile-fit measurement of the source was possible (\texttt{w?m}).
We combine this with \texttt{w?nm}, which gives the number of times it was detected with a $\mathrm{S/N}>3$. We require this should be at least 20\% when compared to \texttt{w?m}:
\begin{equation} \label{equ:wnm}
\mathtt{w?nm} / \mathtt{w?m} \geq 0.2 
\end{equation}
This cut excludes mainly faint sources which are mostly already excluded by the S/N-RCHI2-cut.  
Finally, we exclude sources listed as contaminated or confused by artifacts (\texttt{cc\_flags}\footnote{Contamination and confusion flag due to proximity to an image artifact indicated by four flags: diffraction spike (d, D), persistence of a short-term latent image (p, P), halo (h, H), and optical ghost (o, O). 0 indicates not-confused photometry. Upper-case letters denote spurious detection of an artifact and lower-case letters denote that the measurement may be real but might be contaminated by the artifact.}), allowing only 0 or lower-case letters (d, p, h, o).
In the following \emph{WISE} based selections we apply the discussed cuts only to those bands used for a selection. 

\begin{figure*}[!ht]
	\centering
	\includegraphics[width=0.85\linewidth]{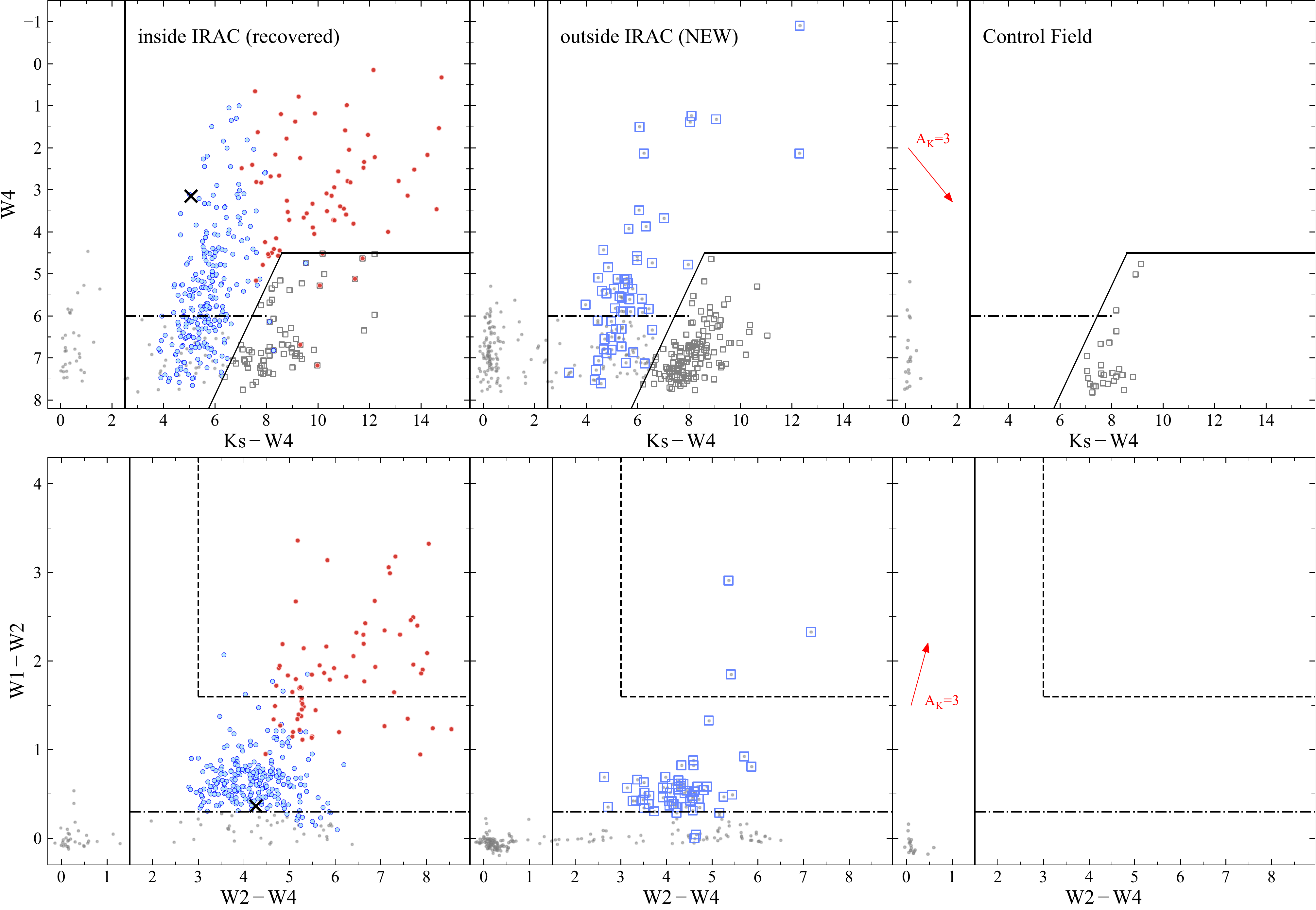}
	\caption{W124-selection. See caption of Fig.~\ref{fig:w123} for more explanations. }
	\label{fig:w124}
\end{figure*}

Further we over-plot known objects from the already analyzed IRAC region to get an idea of the color-spaces of YSOs and contaminating objects in the diagrams (see left plots in Figs.~\ref{fig:w123}, \ref{fig:w124}, \ref{fig:HKW12}), as it was done for the \emph{Spitzer} based selections above. However, we use only known objects from the L1641 region ($l > \SI{210}{\degree}$), which is a region less influenced by fuzzy nebulous contamination. Otherwise, if including the ONC region, one needs to apply more conservative error cuts.


\subsubsection{VISTA and W123} \label{W123}

First we use a combination of the first three \emph{WISE} bands and VISTA $K_S$ to investigate the CCD $W2-W3$ vs $W1-W2$ and the CMD $K_S-W3$ vs $W3$  (W123-selection, Fig.~\ref{fig:w123}). Including the $K_S$ band reduces contamination due to fake \emph{WISE} point-sources. 
Additional to the \emph{WISE} error-cuts presented above we further apply the following: 
\begin{equation}
\begin{aligned}
K_S\mathit{err} &< \SI{0.1}{mag}  \\
W1, \, W2 &< \SI{14}{mag}
\end{aligned}
\end{equation}
With this there are 1,910 sources left in the IRAC region at L1641, 2,276 outside the IRAC region, and 350 in the control field. 
%
To eliminate extra-galactic contamination we exclude sources in the CMD, where AGNs and galaxies with PAH emission are located at the lower right of this diagram at about $K_S-W3 \approx 5$ (see control field, AGN-cut, Fig.~\ref{fig:w123}, gray boxes). Therefore all source that do not fulfill the following conditions are removed from the CCD.
\begin{equation} \label{equ:kw3-agn}
\begin{aligned}
W3 &\leq \SI{7.4}{mag} \ \lor \\ 
W3 &\leq -1.2 \cdot (K_S - W3 - 5.95) + 7.4 
\end{aligned}
\end{equation} 
%
\begin{figure}[!ht]
	\centering
	\includegraphics[width=0.63\linewidth]{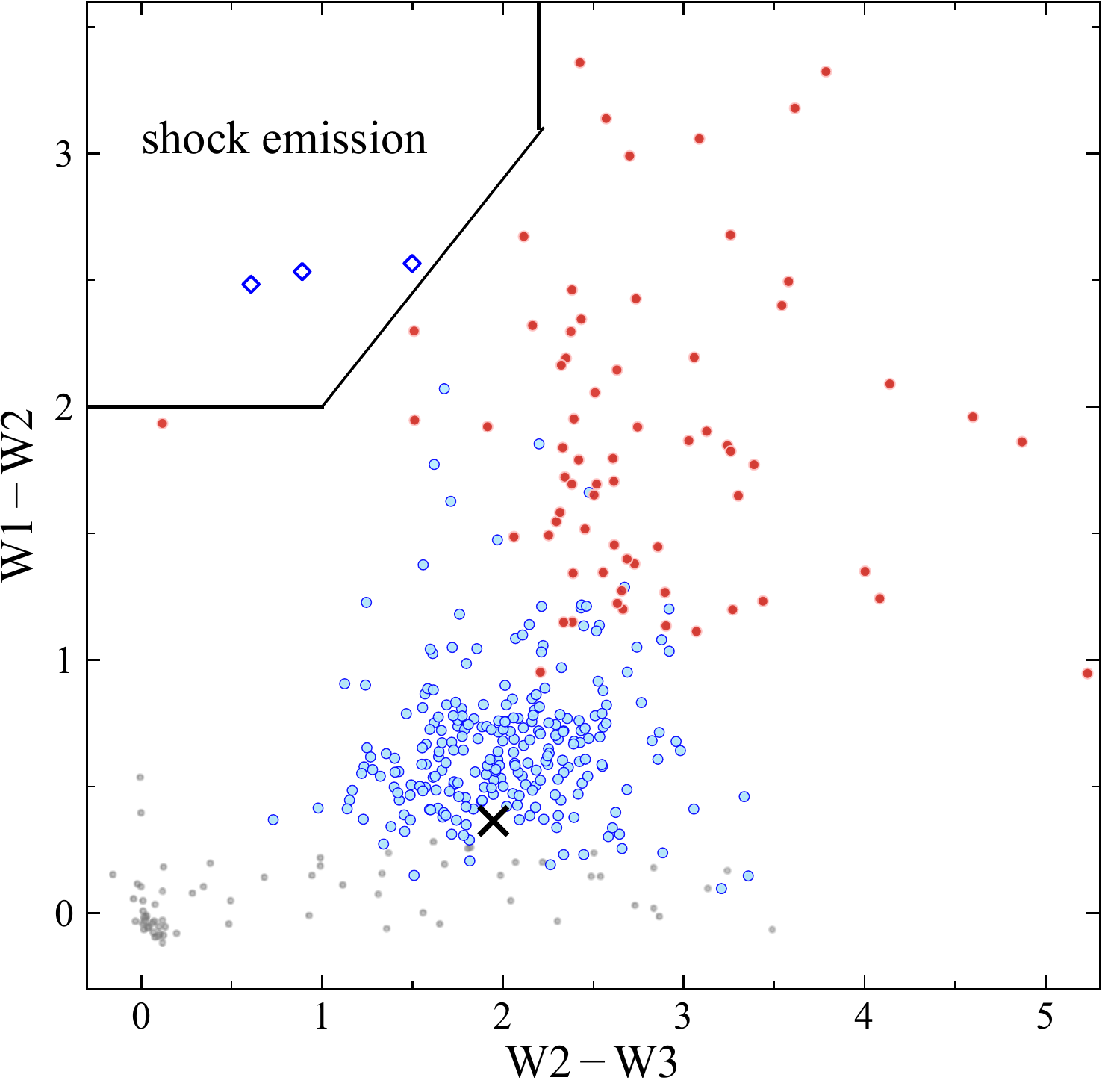}
	\caption{W123 CCD showing the cut to exclulde shock emission from the W124 selection.}
	\label{fig:w124-w123-fuzz}
\end{figure}
To exclude MS stars and reddened sources we apply the following conditions, which are the solid vertical lines on the left in the CMD and CCD plots,
\begin{equation}
\begin{aligned}
K_S-W3 &\geq 1.54 \ \land \\
W1-W2 &\geq 0.75
\end{aligned}
\end{equation}
and we cut sources at the very bottom of the CCD, which is a color region highly contaminated by fake $W3$ point-sources,
\begin{equation}
W1-W2 > 0.
\end{equation}
To further mitigate such contaminants we apply a special cut for faint sources, shown by the horizontal dash-dotted lines in the CMD and CCD; for faint sources with $W3>\SI{9}{mag}$ we require $W1-W2 \geq 0.3$, since this is a color region less influence by contaminated photometry.
The black dashed borders at the top right of the CCD are used to re-add candidates. Similar to the HKI12-selection, this region is found not to be effected by galaxy contamination, and the following conditions are applied without the AGN-cut (Equ.~\ref{equ:kw3-agn}):
\begin{equation}
\begin{aligned}
W1-W2 &\geq 1.55 \ \land \\ 
W2-W3 &\geq 1.9 
\end{aligned}
\end{equation}
With these conditions we select 97 new YSO candidates outside the IRAC region, as shown by the blue open boxes in the middle plots of Figure \ref{fig:w123}.
Inside the IRAC region at L1641 (left plots) we select 456 YSO candidates, of which 454 are recovered known YSOs, and two are contaminants due to extended emission in the $W3$ band.
In the whole Orion\,A IRAC region the W123-selection recovers about 28\% of known YSO candidates, while in L1641 the recovery rate lies at about 51\%.


\subsubsection{VISTA and W124} \label{W124}

\begin{figure*}[!ht]
	\centering
	\includegraphics[width=0.85\linewidth]{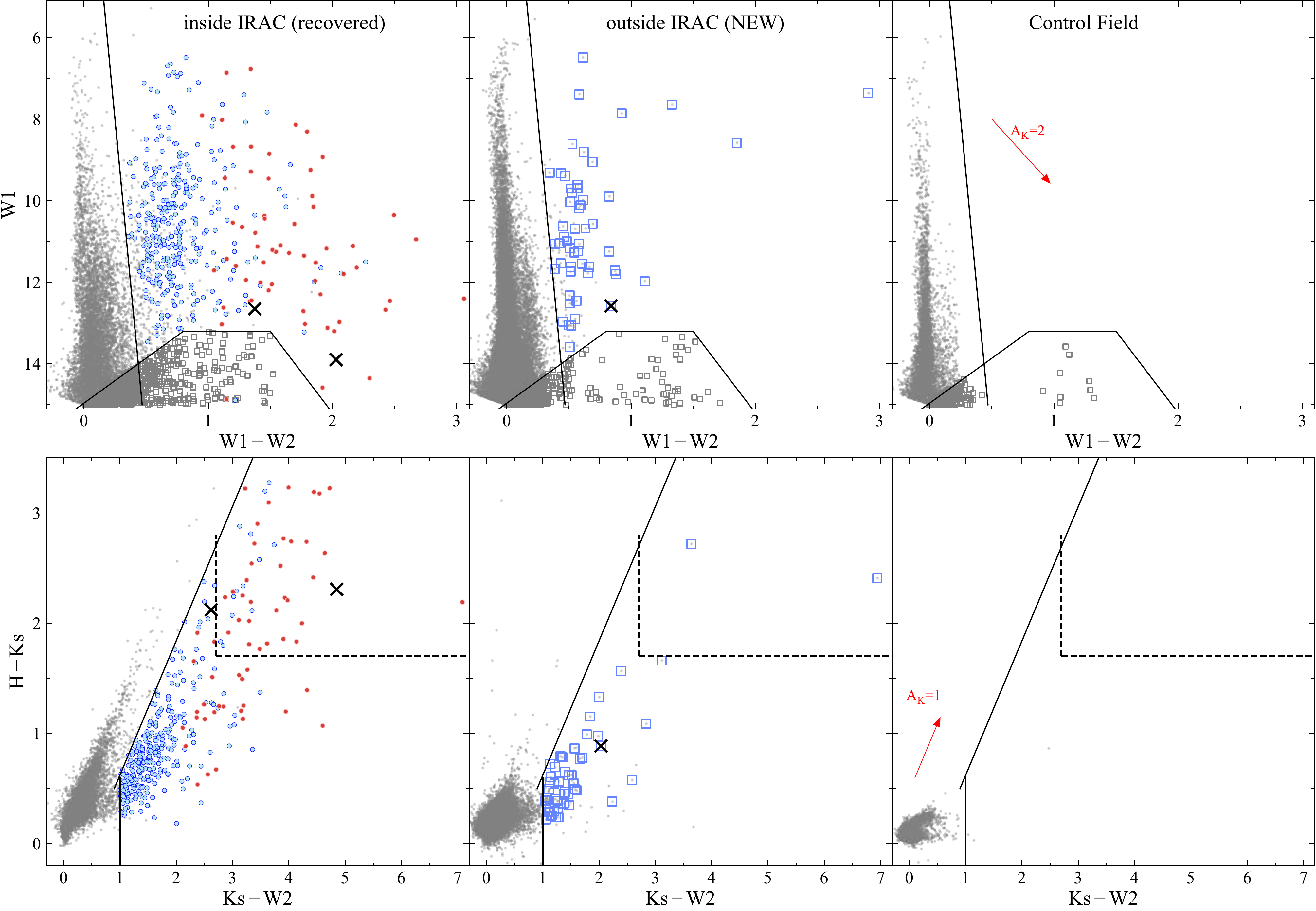}
	\caption{HKW12-selection. See also caption of Fig.~\ref{fig:w123}. Top: $W1-W2$ vs $W1$ CMD. To clean the sample of extra-galactic contamination we exclude sources below the black bottom lines, indicating the AGN-cut (gray boxes). 
	Bottom: $K_S-W2$ vs $H-K_S$ CCD. The black lines show the selection conditions, and the black dashed lines the additional condition for a region with no AGN contamination. This gives a selection of 52 new YSO candidates outside the IRAC region.}
	\label{fig:HKW12}
\end{figure*}

Next we add sources by investigating the CCD $W2-W4$ vs $W1-W2$ and the CMD $K_S-W4$ vs $W4$ (W124-selection, Fig.~\ref{fig:w124}), with a very similar approach to the W123-selection. Beside the main basic error cuts we require:
\begin{equation}
\begin{aligned}
K_S\mathit{err} &< \SI{0.1}{mag} \\
W1, \, W2 &< \SI{15}{mag}
\end{aligned}
\end{equation}
With this there are 449 sources left in the IRAC region at L1641, 363 outside the IRAC region, and only 44 in the control field. 
We apply again an AGN-cut using the CMD (Fig.~\ref{fig:w124}, gray boxes):
\begin{equation} \label{equ:kw4-agn}
\begin{aligned}
W4 &\leq 4.5 \ \lor \\
W4 &\leq -1.6 \times (K_S - W4 - 8.6) + 4.5
\end{aligned}
\end{equation}
as shown by the solid black lines in the bottom right of the CMD. The sources beyond these borders (gray boxes) will be removed from the CCD.
To exclude MS stars and reddened sources due to extinction we apply the following conditions:
\begin{equation}
\begin{aligned}
K_S-W4 &\geq 2.5 \ \land \\
W2-W4 &\geq 1.5
\end{aligned}
\end{equation}
Again we apply a special cut for faint source to mitigate contaminated $W4$ photometry; we require for faint sources with $W4 \geq \SI{6}{mag}$ that they should be redder than $W1-W2 \geq 0.3$, shown by the horizontal dash-dotted lines in the CMD and CCD. 
Furthermore, we re-add again sources in the top right corner of the CCD (black dashed borders), therefore, the following conditions are applied without the AGN-cut (Equ.~\ref{equ:kw4-agn}):
\begin{equation}
\begin{aligned}
W1-W2 &\geq 1.6 \ \land \\ 
W2-W4 &\geq 3 
\end{aligned}
\end{equation} 
This region is not contaminated by extra-galactic sources, however, there is still some contamination due to shock emission. Similar to the HKI12-selection we require an additional cut to exclude this fuzzy contamination, which can be identified using the W123 CCD (Fig.~\ref{fig:w124-w123-fuzz}). Such sources are located in the top left corner of this diagram. We will apply this only to sources with a valid $W3$ measurement ($W3$ not \texttt{NULL}). 
\begin{equation}
\begin{aligned}
&W1-W2 \leq 2 \ \lor \\
&W1-W2 \leq 0.9 \times (W2-W3 - 1) + 2 \ \lor \\
(&W1-W2 \geq 2 \ \land \ W2-W3 > 2.2)
\end{aligned}
\end{equation}
With the W124-selection we get 59 new YSO candidates outside IRAC (middle plots). Inside the IRAC region (left plots) we select 321 YSO candidates, of which 320 are recovered known YSOs. One was found to be contamination due to a double-star, which was erroneously cross-matched. In the whole Orion\,A region this selection recovers only about 18\% of known YSO candidates, while for L1641 the recovery rate is about 36\%. 
The combined W123-W124 selection recovers about 54\% in L1641.

\begin{table*}[!ht]
\small
\begin{center} 
\caption{Recovery rates for the six individual YSO selection conditions as presented in this Appendix.} 
\begin{tabular}{lcccccccccc}
\hline \hline
 \multicolumn{1}{c}{} &
 \multicolumn{1}{c}{L1641\tablefootmark{a}} &
 \multicolumn{4}{c}{VISTA/\emph{Spitzer} (VS)} &
 \multicolumn{4}{c}{VISTA/\emph{WISE} (VW)} &
 \multicolumn{1}{c}{Total} \\
\cmidrule(lr){3-6}
\cmidrule(lr){7-10}
 \multicolumn{1}{c}{Class} &
 \multicolumn{1}{c}{known} &
 \multicolumn{1}{c}{HKI12} &
 \multicolumn{1}{c}{HKI24} &
 \multicolumn{1}{c}{HKM}   &
 \multicolumn{1}{c}{total} &
 \multicolumn{1}{c}{W123}  &
 \multicolumn{1}{c}{W124}  &
 \multicolumn{1}{c}{HKW12} &
 \multicolumn{1}{c}{total} &
 \multicolumn{1}{c}{VS+VW} \\
\hline
 ALL      & 880 & 47\% & 70\% & 68\% & 82\% &   51\% & 36\% & 41\% & 59\% &   86\%  \\
\hline 
 D        & 665 & 45\% & 81\% & 78\% & 91\% &   55\% & 35\% & 42\% & 62\% &   92\%  \\
 F        & 101 & 81\% & 63\% & 61\% & 83\% &   51\% & 51\% & 65\% & 69\% &   88\%  \\
 P        & 114 & 30\% & 13\% & 14\% & 31\% &   30\% & 33\% & 21\% & 38\% &   48\%  \\
\hline
\end{tabular}
\label{tab:recovery}
\end{center}
\tablefoot{
        The recovery check is applied only on sources located in L1641 inside the IRAC coverage at $l > \SI{210}{\degree}$, to avoid high contaminated regions close to the ONC. It is given in \% compared to known YSO candidates.
        \tablefoottext{a}{Number of known YSO candidates (revisited MGM and \citetalias{Furlan2016} sample) in L1641.}
           }
\end{table*}

\begin{figure*}[!ht]
	\centering
	    \includegraphics[width=1\linewidth]{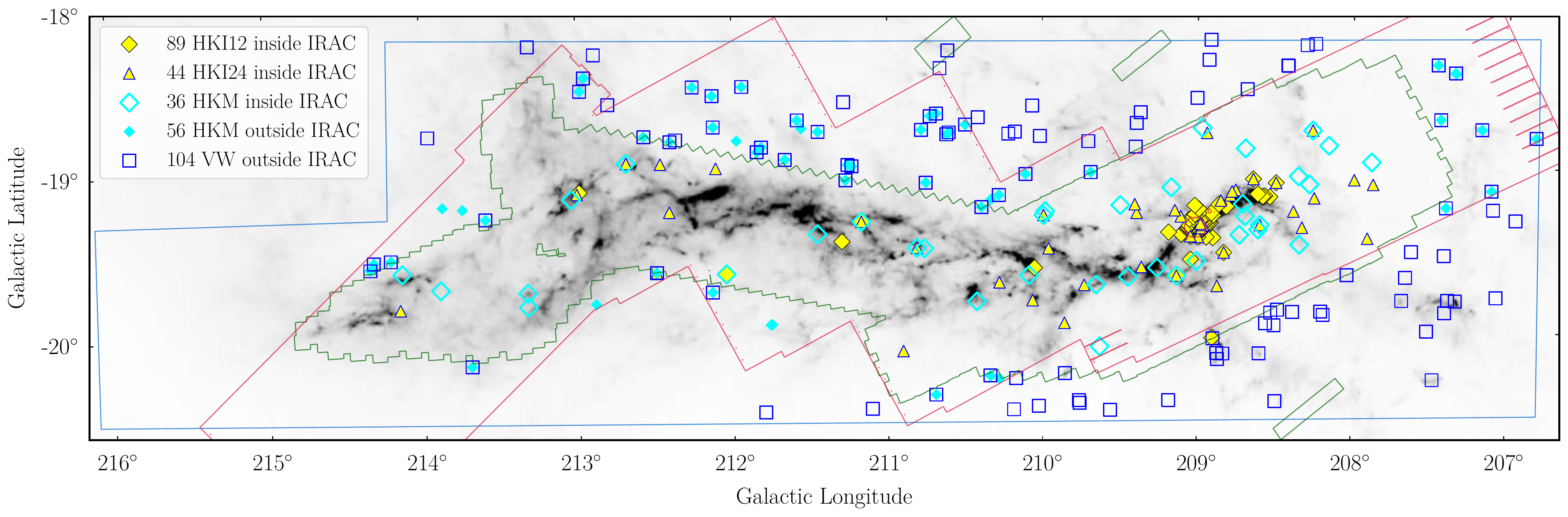}
	\caption{The distribution of the 268 new YSO candidates selected with the VISTA/WISE/Spitzer methods. 
	The sources found in- and outside the IRAC coverage are shown separately with different symbols (see legend). 
	Outside the IRAC region: blue open box symbols - new VISTA/\emph{WISE} (VW) YSO candidates; cyan filled diamonds - new VISTA/$M1$ (HKM) YSO candidates. 
	Inside the IRAC region: yellow diamonds - new VISTA/$I12$ (HKI12) YSO candidates; yellow triangles - new VISTA/$I24$ (HKI24) YSO candidates; cyan open diamonds - new VISTA/$M1$ (HKM) YSO candidates.}
	\label{fig:mapnew2}
\end{figure*}


\subsubsection{VISTA and W12} \label{HKW12}

Finally we use the CCD $K_S-W2$ vs $H-K_S$ and the CMD $W1-W2$ vs $W1$  (HKW12-selection, Fig.~\ref{fig:HKW12}), which is a similar selection as the HKI12-selection. It aims to add sources that might have been missed by the W123 or W124 selections due to the higher contamination in the longer wavelength bands. Additionally to the basic \emph{WISE} cuts we apply the following error-cuts:
\begin{equation}
\begin{aligned}
K_S\mathit{err},\, H\mathit{err} &< 0.1 \\
W1, \, W2 &< \SI{14.5}{mag} \\
\mathtt{ClassSex} &> 0.1
\end{aligned}
\end{equation}
There are 8,325 sources left inside the IRAC region in L1641 (14,641 in whole Orion\,A), 27,610 outside, and 4,611 in the control field.
To get rid of faint uncertain sources and extra-galactic contamination we again apply a cut in the CMD (AGN-cut, Fig.~\ref{fig:HKW12}), as shown by black solid lines at the bottom (trapezoidal shape). 
\begin{equation} \label{equ:kw12-agn}
\begin{aligned}
W1 &\leq 13.2 \ \lor \\
W1 &\leq -2.2 \times (W1 - W2 - 0.8) + 13.2 \ \lor \\
W1 &\leq 4 \times (W1 - W2 - 1.5) + 13.2
\end{aligned}
\end{equation}
Again, the sources beyond these borders (gray boxes) will be removed from the CCD.
To exclude MS stars we apply the following conditions: 
\begin{equation}
\begin{aligned}
W1 &\geq 32 \times (W1 - W2) \ \land \\
K_S-W2 &\geq 1  \ \land \\
H-K_S &\leq 0.546/0.448 \times (K_S - W2 - 0.9) + 0.5 
\end{aligned}
\end{equation}
These conditions are shown by the solid lines on the left in the CMD and CCD.
The latter is a slope parallel to the reddening vector in the CCD, excluding reddened MS-stars due to extinction,
To re-add YSO candidates which fall in the AGN-cut (Equ.~\ref{equ:kw12-agn}) we again introduce a selection at the top right part of the CCD, indicated by the black dashed lines:
\begin{equation}
\begin{aligned}
H-K_S &> 1.7 \ \land \\
K_S-W2 &> 2.7
\end{aligned}
\end{equation}

With the HKW12 selection we select 52 new YSO candidates outside the IRAC region, as shown by the blue open boxes in the middle plots of Figure \ref{fig:w124}.
Inside the IRAC region (left plots) we select 364 sources, of which 362 are recovered known YSO candidates, and two are uncertain or contaminating sources. 
Inside the whole Orion\,A IRAC region we recover about 27\% of the known YSO candidates with the HKW12 selection, while the recovery rate for the L1641 region is about 41\%. 
There are 47 overlaps with the W123-W124 selections outside IRAC, therefore we find only five new candidates in the surroundings with the HKW12 selection. However, the recovery rate in L1641 improves to 59\% when combining this selection with the two previous ones. 
In total we are able to add 104 new YSO candidates with the \emph{WISE} based selections (W123-W124-HKW12) outside IRAC (see blue open squares on the map in Fig.~\ref{fig:mapnew2}.).


\subsection{Recovery rate of the YSO selections} \label{app:recovery}

We test the recovery rate for each individual selection by comparing with known YSO candidates (from Sect.~\ref{revisit}) in L1641 ($l > \SI{210}{\degree}$) inside the IRAC region. The low-mass star forming region L1641 is a more fair comparison to the surrounding regions (outside IRAC), being overall less affected by contamination due to extended emission or crowding. Especially for a \emph{WISE} based selection this comparison makes more sense, since \emph{WISE} shows large saturated regions near the ONC.
The individual recovery rates are summarized in Table~\ref{tab:recovery}. We also give individual recovery rates for the three YSO classes as classified in this work.

The combined \emph{Spitzer} based selections (VS-selection) recover about 82\% of the known YSO candidates in L1641, or about 80\% when including the ONC.
The rest 20\% are likely missed due to the chosen error cuts and selection conditions. We like to note that these conditions were not designed to redo the selection for the IRAC region, but rather to choose color spaces that can provide additional candidates. 
The combined \emph{WISE} selections (VW-selection) recover together about 59\% in the region of L1641 inside IRAC. Including the ONC the recovery rate drops to about 38\%, highlighting the influence of massive star-forming regions on low resolution MIR data.
The whole selection (combining VISTA/\emph{WISE}/\emph{Spitzer}) recovers about 86\% in L1641. 
When looking at the individual recovery rates for the three YSO classes, we see that our methods better recover disks and flat-spectrum sources, while they recover less than 50\% of the protostars. By including the extension flags from VISTA, we are likely loosing sources that are connected to outflows or nebulosities. Also, by including NIR data, highly embedded sources can be missed. Nevertheless, such sources are mostly already known, or are added by FIR data or visually (e.g., when there are prominent outflows).


\section{Table} \label{Catalogs}
In Table~\ref{tab:master} we provide the column information for the final table, which contains all sources discussed in this work (3117). The table gives basic information, including RA/Dec J2000 positions, the VISION\,ID and other relevant identification numbers. The VISTA, \emph{Spitzer}, and \emph{WISE} magnitudes are listed with errors, and the spectral indices, which were used to classify the YSO candidates. Columns~87 to 91 give classifications from previous works when available. The latter two classifications from \citet{Fang2009, Fang2013} and \citet{Pillitteri2013} are not discussed in this work but are listed for completeness ($C_\mathrm{Fang}$, $C_\mathrm{P13}$). Column~93 ({\tt Class\_flag}) separates the sources in four categories, labeled with the numbers ``1,2,3,4'': 
\begin{flushleft}
1 = revisited YSO candidates which were previously selected by MGM or \citetalias{Furlan2016} (2706) \\ 
2 = new YSO candidates (274) \\
3 = rejected candidates (96) \\
4 = uncertain candidates (41)
\end{flushleft}
The number counts of each sample are given in parenthesis. We like to note that the rejected candidates include four sources (92+4) which were actually not listed as YSO candidates previously but as contaminating objects by \citetalias{Furlan2016}. We keep them in this final catalog for completeness.

\onecolumn
\begin{landscape}
\small
\begin{longtable}{clcl}
\caption{Description of columns of the final catalog containing all sources discussed in this paper. The catalog is only available in electronic form at the CDS.} \\ 
\label{tab:master}\\
\hline \hline
  \makecell[c]{Column\\Number} &
  \makecell[c]{Column\\Name} &
  \multicolumn{1}{c}{Units} &
  \multicolumn{1}{c}{Description} \\
\hline
\endfirsthead %

\caption{ --- Continued} \\
\hline \hline
  \makecell[c]{Column\\Number} &
  \makecell[c]{Column\\Name} &
  \multicolumn{1}{c}{Units} &
  \multicolumn{1}{c}{Description} \\
\hline
\endhead

\hline
\endfoot

  1 & RAJ2000 & hh:mm:ss & Right Ascension taken from the reference as given in Col.~3. \\
  2 & DEJ2000 & dd:mm:ss & Declination taken from the reference as given in Col.~3. \\
  3 & Ref & & The Reference tells the origin of the coordinates. (1) VISTA coordinates, \citet{Meingast2016}; \\
    &     & & (2) \emph{Spitzer} coordinates, \citet{Megeath2012}; (3) \emph{Herschel}/PACS point source catalog coordinates, \citet{Pilbratt2010}. \\
\hline

  4 & VISTA & & VISTA identification number \\
  5 & WISE  & & WISE identification number \\
  6 & PACS  & & \emph{Herschel}/PACS point source catalog (HPPSC) identification number \\
  7 & ID    & & Running identification number from this work \\
  8 & MGM   & & Source index from \citet{Megeath2012} or \citet{Megeath2016} for MGM sources  \\
  9 & HOPS  & & Source index from \citet{Furlan2016} for HOPS sources \\
 10 & Simbad\_Name  & & Main identification as given in the SIMBAD Astronomical Database \citep{Wenger2000} \\
 11 & Otype & & Object Type as given in SIMBAD \\
\hline  
  
 12 & Jmag        & mag & J magnitude from VISTA \\
 13 & e\_Jmag     & mag & 1$\sigma$ error of J magnitude from VISTA \\
 14 & Hmag        & mag & H magnitude from VISTA \\
 15 & e\_Hmag     & mag & 1$\sigma$ error of H magnitude from VISTA \\
 16 & Ksmag    & mag & K$_S$ magnitude from VISTA \\
 17 & e\_Ksmag & mag & 1$\sigma$ error of K$_S$ magnitude from VISTA \\
 
 18 & IRAC1    & mag & IRAC1 magnitude from \emph{Spitzer}/IRAC \\
 19 & e\_IRAC1 & mag & error of IRAC1 magnitude from \emph{Spitzer}/IRAC \\
 20 & IRAC2    & mag & IRAC2 magnitude from \emph{Spitzer}/IRAC \\
 21 & e\_IRAC2 & mag & error of IRAC2 magnitude from \emph{Spitzer}/IRAC \\ 
 22 & IRAC3    & mag & IRAC3 magnitude from \emph{Spitzer}/IRAC \\
 23 & e\_IRAC3 & mag & error of IRAC3 magnitude from \emph{Spitzer}/IRAC \\
 24 & IRAC4    & mag & IRAC4 magnitude from \emph{Spitzer}/IRAC \\
 25 & e\_IRAC4 & mag & error of IRAC4 magnitude from \emph{Spitzer}/IRAC \\
 26 & MIPS1    & mag & MIPS1 magnitude from \emph{Spitzer}/MIPS \\
 27 & e\_MIPS1 & mag & error of MIPS1 magnitude from \emph{Spitzer}/MIPS \\
 
 28 & W1    & mag & W1 magnitude given as \textit{w1mpro} in the AllWISE catalog \\
 29 & e\_W1 & mag & error of W1 magnitude given as \textit{w1sigmpro} in the AllWISE catalog \\
 30 & W2    & mag & W2 magnitude given as \textit{w2mpro} in the AllWISE catalog \\
 31 & e\_W2 & mag & error of W2 magnitude given as \textit{w2sigmpro} in the AllWISE catalog \\
 32 & W3    & mag & W3 magnitude given as \textit{w3mpro} in the AllWISE catalog \\
 33 & e\_W3 & mag & error of W3 magnitude given as \textit{w3sigmpro} in the AllWISE catalog \\
 34 & W4    & mag & W4 magnitude given as \textit{w4mpro} in the AllWISE catalog \\
 35 & e\_W4 & mag & error of W4 magnitude given as \textit{w4sigmpro} in the AllWISE catalog \\
\hline 
 
 36 & alpha\_KM & & Observed spectral index $\alpha_\mathrm{KM}$ from 2.15 to $\SI{24}{\micro \meter}$ covering VISTA/K$_S$, the four IRAC bands, and MIPS1 \\
 37 & e\_alpha\_KM & & fitting error of $\alpha_\mathrm{KM}$  \\
 38 & alpha\_IM & & Observed spectral index $\alpha_\mathrm{IM}$ from 3.6 to $\SI{24}{\micro \meter}$ covering the four IRAC bands and MIPS1 \\
 39 & e\_alpha\_IM & & fitting error of $\alpha_\mathrm{IM}$  \\
 40 & alpha\_IRAC & & Observed spectral index $\alpha_\mathrm{IRAC}$ from 3.6 to $\SI{8}{\micro \meter}$ covering the four IRAC bands \\
 41 & e\_alpha\_IRAC & & fitting error of $\alpha_\mathrm{IRAC}$  \\
 42 & alpha\_I2M & & Observed spectral index $\alpha_\mathrm{I2M}$ from 4.5 to $\SI{24}{\micro \meter}$ covering three IRAC bands and MIPS1 \\
 43 & e\_alpha\_I2M & & fitting error of $\alpha_\mathrm{I2M}$  \\
 44 & alpha\_I3M & & Observed spectral index $\alpha_\mathrm{I3M}$ from 5.8 to $\SI{24}{\micro \meter}$ covering two IRAC bands and MIPS1 \\
 45 & e\_alpha\_I3M & & fitting error of $\alpha_\mathrm{I3M}$  \\
 46 & alpha\_KI3 & & Observed spectral index $\alpha_\mathrm{KI3}$ from 2.15 to $\SI{5.8}{\micro \meter}$ covering VISTA/K$_S$ and three IRAC bands \\
 47 & e\_alpha\_KI3 & & fitting error of $\alpha_\mathrm{KI3}$  \\
 
 48 & alpha\_KW3 & & Observed spectral index $\alpha_\mathrm{KW3}$ from 2.15 to $\SI{12}{\micro \meter}$ covering VISTA/K$_S$ and the first three WISE bands \\
 49 & e\_alpha\_KW3 & & fitting error of $\alpha_\mathrm{KW3}$  \\
 50 & alpha\_KW & & Observed spectral index $\alpha_\mathrm{KW}$ from 2.15 to $\SI{22}{\micro \meter}$ covering VISTA/K$_S$ and all four WISE bands \\
 51 & e\_alpha\_KW & & fitting error of $\alpha_\mathrm{KW}$  \\
 52 & alpha\_W13 & & Observed spectral index $\alpha_\mathrm{W13}$ from 3.4 to $\SI{12}{\micro \meter}$ covering the first three WISE bands \\
 53 & e\_alpha\_W13 & & fitting error of $\alpha_\mathrm{W13}$  \\
 54 & alpha\_WISE & & Observed spectral index $\alpha_\mathrm{WISE}$ from 3.4 to $\SI{22}{\micro \meter}$ covering the four WISE bands \\
 55 & e\_alpha\_WISE & & fitting error of $\alpha_\mathrm{WISE}$  \\
 56 & alpha\_KW12M & & Observed spectral index $\alpha_\mathrm{KW12M}$ from 2.15 to $\SI{24}{\micro \meter}$ covering VISTA/K$_S$, the first two WISE bands, and MIPS1 \\
 57 & e\_alpha\_KW12M & & fitting error of $\alpha_\mathrm{KW12M}$  \\
\hline

 58 & alpha\_KM\_0 & & De-reddened spectral index $\alpha_\mathrm{KM}$ from 2.15 to $\SI{24}{\micro \meter}$ covering VISTA/K$_S$, the four IRAC bands, and MIPS1 \\
 59 & e\_alpha\_KM\_0 & & fitting error of $\alpha_\mathrm{KM}$  \\
 60 & alpha\_IM\_0 & & De-reddened spectral index $\alpha_\mathrm{IM}$ from 3.6 to $\SI{24}{\micro \meter}$ covering the four IRAC bands and MIPS1 \\
 61 & e\_alpha\_IM\_0 & & fitting error of $\alpha_\mathrm{IM}$  \\
 62 & alpha\_IRAC\_0 & & De-reddened spectral index $\alpha_\mathrm{IRAC}$ from 3.6 to $\SI{8}{\micro \meter}$ covering the four IRAC bands \\
 63 & e\_alpha\_IRAC\_0 & & fitting error of $\alpha_\mathrm{IRAC}$  \\
 64 & alpha\_I2M\_0 & & De-reddened spectral index $\alpha_\mathrm{I2M}$ from 4.5 to $\SI{24}{\micro \meter}$ covering three IRAC bands and MIPS1 \\
 65 & e\_alpha\_I2M\_0 & & fitting error of $\alpha_\mathrm{I2M}$  \\
 66 & alpha\_I3M\_0 & & De-reddened spectral index $\alpha_\mathrm{I3M}$ from 5.8 to $\SI{24}{\micro \meter}$ covering two IRAC bands and MIPS1 \\
 67 & e\_alpha\_I3M\_0 & & fitting error of $\alpha_\mathrm{I3M}$  \\
 68 & alpha\_KI3\_0 & & De-reddened spectral index $\alpha_\mathrm{KI3}$ from 2.15 to $\SI{5.8}{\micro \meter}$ covering VISTA/K$_S$ and three IRAC bands \\
 69 & e\_alpha\_KI3\_0 & & fitting error of $\alpha_\mathrm{KI3}$  \\
 
 70 & alpha\_KW3\_0 & & De-reddened spectral index $\alpha_\mathrm{KW3}$ from 2.15 to $\SI{12}{\micro \meter}$ covering VISTA/K$_S$ and the first three WISE bands \\
 71 & e\_alpha\_KW3\_0 & & fitting error of $\alpha_\mathrm{KW3}$  \\
 72 & alpha\_KW\_0 & & De-reddened spectral index $\alpha_\mathrm{KW}$ from 2.15 to $\SI{22}{\micro \meter}$ covering VISTA/K$_S$ and all four WISE bands \\
 73 & e\_alpha\_KW\_0 & & fitting error of $\alpha_\mathrm{KW}$  \\
 74 & alpha\_W13\_0 & & De-reddened spectral index $\alpha_\mathrm{W13}$ from 3.4 to $\SI{12}{\micro \meter}$ covering the first three WISE bands \\
 75 & e\_alpha\_W13\_0 & & fitting error of $\alpha_\mathrm{W13}$  \\
 76 & alpha\_WISE\_0 & & De-reddened spectral index $\alpha_\mathrm{WISE}$ from 3.4 to $\SI{22}{\micro \meter}$ covering the four WISE bands \\
 77 & e\_alpha\_WISE\_0 & & fitting error of $\alpha_\mathrm{WISE}$  \\
 78 & alpha\_KW12M\_0 & & De-reddened spectral index $\alpha_\mathrm{KW12M}$ from 2.15 to $\SI{24}{\micro \meter}$ covering VISTA/K$_S$, the first two WISE bands, and MIPS1 \\
 79 & e\_alpha\_KW12M\_0 & & fitting error of $\alpha_\mathrm{KW12M}$  \\
\hline

 80 & AK\_Herschel &  mag & The extinction extracted from the Herschel map at the position of each source.  \\
 81 & AK\_IR &  mag & The line-of-sight foreground extinction towards each source, mainly obtained by the NICER technique using VISTA NIR data. \\
 & & & See column ``AK\_method'' for more details. \\
 
 82 & AK\_method & & The method, which was used to infer the line-of-sight extinction towards each source. The used methods are: \\
 & & & {\tt LIT(Ref)}: Extinction taken from spectral surveys from the Literature with the Reference in brackets: (4) \citet{Hillenbrand1997}, \\
 & & & (5) \citet{Fang2009,Fang2013}, (6) \citet{Furlan2016};  \\
 & & & {\tt NICER} \citep{Lombardi2001}; {\tt JH} (reddening $E(J-H)$); {\tt HK} (reddening $E(H-K)$); {\tt JK} (reddening $E(J-K)$); {\tt PNICER} \citep{Meingast2017}; \\
 & & & {\tt HERSCHEL} (If the individual line-of-sight extinction is significantly larger than that of the cloud, measured by {\it Herschel} ($A_\mathrm{K,IR} > A_\mathrm{K,Herschel}$), \\
 & & & or when too few bands were available for an extinction calculation, then $A_\mathrm{K,Herschel}$ was used for an estimate of line-of-sight extinction \\
 & & & towards that source, while larger values than $A_\mathrm{K} > \SI{9}{mag}$ were not allowed); \\
 \hline
 
 83 & ClassCog & & VISTA extension flag \{0,1\}. Source morphology derived from variable aperture photometry in combination with machine learning techniques. \\
 & & & 0 indicates an extended object, 1 indicates point-like morphology.  \\
 84 & ClassSex & & VISTA extension flag [0,1]. Determined by the source extraction algorithm SExtractor. Values close to 0 indicate an extended object, \\
 & & & values close to 1 point-like morphology. \\
 
 85 & X & & A reference is given if the source was detected as X-ray source: (7) COUP, \citet{GetmanA2005}; (8) SFINCS, \citet{Getman2017}; \\
 & & & (9) XMM-Newton L1641, \citet{Pillitteri2013}; (10) XMM-Newton $\kappa$-Ori, \citet{Pillitteri2016}. \\
 
 86 & TTS & & Flag if the source was classified as T-Tauri star: C = CTTS, W = WTTS, Ha = H$\alpha$ emission line star, with the reference in brackets: \\
 & & &  (11) \citet{Elek2013}; (12) \citet{Fang2009, Fang2013, Fang2017}; (13) \citet{Hsu2012, Hsu2013}; (14) \citet{DaRio2009};  \\
 & & & (15) \citet{Pettersson2014}. \\
\hline

 87 & C$_\mathrm{MGM}$ & & Classification from MGM including disk sources (D), protostars (P), faint protostar candidates (FP), red protostar candidates (RP). \\
 
 88 & C$_\mathrm{FF16}$ & & Classification for HOPS sources from \citet{Furlan2016} including Class\,0 (0), Class\,I (I), flat spectrum (F), Class\,II (II), \\
 & & & galaxies (G), and unclear objects (U). \\
    
 89 & C$_\mathrm{LL16}$ & & Classification from \citet{Lewis2016} for the 44 low--$A_\mathrm{K}$ MGM protostars. \\
 90 & C$_\mathrm{Fang}$ & & Classification from \citet{Fang2009, Fang2013} for the L1641 region. \\
 91 & C$_\mathrm{P13}$ & & Classification from \citet{Pillitteri2013} for the L1641 region. \\
\hline
 92 & Class & & Classification as proposed in this work including YSO candidates with IR-excess and rejected sources: \\
 & & & Class\,0 (0) or Class\,I protostars (I), flat spectrum sources (F), Class\,II/III pre-main-sequence stars with disks (D), anemic disks (AD), \\
 & & & or transition disks (TD).\\
 & & & For rejected candidates the type of contamination (false positive) is given: \\
 & & & galaxy (G), nebulosity and fuzzy contamination (fuzz), main-sequence star (star), Class\,III candidate without IR-excess (III), \\
 & & & or photometric contamination like image artifact (C). \\
 & & & For uncertain candidates we indicate the suggested uncertain object type as  follows: galaxy candidate ``UG'', or uncertain YSO candidate: \\
 & & & ``UP'', ``UF'', and ``UD'' (depending on the spectral index). \\

 93 & Class\_flag & & Flag to distinguish between: 1 = revisited YSO candidates, 2 = new YSO candidates, 3 = rejected candidates, and 4 = uncertain candidates.\\

\end{longtable}
\end{landscape}

\end{appendix}
\end{document}